\documentclass[
aps, prd, reprint,
10pt, notitlepage, a4paper,
floats, floatfix,
amsmath, amssymb, amsfonts, eqsecnum,
superscriptaddress,
showpacs, showkeys,
nofootinbib,
longbibliography,
]{revtex4-1}

\usepackage{graphicx} 
\usepackage[usenames,dvipsnames]{xcolor}
\usepackage{xspace} 
\usepackage{bm} 
\usepackage{amsmath,amsfonts,amssymb,amsthm}
\usepackage[utf8]{inputenc} 

\xdefinecolor{mylinkcolor}{rgb}{0,0,0.5}
\usepackage[
	bookmarksnumbered, bookmarksopen, bookmarksopenlevel=2,
	breaklinks=true,
	colorlinks=true, filecolor=mylinkcolor, citecolor=mylinkcolor,
	linkcolor=mylinkcolor, urlcolor=mylinkcolor, menucolor=mylinkcolor,
]{hyperref}

\DeclareMathOperator{\Order}{\mathcal{O}}

\allowdisplaybreaks

\newcommand{\AEI}{\affiliation{Max Planck Institute for Gravitational Physics (Albert Einstein Institute), Am M\"uhlenberg 1, Potsdam 14476, Germany}}
\newcommand{\Maryland}{\affiliation{Department of Physics, University of Maryland, College Park, MD 20742, USA}}

\begin{document}

\title{Hairy binary black holes in Einstein-Maxwell-dilaton theory and their effective-one-body description}
\hypersetup{pdftitle={TITLE}}

\author{Mohammed Khalil}\email{mohammed.khalil@aei.mpg.de}\Maryland\AEI
\author{Noah Sennett}\email{noah.sennett@aei.mpg.de}\AEI\Maryland
\author{Jan Steinhoff}\email{jan.steinhoff@aei.mpg.de}\AEI
\author{Justin Vines}\email{justin.vines@aei.mpg.de}\AEI
\author{Alessandra Buonanno}\email{alessandra.buonanno@aei.mpg.de}\AEI\Maryland

\date{\today}

\begin{abstract}

In General Relativity and many modified theories of gravity, isolated black holes (BHs) cannot source massless scalar fields.
Einstein-Maxwell-dilaton (EMd) theory is an exception: through couplings both to electromagnetism and (non-minimally) to gravity, a massless scalar field can be generated by an electrically charged BH.
In this work, we analytically model the dynamics of binaries comprised of such scalar-charged (``hairy'') BHs.  While BHs are not expected to have substantial electric charge within the Standard Model of particle physics, nearly-extremally charged BHs could occur in models of minicharged dark matter and dark photons.
We begin by studying the test-body limit for a binary BH in EMd theory, and we argue that only very compact binaries of nearly-extremally charged BHs can manifest non-perturbative phenomena similar to those found in certain scalar-tensor theories.
Then, we use the post-Newtonian approximation to study the dynamics of binary BHs with arbitrary mass ratios.
We derive the equations governing  the conservative and dissipative sectors of the dynamics at next-to-leading order, use our results to compute the Fourier-domain gravitational waveform in the stationary-phase approximation, and compute the number of useful cycles measurable by the Advanced LIGO detector.
Finally, we construct two effective-one-body (EOB) Hamiltonians for binary BHs in EMd theory: one that reproduces the exact test-body limit and another whose construction more closely resembles similar models in General Relativity, and thus could be more easily integrated into existing EOB waveform models used in the data analysis of gravitational-wave events by the LIGO and Virgo collaborations.
\end{abstract}


\maketitle

\section{Introduction}
\label{sec:Intro}

The first observations of gravitational waves (GWs) from coalescing binary black holes  (BHs) \cite{Abbott:2016blz, Abbott:2016nmj, Abbott:2017vtc, Abbott:2017gyy, Abbott:2017oio} and neutron stars \cite{TheLIGOScientific:2017qsa} offer unprecedented opportunities to test the highly dynamical, strong-field regime of General Relativity (GR) \cite{Will:2014kxa, Berti:2015itd, Yunes:2016jcc}.
Leveraging the extraordinary precision of GW detectors to test gravity requires waveform models that incorporate potential deviations from GR.
One can construct such models in a theory-independent way by considering phenomenological deviations to waveform models in GR and then constraining the magnitude of these corrections, see, e.g., the constructions of \cite{Arun:2006hn, Yunes:2009ke, Mishra:2010tp, Li:2011cg}.
Such an approach has been used by the LIGO and Virgo collaborations to test GR with binary BHs \cite{TheLIGOScientific:2016src,TheLIGOScientific:2016pea,Abbott:2017vtc}.
Alternatively, one can compute the waveform produced in a particular alternative theory, which can then be used to measure directly the fundamental quantities that define that modified theory of gravity~\cite{Will:2014kxa}.

In this paper, we adopt the latter approach, focusing on the dynamics of binary BHs in Einstein-Maxwell-dilaton (EMd) theory.
This theory originated as a low-energy limit of string theory \cite{Gibbons:1987ps, Garfinkle:1990qj}.
In EMd theory, a scalar field (the dilaton) couples to a vector field (the photon) such that BHs with electric charge also source the scalar; the BH develops a scalar charge, or hair.
It has been shown that in GR (and some scalar extensions)  isolated BHs cannot carry such a charge \cite{Hawking:1972qk,Sotiriou:2011dz}; these results are often referred to as ``no-hair theorems.'' 
Analytic solutions exist in EMd theory for  spherically symmetric BHs parameterized by the dilaton coupling constant $a$ [see Eqs.~\eqref{eq:JordanFrameAction} and~\eqref{EMdaction} below for the action].
For $a=0$, the theory reduces to Einstein-Maxwell (EM) theory and the BH solution is the Reissner-Nordstr\"{o}m metric. 
For $a=1$, the solution corresponds to the low energy limit of heterotic string theory.
For $a=\sqrt{3}$, the solution corresponds to Kaluza-Klein BHs \cite{Horne:1992zy}, and an analytic solution for charged spinning BHs in EMd theory is only known for that value of $a$ \cite{Frolov:1987rj}.

In the absence of electric charge, isolated BHs in EMd theory behave as in GR.
Within the Standard Model, astrophysical BHs are expected to be electrically neutral; however, there exist various theoretical mechanisms beyond the Standard Model that would allow BHs to accumulate non-negligible charge.
For a BH with charge $Q$ and mass $M$ to accrete a particle with the same-sign charge $q$ and mass $m$, gravitational attraction between the two bodies must overpower their electrostatic repulsion, i.e., $q\,Q\lesssim m\,M$, or equivalently  $Q/M\lesssim m/q$.\footnote{Throughout this work, we use geometric units, in which $G=c=4\pi \epsilon_0=1$, where $G$ is the bare gravitational constant.}
Furthermore, a charged BH will neutralize via spontaneous pair production~\cite{Gibbons:1975kk} or interactions with astrophysical plasmas~\cite{Eardley:1975kp} over timescales that grow with the mass-to-charge ratio of the available fundamental particles.
For electrons, the dimensionless mass-to-charge ratio $m_e/q_e\sim 10^{-22}$ severely limits the charge that BHs can develop through accretion, and guarantees that any BH charged through other means will discharge quickly.
However, particles with much larger mass-to-charge ratios are predicted in models of minicharged dark matter \cite{DeRujula:1989fe, Perl:1997nd, McDermott:2010pa} and would allow BHs to acquire and retain a much larger electric charge \cite{Cardoso:2016olt}.
Similarly, models in which dark matter is charged under a hidden U(1) gauge field \cite{Ackerman:mha,Feng:2008mu,Feng:2009mn}, a ``dark photon,'' would allow for BHs to develop significant hidden charge, provided that the ratio of the dark-matter particle's mass to its (hidden) charge is sufficiently large~\cite{
Cardoso:2016olt}.
These two types of dark matter models are consistent with laboratory experiments and cosmological observations \cite{Davidson:2000hf, Burrage:2009yz, Ahlers:2009kh}; current constraints restrict the new particles' mass to 1 GeV $\lesssim m \lesssim$ 10 TeV \cite{Feng:2009mn} and its charge to $\lesssim 10^{-14} (m/\text{GeV}) q_e$ \cite{Kadota:2016tqq} (see also Fig. 1 in Ref.~\cite{Cardoso:2016olt}).

The dynamical evolution of binary BHs in EMd theory has been studied in various contexts.
Numerical-relativity simulations of single and binary BHs were performed in Ref.~\cite{Hirschmann:2017psw}.
The authors considered small electric charges and  found that the resulting gravitational waveforms are difficult to distinguish from those in GR.
Numerical-relativity simulations of the collision of charged BHs with large electric charges in EM theory were performed in Refs.~\cite{Zilhao:2012gp, Zilhao:2013nda}, where it was found that a significant fraction of the energy is carried away by electromagnetic radiation.

In this work, we compute the conservative and dissipative dynamics of
a binary BH, and the resulting gravitational waveform, in EMd theory,
to first order in the (weak-field and slow-motion) post-Newtonian (PN)
approximation.  We also construct an effective-one-body (EOB)
Hamiltonian description \cite{Buonanno:1998gg,Buonanno:2000ef} of the
conservative dynamics, which provides an analytical resummation of the
PN dynamics to exactly recover the test-body limit. In late 2017, the 1PN Lagrangian for a two-body system in
EMd theory was derived independently in Ref.~\cite{Julie:2017rpw}
using a method different from our own.  In that work, the author also
discussed an abrupt transition in the scalar charge of a BH as the
external scalar field is varied.  However, we show here that this
transition occurs only in binaries composed of nearly-extremal 
charged BHs and only near the end of their coalescence. Although extremally 
charged BHs are excluded when restricting to the Standard Model of particle physics, they 
are still viable in minicharged dark matter and dark photons models, as we have discussed above.
 
The paper is structured as follows.
In Sec.~\ref{sec:singleBH}, we study the behavior of a small BH in the background of a much more massive companion.
By exploring the response of this test BH to its external environment, we discuss whether non-perturbative, strong-field phenomena, akin to those seen in binary neutron stars in scalar-tensor (ST) theories, can occur in binary BHs in EMd theory. In Sec.~\ref{sec:PNapprox}, we use the PN approximation to study the dynamics of a binary system with an arbitrary mass ratio.
We derive the two-body 1PN Lagrangian and Hamiltonian (with details relegated to Appendix~\ref{app:lag}) and calculate the scalar charge of the two bodies.
Further, we derive (with details in Appendix \ref{app:flux}) the next-to-leading order PN scalar, vector, and tensor energy fluxes emitted by the binary.
Restricting our attention to quasi-circular orbits, we compute the Fourier-domain gravitational waveform at next-to-leading-order using the stationary-phase
approximation. In Sec.~\ref{sec:EOB}, we work out an EOB description of the PN Hamiltonian in EMd theory.
We construct two EOB Hamiltonians: one based on the exact BH solution, and the other based on an approximation to that solution.
The former is more physical in the strong-gravity regime because it exactly reproduces the dynamics in the test-body limit; the latter uses the same gauge as EOB models in GR, and thus would be easier to integrate into existing data-analysis infrastructure.
We compare these two EOB Hamiltonians by calculating the binding energy and the innermost stable circular orbit to determine the region of the parameter space in which they agree. Finally, we present some concluding remarks in Sec.~\ref{sec:conclusions}.

\section{Einstein-Maxwell-dilaton theory}
\label{sec:singleBH}

\subsection{Setup}
We consider a generalization of EMd theory presented in Refs.~\cite{Garfinkle:1990qj, Gibbons:1987ps} in the Jordan frame
\begin{align}\label{eq:JordanFrameAction}
S&=\int d^4x \, \frac{\sqrt{-\tilde{g}}}{16\pi}e^{-2 a \varphi} \left(\tilde{R} + (6 a^2 -2) \tilde{g}^{\mu\nu}\tilde{\nabla}_\mu\varphi\tilde{\nabla}_\nu\varphi \right. \nonumber\\
&\qquad \left. -F_{\mu\nu}F^{\mu\nu}\right) +S_m(\tilde{g}_{\mu\nu},A_\mu, \psi),
\end{align}
where $\varphi$ is a scalar field (the dilaton), $a$ is the dilaton coupling constant, $F_{\mu\nu}\equiv \tilde{\nabla}_\mu A_\nu - \tilde{\nabla}_\nu A_\mu$ is the electromagnetic field tensor, and tildes signify quantities in the Jordan frame.
We also include some matter fields $\psi$, which couple minimally to $\tilde{g}_{\mu \nu}$ and, through some fundamental electric charge, to $A_\mu$; we represent this total matter action schematically with $S_m$.
By construction, electrically neutral, non-self-gravitating matter configurations will follow geodesics of $\tilde{g}_{\mu \nu}$, and thus this theory respects the weak equivalence principle.
However, self-gravitating systems  are bound (in part) through non-linear interactions of the scalar field.
The back-reaction of the scalar field on the metric exerts an additional force on such systems, causing them to no longer follow geodesics; thus, this theory violates the strong equivalence principle.

The Einstein frame provides a more convenient representation of EMd theory.
Performing the conformal transformation $g_{\mu\nu}=\mathcal{A}^{-2}(\varphi)\tilde{g}_{\mu\nu}$ with $\mathcal{A}=e^{a\varphi}$, the action becomes
\begin{align}\label{EMdaction}
S&=\int d^4x \, \frac{\sqrt{-g}}{16\pi} \left(R -2g^{\mu\nu}\partial_\mu\varphi\partial_\nu\varphi  -e^{-2a\varphi}F_{\mu\nu}F^{\mu\nu}\right) \nonumber\\
&\quad +S_m(\mathcal{A}^2(\varphi)g_{\mu\nu},A_\mu, \psi),
\end{align}
where $g_{\mu \nu}$ is the Einstein-frame metric.
In this paper, we primarily work in the Einstein frame, but occasionally use quantities in the Jordan frame, denoted with tildes.
For a discussion of the equivalence between the two frames see Ref.~\cite{Flanagan:2004bz}.

For the matter action $S_{m}$, we adopt the approach introduced by Eardley \cite{eardley1975observable}, in which each body is treated as a delta function and the dependence on the scalar field is incorporated into the masses. 
For charged monopolar point particles, neglecting dipoles/spins and higher multipoles,
the matter action in the Einstein frame can be written as \cite{Damour:1992we}
\begin{equation} \label{Smatter}
S_{m} = -\sum_A \int dt \left[ \mathfrak{m}_A(\varphi) \sqrt{-g_{\mu\nu}\,v_A^{\mu}v_A^{\nu}}
- q_A A_\mu v^\mu_A \right] ,
\end{equation}
where $\mathfrak{m}_A(\varphi)$ is the field-dependent mass of particle $A$, $q_A$ is the electric charge, $v_A^\mu \equiv u^\mu_A/u^0_A$ where $u^\mu_A$ is its four-velocity, and the fields are evaluated at the  particle's location.
The mass in the Einstein frame $\mathfrak{m}(\varphi)$ is related to the mass in the Jordan-Fierz frame $\tilde{\mathfrak{m}}(\varphi)$ by
\begin{equation}
\label{mJordan}
\mathfrak{m}(\varphi)=\mathcal{A}(\varphi)\tilde{\mathfrak{m}}(\varphi),
\end{equation} 
where $\tilde{\mathfrak{m}}(\varphi)$ is generally not a constant except for bodies with negligible self-gravity. 

In most cases, a closed-form expression for the field-dependent mass $\mathfrak{m}(\varphi)$ cannot be found. Instead, one expands the mass about the external/background value $\varphi_0$ of the scalar field
\begin{align}
\ln \mathfrak{m}(\varphi)&=\ln \mathfrak{m}(\varphi_0) + \left. \frac{d\ln \mathfrak{m}(\varphi)}{d\varphi}\right|_{\varphi_0} \delta\varphi   \nonumber\\
&\quad 
+\frac{1}{2} \left.\frac{d^2\ln \mathfrak{m}(\varphi)}{d\varphi^2}\right|_{\varphi_0} \delta\varphi^2+ \Order\left(\frac{1}{c^6}\right)\,,
\end{align}
where $\delta\varphi\equiv \varphi-\varphi_0$. 
The mass expansion can be parameterized in terms of
\begin{equation} \label{defalpha}
\alpha(\varphi)\equiv \frac{d\ln \mathfrak{m}(\varphi)}{d\varphi}\,, \qquad \beta(\varphi)\equiv \frac{d\alpha(\varphi)}{d\varphi}\,,
\end{equation}
where $\alpha$ is referred to as the (dimensionless) scalar charge.
With these parameters, the mass expansion can be written as
\begin{equation} \label{Mphi} 
\mathfrak{m}(\varphi)=m\left[1+\alpha\delta\varphi+\frac{1}{2}(\alpha^2+\beta)\delta\varphi^2+ \Order \left (\frac{1}{c^6} \right )\right], 
\end{equation}
where the field-dependent mass is denoted by the Gothic script $\mathfrak{m}$, while the mass evaluated at the background value of the scalar field is denoted by $m$.
We also drop the dependence of the parameters on the background value to simplify the notation, i.e., $\alpha\equiv\alpha(\varphi_0)$, and $\beta\equiv\beta(\varphi_0)$. 
For the field-dependent parameters, we always explicitly write $\alpha(\varphi)$ and $\beta(\varphi)$.
The expression for $\alpha(\varphi)$ depends on the structure of the body; for static BHs, it depends only on the charge-to-mass ratio, whereas for baryonic matter, it also depends on the body's composition.

We note that Eq.~\eqref{Smatter} together with the expansion of the mass \eqref{Mphi} provide a systematic construction of an effective source or action for an extended object in a  PN expansion.
We neglect couplings to derivatives of the field, which would  correspond to dipole/spin and higher multipole interactions. 
Due to invariance under gauge transformations $A_\mu \rightarrow A_\mu + \partial_\mu \epsilon$, the charges $q_A$ must be constant; they cannot depend on the scalar field like the masses.

\subsection{Black-hole solution}

The metric for an electrically-charged non-rotating BH in EMd theory is given by \cite{Garfinkle:1990qj, Gibbons:1987ps} 
\begin{equation} \label{EMdmetric}
ds^2=-A(r)dt^2+B(r)dr^2+r^2C(r)d\Omega^2,
\end{equation}
with
\begin{subequations}
\begin{align}
A(r) &= \left(1-\frac{r_+}{r}\right)\left(1-\frac{r_-}{r}\right)^{\frac{1-a^2}{1+a^2}},\\ 
B(r) &= \frac{1}{A(r)},\\
C(r) &= \left(1-\frac{r_-}{r}\right)^{\frac{2a^2}{1+a^2}},
\end{align}
\end{subequations}
where the constants $r_+$ and $r_-$ are given in terms of the Arnowitt-Deser-Misner mass $M$ and electric charge $Q$ by 
\begin{align}
M&=\frac{r_+}{2}+\left(\frac{1-a^2}{1+a^2}\right)\frac{r_-}{2}\,, \\
Q^2&=\frac{r_+r_-}{1+a^2} \, e^{-2a\varphi_0}\,.
\end{align}
The constant $r_+$ corresponds to the outer horizon, and $r_-$ corresponds to the inner horizon. The surface area of the horizon (entropy of the BH) is proportional to $r_+^2 C(r_+)$. Here, we refer to the metric~\eqref{EMdmetric} as the GHS metric, 
after Garfinkle, Horowitz and Strominger who found the solution in that form in Ref.~\cite{Garfinkle:1990qj}.

The electromagnetic four-potential $A_\mu$, for an electrically-charged BH, is given by
\begin{equation}
A_0(r) = -\frac{Q}{r}e^{2a\varphi_0}, \qquad 
A_i(r) = 0\,,
\end{equation}
and the scalar field $\varphi$ is given by
\begin{equation}\label{phiBH}
\varphi(r)=\varphi_0+\frac{a}{1+a^2}\ln\left(1-\frac{r_-}{r}\right).
\end{equation}
While we consider only electric charges in this paper, we note that the solution for a magnetically charged BH can be obtained from the above solution via the duality rotation that sends $F_{\mu\nu}\to \frac{1}{2}e^{-2a\varphi}{\epsilon_{\mu\nu}}^{\lambda\rho}F_{\lambda\rho}$ and $\varphi\to -\varphi$.~\footnote{
The results of Sec.~\ref{sec:singleBH} hold also for magnetic charges if we flip the sign of $\varphi$. However, the PN and EOB results in the following sections would change in non-trivial ways for the magnetic case, since the BH's $F_{\mu\nu}$ is given by $F_{\theta\phi}=Q_m\sin\theta$ with a magnetic charge $Q_m$, as opposed to $F_{tr}=Q/r^2$ with an electric charge $Q$ (all other components being zero in each case).}
In addition to the electric charge, BHs in EMd theory can acquire scalar charge, also called dilaton charge, defined by \cite{Garfinkle:1990qj}
\begin{equation}
D\equiv\frac{1}{4\pi}\int d^2\Sigma^\mu\nabla_\mu\varphi,
\end{equation}
where the integral is over a two-sphere at spatial infinity, leading to
\begin{equation}
D = \frac{a}{1+a^2}r_-\,.
\end{equation}
Far from the BH, we have $\varphi(r)\simeq\varphi_0-D/r+\mathcal{O}(1/r^2)$, which means that 
$D$ acts as the monopole charge sourcing the scalar field.

The constants $r_+$ and $r_-$ can be expressed in terms of the mass and the dilaton charge, or the mass and electric charge, as
\begin{subequations}
\begin{align}
r_-&= \frac{1+a^2}{a} D \nonumber\\
&= \frac{1+a^2}{1-a^2} \left( M - \sqrt{M^2 - (1-a^2) Q^2 e^{2a\varphi_0}} \right),  
\label{rminus} \\
r_+&=2M - \frac{1-a^2}{a}D \nonumber\\ 
&= M+\sqrt{M^2-(1-a^2)Q^2e^{2a\varphi_0}}\,.
\label{rplus} 
\end{align}
\end{subequations}
Expressing quantities in terms of the dilaton charge $D$, rather than the electric charge $Q$, makes most equations simpler as it avoids the square root. Therefore, in most of the equations below, we use $D$ instead of $Q$.
The relation between $Q$ and $D$ can be read off from Eq.~\eqref{rminus}, or Eq.~\eqref{rplus},
\begin{equation}
\label{QD}
Q^2 \,e^{2a\varphi_0} = \frac{2M}{a} D - \frac{1-a^2}{a^2} D^2\,.
\end{equation}
The maximum electric charge of the BH occurs when $r_+=r_-$, which leads to
\begin{equation}
\label{Qmax}
Q_\text{max}\,e^{a\varphi_0} = \sqrt{1+a^2}M.
\end{equation} 
Hence, for nonzero values of $a$, an EMd BH can be more charged than an extremal Reissner-Nordstr\"{o}m BH with the same mass.
Since the dilaton charge is related to the electric charge via Eq.~\eqref{QD}, the maximum electric charge \eqref{Qmax} corresponds to the maximum dilaton charge $D_\text{max} = a M$.

Without loss of generality, we set the background scalar field to zero, i.e., $\varphi_0=0$.
To recover the dependence on $\varphi_0$, one can simply rescale all electric charges by the factor $e^{a\varphi_0}$, and add the constant $\varphi_0$ to the scalar field.\footnote{
To see why this is true, consider the action \eqref{EMdaction} with the transformation $Q\to Qe^{a\varphi_0}$ and $\varphi\to \varphi+\varphi_0$.
The vacuum part of the action is symmetric under that transformation, and in the matter action \eqref{Smatter}, the mass $\mathfrak{m}(\varphi)$ is parameterized in terms of the difference $\varphi-\varphi_0$. 
The electromagnetic part of the matter action is more subtle; it depends on $qv^\mu A_\mu\propto Qqe^{2a\varphi_0}/r$, and hence, one can absorb a factor of $e^{a\varphi_0}$ into each of the two charges. However, since $A_0=-Qe^{2a\varphi_0}/r$, the transformation $Q\to Qe^{a\varphi_0}$, $\varphi\to \varphi+\varphi_0$ is not valid in equations that depend on  $A_\mu$; one first needs to express $A_\mu$ in terms of the charges before performing that transformation.}
We also consider only non-negative values of $a$ since the action~\eqref{EMdaction} is invariant under $a \rightarrow -a$ and $\varphi \rightarrow -\varphi$, so the predictions for negative dilaton couplings are given by changing the sign of the scalar field.

\begin{figure*}[th]
	\centering
	\begin{minipage}[b]{0.49\linewidth}
		\includegraphics[width=\linewidth]{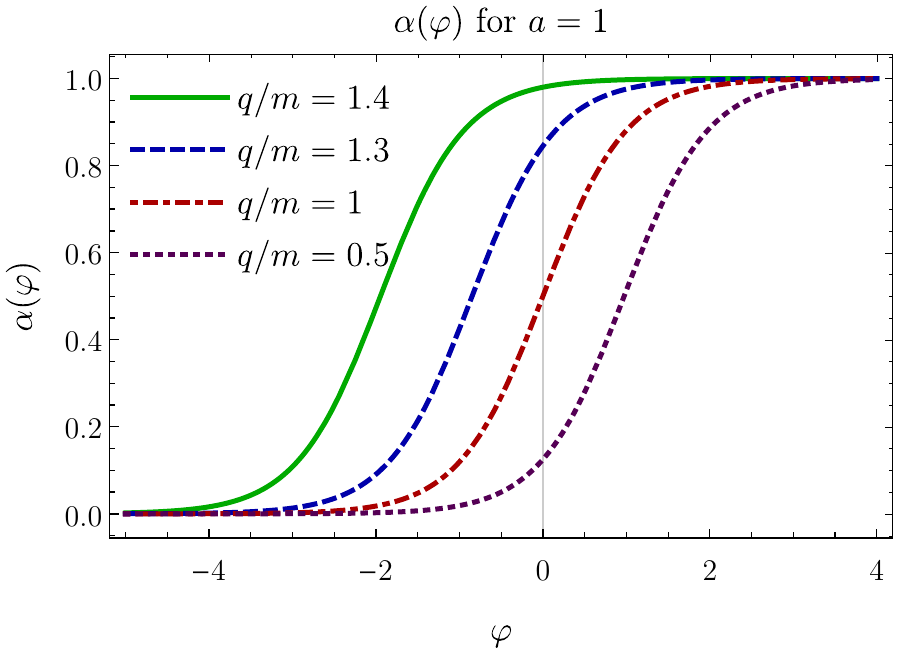} 
	\end{minipage}
	\begin{minipage}[b]{0.49\linewidth}
		\includegraphics[width=\linewidth]{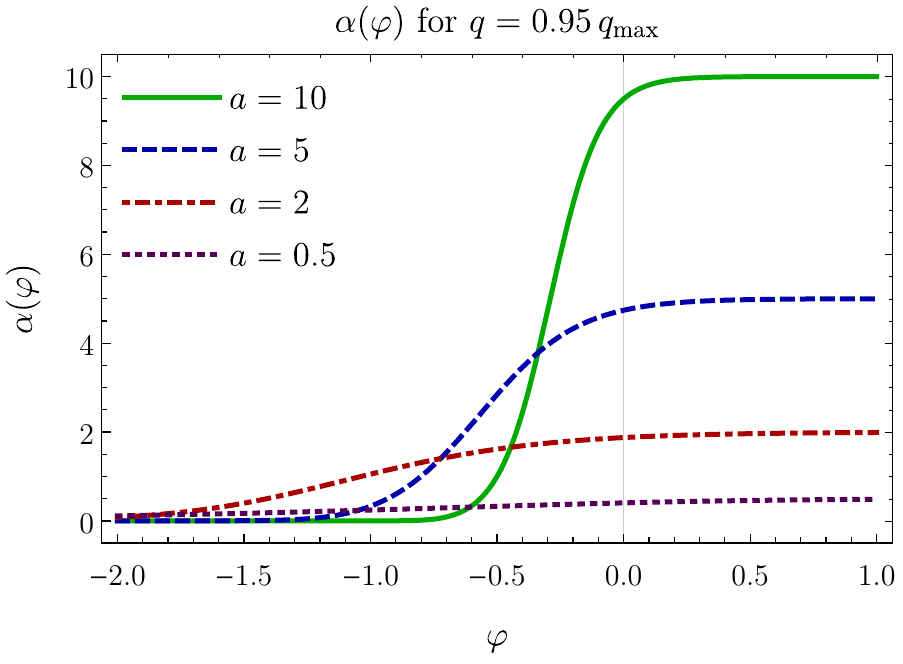}
	\end{minipage}	
	\caption{\label{fig:alpha} 
		$\alpha(\varphi)$ for $a=1$ with different charge-to-mass ratios (left), and for different values of $a$ with $q=0.95\,q_\text{max}$ (right).}
	
	\begin{minipage}[b]{0.49\linewidth}
		\includegraphics[width=\linewidth]{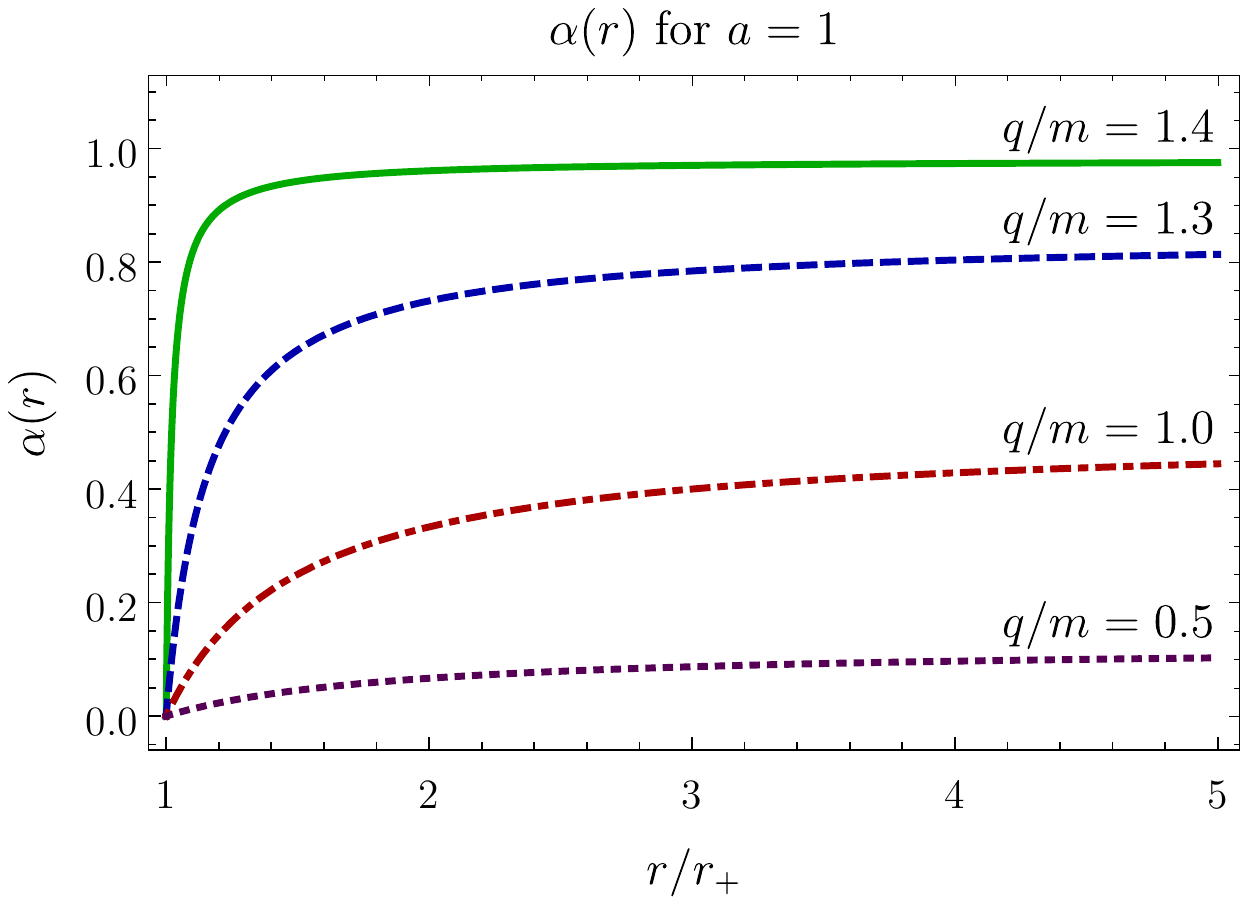} 
	\end{minipage}
	\begin{minipage}[b]{0.49\linewidth}
		\includegraphics[width=\linewidth]{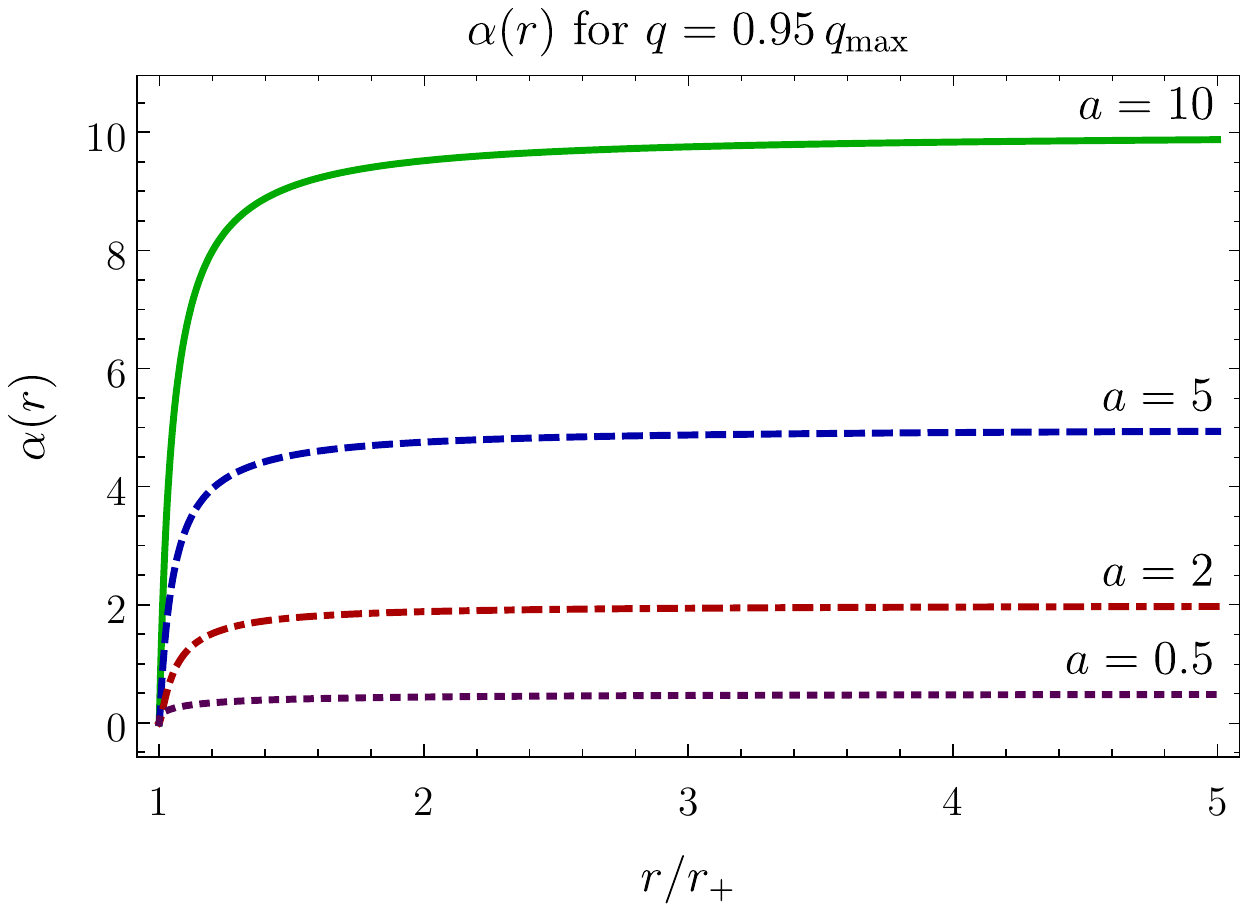}
	\end{minipage}
	\caption{\label{fig:alphar} 
		$\alpha(r)$ for $a=1$ with different charge-to-mass ratios (left), and for different values of $a$ (right). In both plots, the charge of the large BH is extremal $Q=\sqrt{1+a^2}M$, and $r$ is scaled by the horizon radius, which is given by Eq.~\eqref{rplus}. For $a=1$, the horizon radius is $2M$ independently of the charge or the coupling constant.}
\end{figure*}

\subsection{Dynamics of a test black-hole in a background black-hole spacetime}

Before turning to the dynamics of a generic two-BH system in EMd theory, it will be useful to study the test-body limit of such a system, i.e., the limit in which one body's mass is negligible compared to the other's.  In EM theory (without the dilaton), the test-body limit of a charged BH corresponds simply to a monopolar point-mass with constant mass and constant charge.  In EMd theory, however, a BH's mass must retain a dependence on the dilaton field even as its size goes to zero.  In the zero-size limit, we can use the local value of the (background) dilaton field $\varphi$, at the small BH's location, to determine its mass $\mathfrak{m}(\varphi)$ in the same way that a lone finite-size BH's mass would be determined by the asymptotic value of the field (as in the previous subsection).  This defines what we mean by a ``test BH'' in EMd theory.\footnote{
This is not to be confused with some uses of the phrase ``test body" in the context of ST theories, where one means a body with negligible self-gravity (unlike a BH), so that the mass in the Jordan-Fierz frame is constant and the scalar charge is zero.
}

Let us suppose a test BH with mass $\mathfrak{m}(\varphi)$, electric charge $q$, and dilaton charge $d$ moves in the fixed background spacetime of a larger BH with mass $M$, electric charge $Q$, and dilaton charge $D$.
The mass of the test BH $\mathfrak{m}(\varphi)$ depends on the scalar field $\varphi$ generated by the larger BH. The expansion of $\mathfrak{m}(\varphi)$ is given in terms of the parameters $\alpha$ and $\beta$ by Eq.~\eqref{Mphi}, and the scalar field $\varphi$ is given by Eq.~\eqref{phiBH}.

To find how $\alpha$ and $\beta$ depend on the mass and charge of the BH, one needs to find the dependence of the mass  on the scalar field.
We can get a differential equation for $\mathfrak{m}(\varphi)$ from Eq.~\eqref{rminus}, or Eq.~\eqref{rplus}, by identifying the mass $M$ and charge $Q$ with those of the test BH, i.e., $M \rightarrow \mathfrak{m}(\varphi)$ and $Q \rightarrow q$. The background value of the scalar field can be identified with the field from the more massive BH $\varphi_0 \rightarrow \varphi$, and the scalar charge by  $D \rightarrow d \mathfrak{m}(\varphi) / d \varphi$, as was shown by the matching conditions in Ref.~\cite{Julie:2017rpw}. This leads to the equation
\begin{equation} \label{difm}
\frac{d \mathfrak{m}(\varphi)}{d \varphi}
= \frac{a}{1-a^2} \left[\mathfrak{m}(\varphi) - \sqrt{\mathfrak{m}(\varphi)^2 - (1-a^2) q^2 e^{2a\varphi}} \right],
\end{equation}
which, as far as we know, has no analytic solution for arbitrary values of $a$. Nevertheless, we can still obtain an expression for the dimensionless scalar charge, which is defined by  Eq.~\eqref{defalpha},
\begin{equation} \label{alphaphi}
\alpha(\varphi) = \frac{a}{1-a^2} \left[ 1 - \sqrt{1 - (1-a^2) \frac{q^2e^{2a\varphi}}{\mathfrak{m}^2(\varphi)}} \,\right] ,
\end{equation}
and
\begin{equation}  \label{betaphi}
\beta(\varphi)  = \frac{a^2q^2e^{2a\varphi}}{(1-a^2)\mathfrak{m}^2(\varphi)} \left[1-\frac{a^2}{\sqrt{1-(1-a^2)\frac{q^2e^{2a\varphi}}{\mathfrak{m}^2(\varphi)}}}\right],
\end{equation}
in agreement with Ref.~\cite{Julie:2017rpw}.  

It is interesting to note that an exact analytic solution to the differential equation \eqref{difm} can be found when the coupling constant $a=1$, that is
\begin{equation}
\mathfrak{m}(\varphi)=\sqrt{\text{const.} + \frac{1}{2}q^2e^{2\varphi}}\,.
\end{equation}
Since the above expression should give $m$ when $\varphi=0$, the integration constant is found to be $m^2-\frac{1}{2}q^2$. Hence,
\begin{equation}
\label{ma=1}
\mathfrak{m}(\varphi)=\sqrt{m^2 - \frac{1}{2}q^2 +  \frac{1}{2}q^2e^{2\varphi} }\,.
\end{equation}
By differentiating $\mathfrak{m}(\varphi)$, we get the parameters 
\begin{equation}
\label{sensitiva1}
\alpha=\frac{q^2}{2m^2}\,, \qquad \beta=\frac{q^2}{m^2}-\frac{q^4}{2m^4}\,.
\end{equation}

In Fig.~\ref{fig:alpha}, we plot $\alpha(\varphi)$ as a function of $\varphi$.
We see that the test BH's $\alpha(\varphi)$ transitions between two values: zero and $a$.
The function $\alpha(\varphi)$ reaches its maximum value when the quantity $q^2\,e^{2a\varphi}/\mathfrak{m}^2$ approaches $1+a^2$, which means that in the Jordan-Fierz frame, the charge $q$ approaches the extremal value $\sqrt{1+a^2}\tilde{\mathfrak{m}}$, where the mass in the Jordan-Fierz frame $\tilde{\mathfrak{m}}$ is given by Eq.~\eqref{mJordan}.
Changing the charge-to-mass ratio shifts the curve on the horizontal axis, while changing $a$ changes the maximum value of $\alpha$ and determines how quickly this transition occurs.

We emphasize that the scalar field $\varphi$ generated by the more massive BH is always negative, as can be seen from Eq.~\eqref{phiBH}, so the test BH always descalarizes.
Further, because of the logarithm, the magnitude of $\varphi$ increases slowly with decreasing separation until $r$ approaches the inner horizon $r_-$, where it diverges.
For the scalar charge of the test BH to change dramatically before merging with its much larger companion, both BHs must be close to extremally charged.
As discussed in Sec. \ref{sec:Intro}, extremally-charged BHs an exist in minicharged dark matter and dark photon models.
If the test BH is not sufficiently charged, its scalar charge is close to zero when well separated from its companion, and then monotonically decreases toward zero as the binary evolves.
The total shift in the scalar field that the test BH experiences prior to crossing the outer horizon is given by
\begin{equation}
\varphi(r_+)-\varphi(\infty)=\frac{a}{1+a^2}\ln\left[\frac{1-D/D_\text{max}}{1-(1-a^2)D/2D_\text{max}}\right].
\end{equation}
Thus, if the large BH is not also sufficiently charged, then the test BH's scalar charge does not change dramatically.

In Fig.~\ref{fig:alphar}, we substitute the expression for the scalar field of the larger BH $\varphi(r)$ into that for the scalar charge of the test BH $\alpha(\varphi)$, and plot $\alpha(r)$ versus the separation $r$ scaled by the horizon radius.
When setting the charge of the large BH to its extremal value, $Q=\sqrt{1+a^2}M$, we see that the charge of the test BH also needs to be near extremal for the descalarization transition to occur.
Yet, the transition only occurs very close to the horizon of the background BH.
Hence, we expect this descalarization to drastically affect the GW signature only during the late inspiral and plunge of a test BH into a more massive BH and only when the BHs are nearly-extremally charged, when the horizon, the innermost-stable circular orbit, and the divergence in $\varphi$ coincide.
This result is analogous to extremal Kerr BHs, where the plunge occurs at significantly smaller separations \cite{Taracchini:2014zpa}.
However, a comparable-mass binary does not perform many orbits at small separations due to stronger radiation reaction, and thus we expect that the transition in the scalar charge would have a negligible effect on GWs from the inspiral of a comparable-mass binary.

We note that, while the descalarization transition occurs for near-extremal BHs, the largest change in the value of $\alpha$ from infinity until, e.g., $r= 2r_+$ occurs when the electric charge is  $q/m\sim 1$, as can be seen in the left panel of Fig.~\ref{fig:alphar}. 
This is due to the slope of $\alpha(\varphi)$ at the background value of the scalar field $\varphi_0=0$.
So, in order to increase the change in the scalar charge to observe descalarization, it is important to have a maximal $\beta(\varphi)\equiv d\alpha(\varphi)/d\varphi$.

\subsection{Compact objects in Einstein-Maxwell-dilaton and scalar-tensor theories}

Certain ST theories can exhibit non-perturbative phenomena, known as induced or
	dynamical scalarization, in binary systems of neutron
	stars~\cite{Barausse:2012da,Shibata:2013pra,Palenzuela:2013hsa,Taniguchi:2014fqa}.  Having
	established how a BH responds to its scalar environment, we now investigate whether such
	effects could arise in binary BHs in EMd theory. In Ref.~\cite{Hirschmann:2017psw}, the authors
	suggested that dynamical and induced scalarization are much less significant in EMd theory than in
	ST theories. In this subsection, we support this claim using more quantitative arguments by directly comparing the
	behavior of BHs and neutron stars in the respective theories.

In Ref.~\cite{Sennett:2016rwa}, the authors argued that the onset of induced and dynamical
	scalarization coincide with a breakdown of the PN approximation. Specifically, these
	non-perturbative phenomena indicate that the scalar field has grown beyond the validity of
	a PN expansion of $\mathfrak{m}$, e.g., Eq.~\eqref{Mphi}. A useful diagnostic for determining the
	onset of such phenomena is to compare the relative size of the coefficients of such a power series
	to the small parameter with which one constructs the expansion.

While both EMd theory and ST theories include an additional scalar field, the
	non-minimal coupling of that field to the Jordan-Fierz  (physical) metric can differ
	substantially. To facilitate comparisons between these theories, we consider an expansion of
	$\mathfrak{m}$ in $G_\text{N}(\varphi)$, the parameter that characterizes the gravitational force
	felt between two test bodies placed in the scalar background $\varphi$. In both EMd theory and
	ST theories, this Newton's ``constant'' is given by
	\begin{align}
	G_\text{N}(\varphi)\equiv \mathcal{A}^2(\varphi)\left[1+\left(\frac{d \log \mathcal{A}}{d \varphi}\right)^2\right].\label{eq:GNdef}
	\end{align}
	We expand $\mathfrak{m}$ in terms of this quantity
	\begin{align}
	\mathfrak{m}(G_\text{N})=m\left[1+C_1\left(\frac{G-G_\text{N}^0}{G_\text{N}^0}\right)+C_2\left(\frac{G-G_\text{N}^0}{G_\text{N}^0}\right)^2+\ldots\right],
	\end{align}
	where we have defined
\begin{subequations}
	\begin{align}
	G_\text{N}^0 &\equiv  G_\text{N}(\varphi=0),\\
	C_1&\equiv \left[\frac{d \log \mathfrak{m}}{d \log G_\text{N}}\right]_{G_\text{N}=G_\text{N}^0},\label{eq:c1def}\\
	C_2&\equiv \frac{1}{2}\left[\frac{d^2 \log \mathfrak{m}}{(d \log G_\text{N})^2}+\left(\frac{d \log \mathfrak{m}}{d \log G_\text{N}}\right)^2 - \frac{d \log \mathfrak{m}}{d \log G_\text{N}}  \right]_{G_\text{N}=G_\text{N}^0}.\label{eq:c2def}
	\end{align}
\end{subequations}

\begin{figure}[t]
	\centering
	\includegraphics[width=\columnwidth]{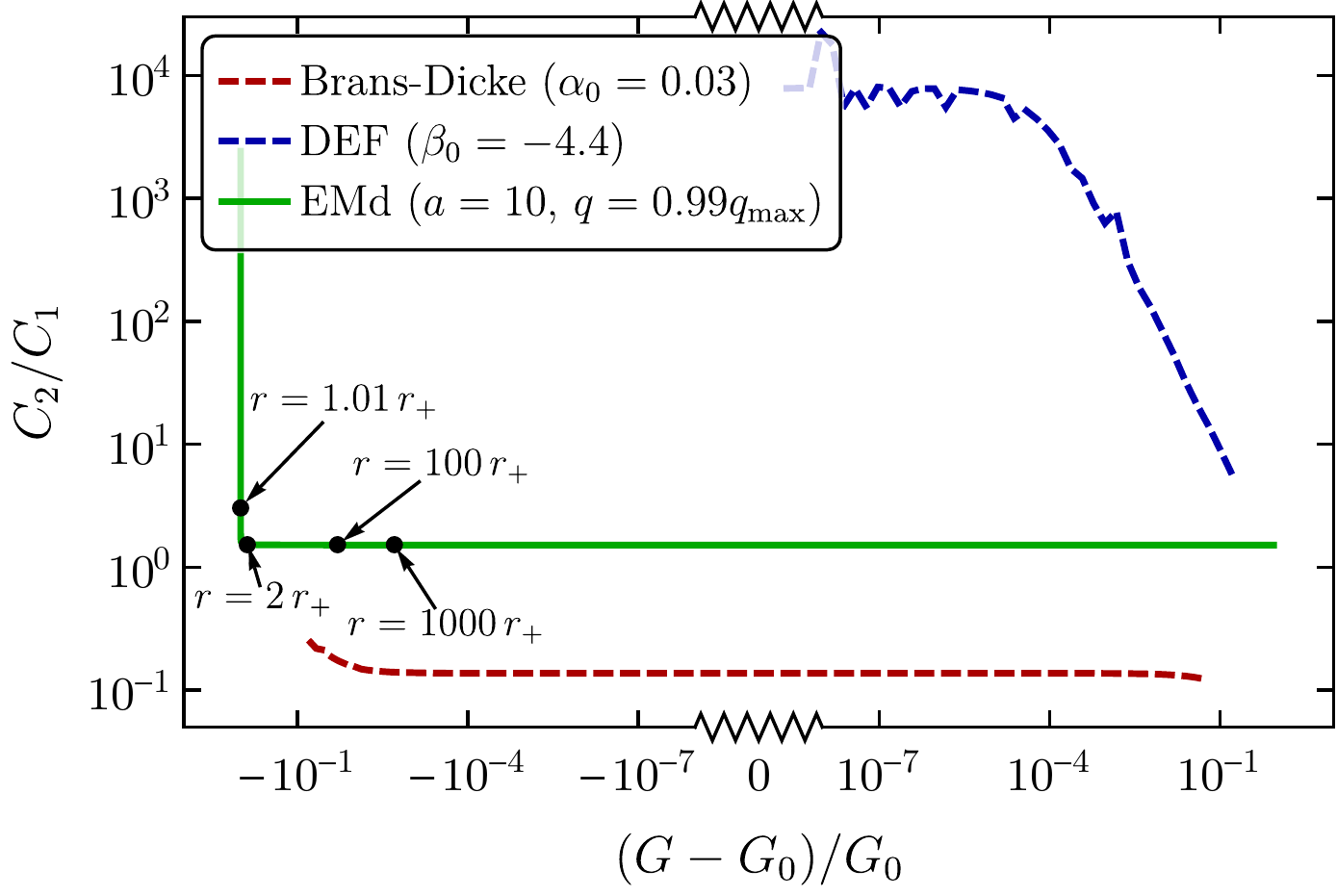}
	\caption{Ratio of the coefficients $C_2/C_1$ defined in Eqs.~\eqref{eq:c1def} and~\eqref{eq:c2def} as a function of $G_\text{N}$ for BHs in EMd theory (solid) and neutron stars in various ST theories (dashed).
          Annotated points depict this ratio at various separations for a test BH with $q=0.99 q_\text{max}$ in the background of a BH with $Q=Q_\text{max}$ in EMd theory ($r_+$ refers to the outer horizon of the background spacetime). }\label{fig:sensitivityRatio}
\end{figure}

We compare these coefficients for BHs in EMd theory to that of neutron stars in Brans-Dicke
	gravity~\cite{Jordan:1955,Fierz:1956zz,Brans:1961sx}, defined by the coupling
	\begin{align}
	\mathcal{A}_\text{BD}(\varphi) = e^{- \alpha_0 \varphi},\label{eq:BDcoupling}
	\end{align}
	and theories first considered by Damour and Esposito-Far\`{e}se (DEF)~\cite{Damour:1992we,Damour:1993hw}
	\begin{align}
	\mathcal{A}_\text{DEF}(\varphi) = e^{- \beta_0 \varphi^2 /2},\label{eq:DEFcoupling}
	\end{align}
	in which induced and dynamical scalarization can occur when $\beta_0$ is sufficiently negative. In
	Fig.~\ref{fig:sensitivityRatio}, we plot the ratio $C_2/C_1$ for compact objects in the various
	theories. For the ST theories, we consider neutrons stars with $m=1.45 M_\odot$ with the
	piecewise polytropic fit to the SLy equation of state constructed in
	Ref.~\cite{Read:2008iy}. The solid curve depicts this ratio for BHs in EMd theory with
	coupling $a=10$. By comparison, this same quantity is shown with red and blue dashed curves
	for neutron stars in Brans-Dicke gravity with $\alpha_0 = 0.03$ and in the theory of Damour and
	Esposito-Far\`{e}se with $\beta_0= -4.4$, respectively. Note that by inserting
	Eq.~\eqref{eq:DEFcoupling} into Eq.~\eqref{eq:GNdef}, one sees that this theory is only defined for
	$G_\text{N}(\varphi)>G_\text{N}^0$. For reference, we indicate with black points the separation at which these values are achieved in EMd theory when the test BH in placed in the background of an extremally charged BH; $r_+$ corresponds to the outer horizon of the background BH.
 We see that the magnitude of the ratio $C_2/C_1$ drastically differs between ST theories
	that manifest induced and dynamical scalarization (DEF) and EMd theories. This result indicates
	that a perturbative expansion of the dynamics has a larger regime of validity, and that
	non-perturbative phenomena are less likely to emerge during the coalescence of binary BHs in EMd theory.

\section{Post-Newtonian  approximation in Einstein-Maxwell-dilaton theory}
\label{sec:PNapprox}

\subsection{Two-body dynamics}

To go beyond the test-body limit, treating two-body systems with
arbitrary mass ratios, we 
employ the PN approximation, which is valid in the weak-field, 
slow-motion regime \cite{,Blanchet:2013haa}.
In Appendix~\ref{app:lag}, we derive results for the conservative dynamics of a
binary BH system in EMd theory, at next-to-leading order in the PN expansion,
i.e., at 1PN order.  We employ the Fokker action method
\cite{Fokker:1929} (see also Ref.~\cite{Damour:1995kt}), which has
been used to treat the 4PN dynamics in GR \cite{Bernard:2015njp}, and
the 2PN \cite{Damour:1995kt} and 3PN \cite{Bernard:2018hta} dynamics
in ST theories.  We begin by considering the PN expansions
of the EMd action in Eq.~\eqref{EMdaction} and the matter action for
point particles in Eq.~\eqref{Smatter}, using the mass expansion in
terms of the $\alpha$ and $\beta$ parameters from Eq.~\eqref{Mphi}.
From the initial full action expanded to 1PN order, we obtain field
equations for the scalar field, the metric potential, and the
electromagnetic 4-potential.  The Fokker action is obtained by
plugging the (regularized) solutions to the field equations back into
the action, eliminating the field degrees of freedom, yielding an
action depending only on the matter variables.  We work in the
harmonic gauge $g^{\mu\nu}\Gamma^\lambda_{\mu\nu}=0$ and the Lorenz
gauge $\partial_\mu A^\mu=0$ throughout.  The final result for the
two-body Lagrangian is given by
\begin{widetext}
\begin{align}
\label{1PNlag}
L &= -m_1 - m_2 + \frac{1}{2}m_1v_1^2+\frac{1}{2}m_2v_2^2+ \left(1+\alpha_1\alpha_2 -\frac{q_1q_2}{m_1m_2}\right)\frac{m_1m_2}{r} \nonumber\\
&\quad
+\frac{1}{8}m_1v_1^4+\frac{1}{8}m_2v_2^4+ \frac{q_1q_2}{2r}\left[\bm{v_1\cdot v_2}+(\bm{n}\cdot\bm{v_1})(\bm{n}\cdot\bm{v_2})\right] \nonumber\\
&\quad +\frac{m_1m_2}{2r}\left[(3-\alpha_1\alpha_2)(v_1^2+v_2^2)- (7-\alpha_1\alpha_2)(\bm{v_1}\cdot\bm{v_2})-(1+\alpha_1\alpha_2)(\bm{n}\cdot\bm{v_1})(\bm{n}\cdot\bm{v_2})\right]\nonumber\\
&\quad
-\frac{m_1m_2}{2r^2}\left[(1+2\alpha_1\alpha_2)(m_1+m_2)+m_1\alpha_1^2(\alpha_2^2+\beta_2)+m_2\alpha_2^2(\alpha_1^2+\beta_1)\right] \nonumber\\
&\quad
 +\frac{q_1q_2}{r^2} \left[m_1(1+a\alpha_1)+m_2(1+a\alpha_2)\right]
- \frac{1}{2r^2} \left[m_1q_2^2(1+a\alpha_1)+m_2q_1^2(1+a\alpha_2)\right]+\mathcal{O}\left(\frac{1}{c^4}\right),
\end{align}
\end{widetext}
where $\bm{r}\equiv \bm{x}_1-\bm{x}_2$ is the separation between the two bodies, and $\bm{n}\equiv \bm{r}/r$.
This Lagrangian agrees with the one derived by Damour and Esposito-Far\`{e}se \cite{Damour:1992we,Damour:1995kt} when the Maxwell fields are zero. 
The standard 1PN Lagrangian in GR is obtained by setting $q_i=\alpha_i=\beta_i=0$, while the Lagrangian in EM theory is obtained when $\alpha_i=\beta_i=0$. 
Note that, since we use the mass expansion in Eq.~\eqref{Mphi} given in terms of generic parameters $\alpha$ and $\beta$, our results are not restricted to BHs in EMd theory, but are applicable to more generic bodies as well.

During the course of this project, the same 1PN Lagrangian for a two-body system in EMd theory was derived independently by Juli\'{e} in Ref.~\cite{Julie:2017rpw}.  While our results agree, our derivation differs from that of Ref.~\cite{Julie:2017rpw} in some notable respects.  
In Ref.~\cite{Julie:2017rpw}, the (unexpanded) field equations were directly obtained from the action \eqref{EMdaction}, and then those equations were expanded and solved for the fields.  The primary difference with our derivation is in how Ref.~\cite{Julie:2017rpw} constructed the two-body Lagrangian: (i) taking (only) the matter action for one body (without the field part of the action, and without the matter action for the other body), which would apply if the body were a test body in some given fields, (ii) inserting for those fields the (regularized) solutions to the field equations resulting from the total (two bodies + fields) action, and (iii) taking the resultant Lagrangian and ``symmetrizing'' it with respect to the two bodies.  While this procedure does produce a correct Lagrangian at 1PN order, it is not justified in general, and it is important to see how the result can be obtained from a consistent treatment of the full action for the two bodies and fields.
In Ref.~\cite{Julie:2017rpw}, it was also found that it is possible to parameterize the 1PN Lagrangian in EMd theory to have the same structure as the 1PN Lagrangian in ST theories, which means that many results in ST theories  can be directly extended to EMd theory at 1PN order. We choose not to use that parameterization to make the dependence on the electric charges more apparent, and because many of our results are specific to EMd theory, such as calculating the vector energy flux and developing the EOB Hamiltonians.

The Hamiltonian in the center-of-mass frame can be derived from the Lagrangian using the Legendre transformation \cite{Blanchet:2002mb}
\begin{equation}
H=  \bm{v}\cdot \bm{p} - L,
\end{equation}
where the relative velocity $\bm{v}\equiv \bm{v}_1-\bm{v}_2$ and the center-of-mass momentum
\begin{equation}
p_i=\frac{\partial L}{\partial v^i}.
\end{equation}
This leads to the energy
\begin{align} \label{energy}
E&= M+\frac{1}{2}\mu v^2 -\frac{G_{12}M\mu}{r}  
+\frac{3}{8}(1-3\nu)\mu v^4 \nonumber\\
&\quad +\frac{G_{12}M\mu}{2r}\left[\left(\frac{3-\alpha_1\alpha_2}{1+\alpha_1\alpha_2-\frac{q_1q_2}{M\mu}} +\nu\right)v^2 +\nu \dot{r}^2\right] \nonumber\\
&\quad
+\frac{M^2\mu}{2r^2}\bigg[
(1+\alpha_1\alpha_2)^2
+X_2\alpha_2^2\beta_1 +X_1\alpha_1^2\beta_2 \nonumber\\
&\quad +X_1\frac{q_2^2}{M\mu}(1+a\alpha_1) +X_2\frac{q_1^2}{M\mu}(1+a\alpha_2)  \nonumber\\
&\quad -2\frac{q_1q_2}{M\mu}\left(1+a \alpha_1X_1 +a\alpha_2X_2\right)
\bigg] +\mathcal{O}\left(\frac{1}{c^4}\right),
\end{align}
where $\dot{r}=\bm{n}\cdot\bm{v}$, and we defined the  total mass $M$, reduced mass $\mu$,  symmetric mass ratio $\nu$, and the mass ratios $X_i$ in terms of the constant masses $m_1$ and $m_2$ by
\begin{gather}
M\equiv m_1+m_2\,, \qquad \mu\equiv \frac{m_1m_2}{M}\,, \qquad \nu \equiv \frac{\mu}{M}\,, \nonumber\\
X_1\equiv \frac{m_1}{M}, \qquad X_2\equiv\frac{m_2}{M}.
\end{gather} 
We also define the coefficient $G_{12}$ by
\begin{equation}
G_{12} \equiv 1+\alpha_1\alpha_2 -\frac{q_1q_2}{M\mu}\,,
\end{equation}
which reduces to the usual definition in ST theories when the electric charges are zero.
The advantage of including the charges in $G_{12}$ is that the Newtonian-order acceleration is simply given by $\bm{a}=-G_{12}M\bm{n}/r^2+\Order(1/c^2)$.

\begin{figure*}[th]
	\centering
	\begin{minipage}[b]{0.49\linewidth}
		\includegraphics[width=\linewidth]{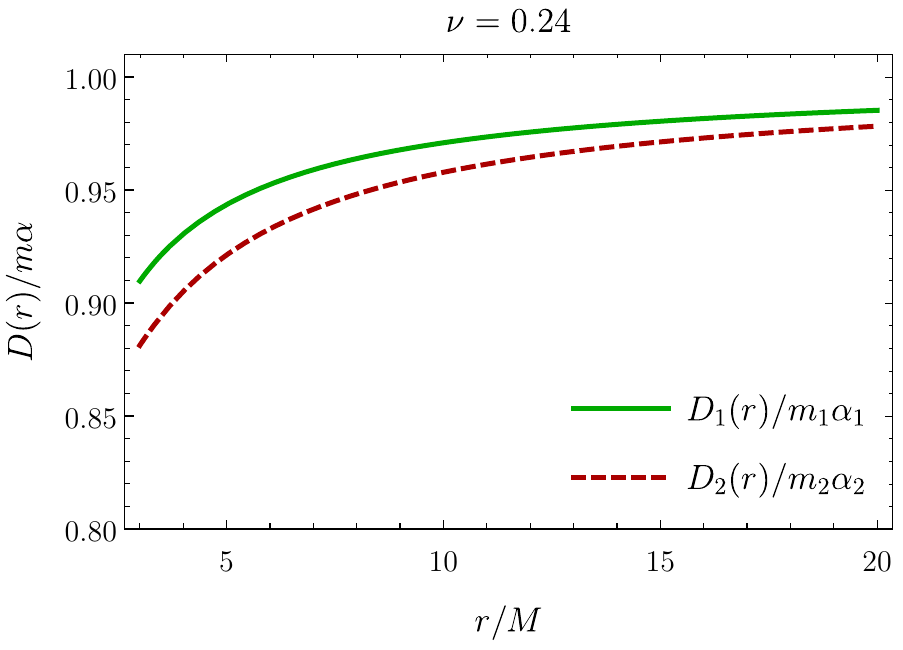}
	\end{minipage}
	\begin{minipage}[b]{0.49\linewidth}
		\includegraphics[width=\linewidth]{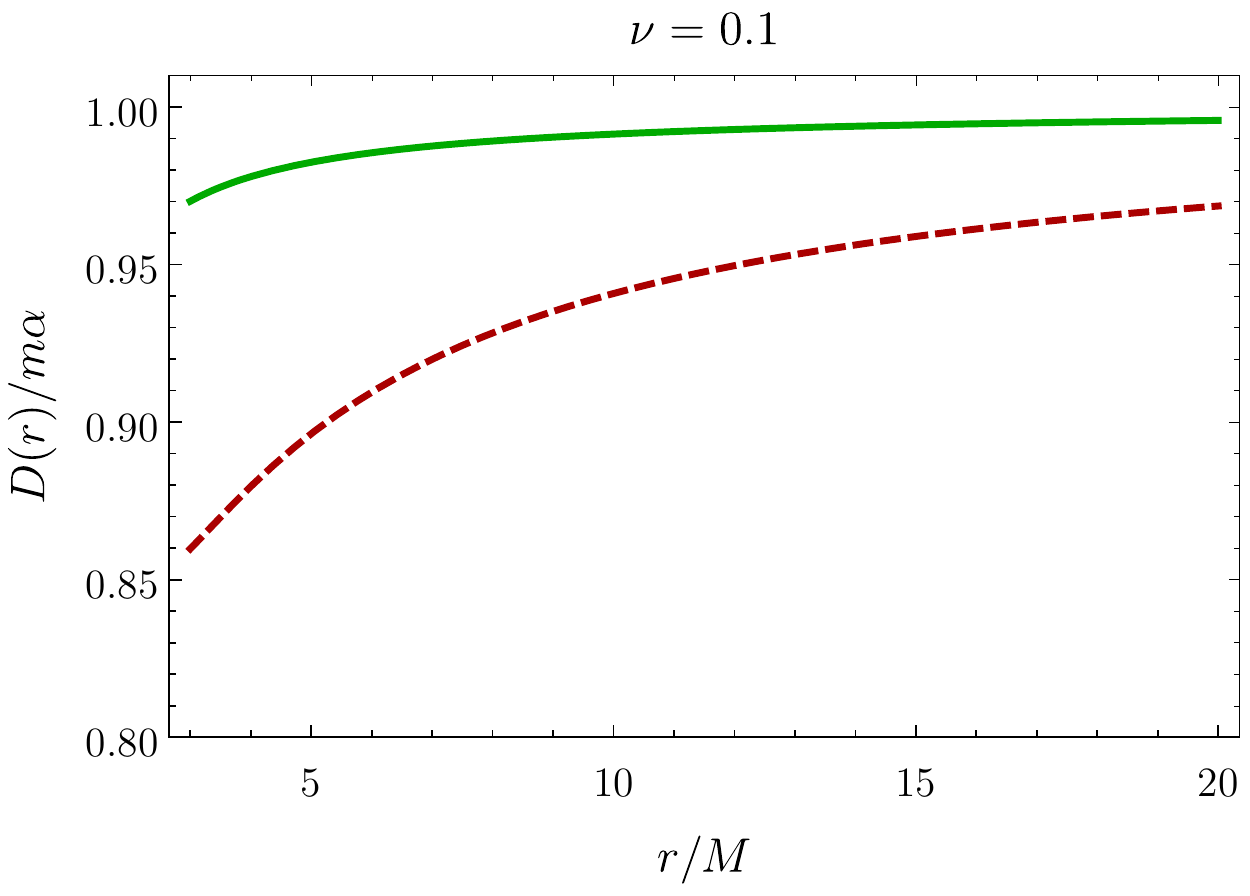}
	\end{minipage}
	\caption{\label{fig:Drnu} Scalar charges scaled by their asymptotic value as a function of the separation $r$ of a binary BH scaled by the total mass. In both plots, the charge-to-mass ratio $q_1/m_1=q_2/m_2=1$ and the dilaton coupling $a=1$; in the left panel $\nu=0.24$, while in the right  $\nu=0.1$.} 
	
	\begin{minipage}[b]{0.49\linewidth}
		\includegraphics[width=\linewidth]{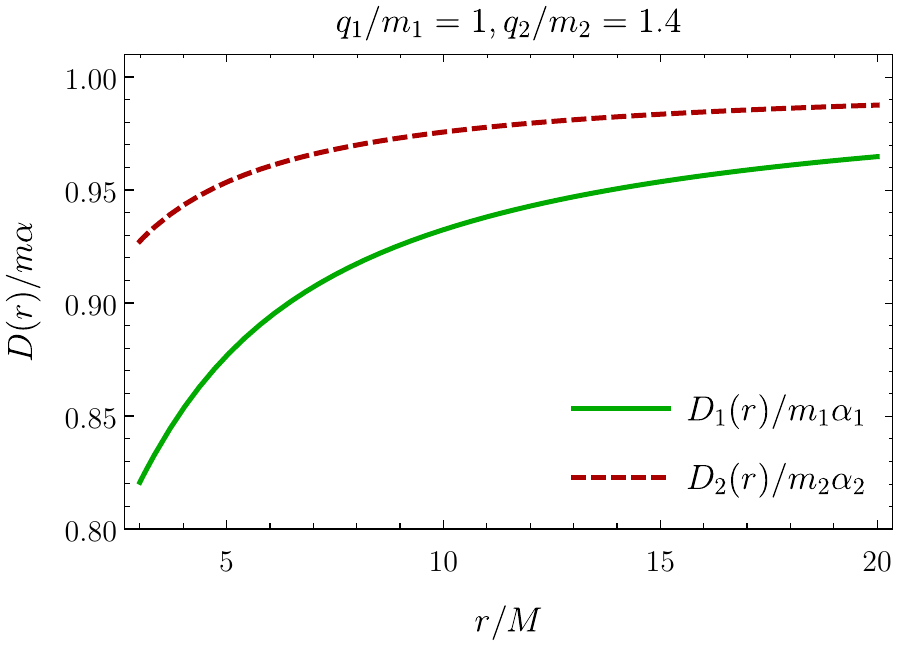}
	\end{minipage}
	\begin{minipage}[b]{0.49\linewidth}
		\includegraphics[width=\linewidth]{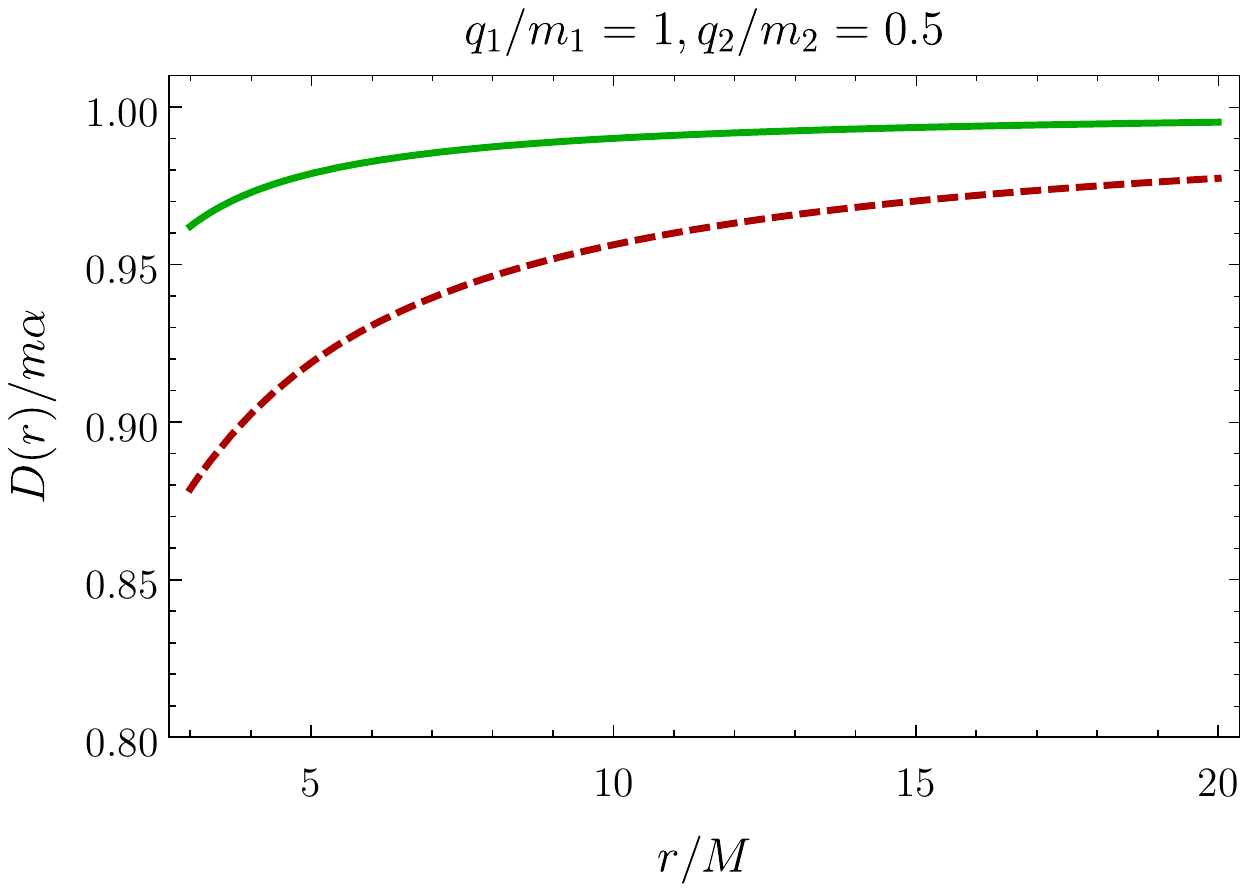}
	\end{minipage}
	\caption{\label{fig:Drqm} Scalar charges of a binary BH as a function of $r$ for equal masses $(\nu=1/4)$, dilaton coupling $a=1$, and charge-to-mass ratios $q_1/m_1=1, ~q_2/m_2=1.4$ (left) and  $q_1/m_1=1, ~q_2/m_2=0.5$  (right).} 
\end{figure*}

Expressing the energy in terms of the center-of-mass momentum $\bm{p}\equiv\bm{p_1}=-\bm{p_2}$, instead of the velocity, we obtain the Hamiltonian
\begin{align}
\label{Hreal}
H &= M + \frac{p^2}{2\mu}- \frac{G_{12}M\mu}{r} 
-\frac{1}{8}(1-3\nu)\frac{p^4}{\mu^3} \nonumber\\
&\quad 
-\frac{G_{12}M}{2\mu r}\left[\left(\frac{3-\alpha_1\alpha_2}{1+\alpha_1\alpha_2-\frac{q_1q_2}{M\mu}} +\nu\right)p^2
+\nu p_r^2\right] \nonumber\\
&\quad
+\frac{M^2\mu}{2r^2}\bigg[(1+\alpha_1\alpha_2)^2 +X_2\alpha_2^2\beta_1 +X_1\alpha_1^2\beta_2  \nonumber\\\ 
&\quad+X_1\frac{q_2^2}{M\mu}(1+a\alpha_1) +X_2\frac{q_1^2}{M\mu}(1+a\alpha_2) \nonumber\\ 
&\quad -2\frac{q_1q_2}{M\mu}\left(1+a \alpha_1X_1 +a\alpha_2X_2\right)\bigg]+\mathcal{O}\left(\frac{1}{c^4}\right),
\end{align}
where $p_r=\bm{n}\cdot\bm{p}$.

Next, we examine how the scalar charges of  the two bodies change with their separation. 
The dilaton charge  is given by
\begin{equation}
D(\varphi) = \frac{d\mathfrak{m}(\varphi)}{d\varphi} = \mathfrak{m}(\varphi)\alpha(\varphi).
\end{equation}
For the two bodies, the dilaton charge as a function of the separation $r$ has the expansion 
\begin{subequations}
\begin{align}
D_1(r) &= m_1\Big[\alpha_1 +(\alpha_1^2+\beta_1)\varphi_1(r)  +\frac{1}{2}\big(3\beta_1\alpha_1 +\alpha_1^3 \nonumber\\
&\quad\qquad +\beta_1'\big)\varphi_1^2(r) +\Order\left(1/c^6\right)\Big], \\
D_2(r) &= m_2\Big[\alpha_2 +(\alpha_2^2+\beta_2)\varphi_2(r)  +\frac{1}{2}\big(3\beta_2\alpha_2 +\alpha_2^3 \nonumber\\
&\quad\qquad +\beta_2'\big)\varphi_2^2(r) +\Order\left(1/c^6\right)\Big],
\end{align}
\end{subequations}
where $\beta'\equiv \left. d\beta(\varphi)/d\varphi \right|_{\varphi_0}$, $\varphi_1$ is the scalar field at the location of body 1, and $\varphi_2$ is the scalar field at the location of body 2.
From the 1PN scalar field in Eq.~\eqref{1PNphi},
\begin{subequations} 
\begin{align}
\varphi_1(r) &= -\frac{\alpha_2 m_2}{r} +\frac{m_1m_2}{r^2} \left(\alpha_2+\alpha_1\alpha_2^2 +\alpha_1\beta_2\right) -\frac{aq_1q_2}{r^2} \nonumber\\
&\quad  +\frac{aq_2^2}{2r^2}+\frac{1}{2}\alpha_2m_2(\bm{n}\cdot\bm{a}_2)+\mathcal{O}\left(1/c^6\right), \\
\varphi_2(r) &= -\frac{\alpha_1 m_1}{r} +\frac{m_1m_2}{r^2} \left(\alpha_1+\alpha_2\alpha_1^2 +\alpha_2\beta_1\right) -\frac{aq_1q_2}{r^2} \nonumber\\
&\quad  +\frac{aq_1^2}{2r^2} -\frac{1}{2}\alpha_1m_1(\bm{n}\cdot\bm{a}_1)+\mathcal{O}\left(1/c^6\right),
\end{align}
\end{subequations}
where, using
$\bm{a}=-G_{12}M\bm{n}/r^2+\Order(1/c^2)$ and Eq.~\eqref{x1x2CM},
\begin{subequations}
\begin{align}
\bm{a}_1 &=\frac{m_2}{M}\bm{a}= -\frac{G_{12}m_2}{r^2}\bm{n} +\mathcal{O}\left(1/c^2\right), \\
\bm{a}_2 &= -\frac{m_1}{M}\bm{a}= \frac{G_{12}m_1}{r^2}\bm{n} +\mathcal{O}\left(1/c^2\right)\,.
\end{align}
\end{subequations}

In Fig.~\ref{fig:Drnu}, we plot $D_1(r)$ and $D_2(r)$  for charge-to-mass ratios $q_1/m_1=q_2/m_2=1$,  dilaton coupling constant $a=1$, and  symmetric mass ratios $\nu=0.24$ and $\nu=0.1$. 
The curves are plotted until $r=3M$ because the PN expansion becomes inaccurate  well before that separation.
From the figure, we see that the scalar charge of both bodies decreases as the separation decreases, with the charge of the lighter body decreasing more quickly. 
Figure~\ref{fig:Drqm}  shows the scalar charge as a function of the separation for equal masses but with different charge-to-mass ratios. We keep $q_1/m_1=1$ while $q_2/m_2$ takes the values 1.4 and 0.5. The scalar charge of the less-charged body decreases more quickly with decreasing separation.
These results are consistent with what was found in the previous section for the scalar charge of a  test BH, but here, we do not see a transition or a divergence near the horizon.

\subsection{Gravitational energy flux}
\begin{figure*}[th]
	\centering
	\begin{minipage}[b]{0.49\linewidth}
		\includegraphics[width=\linewidth]{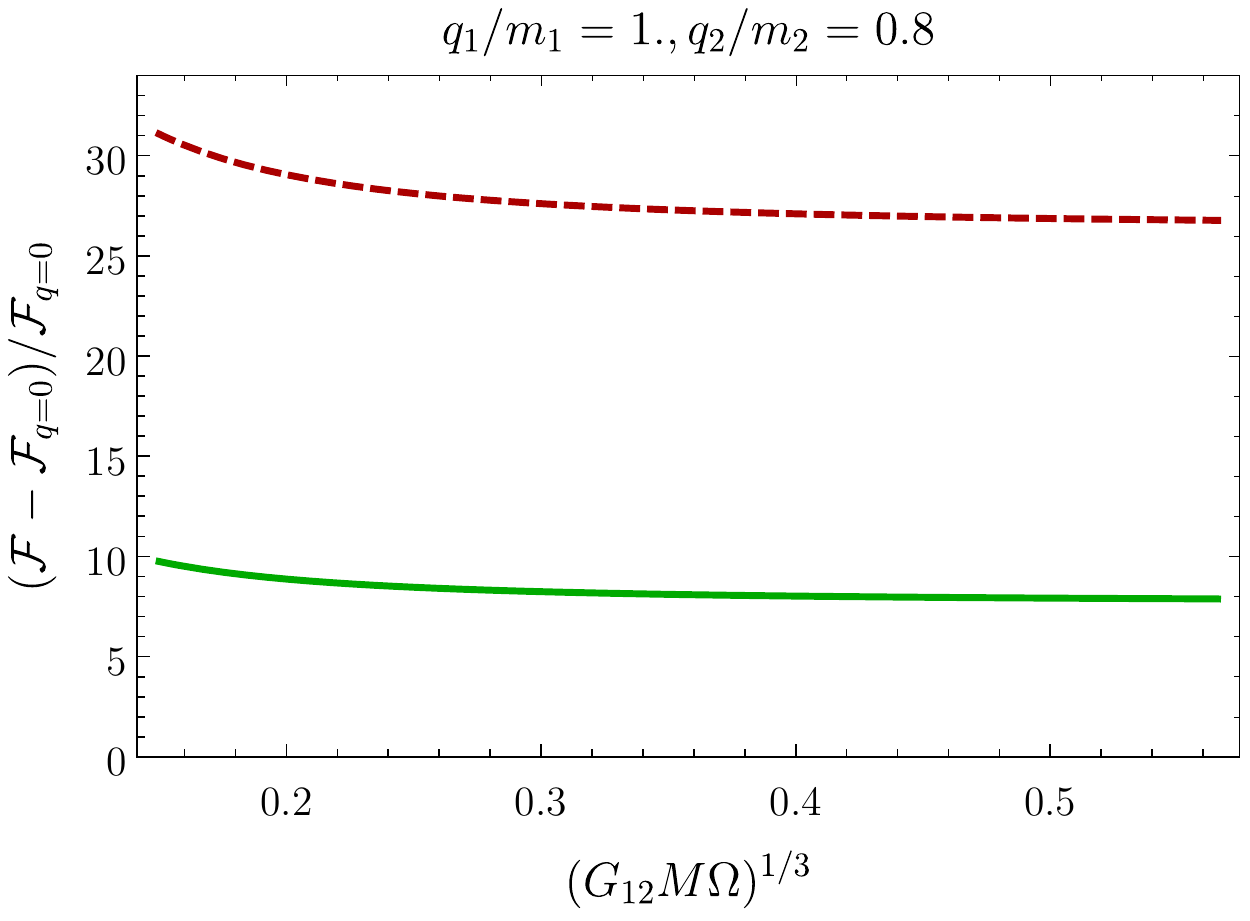} 
	\end{minipage}
	\begin{minipage}[b]{0.49\linewidth}
		\includegraphics[width=\linewidth]{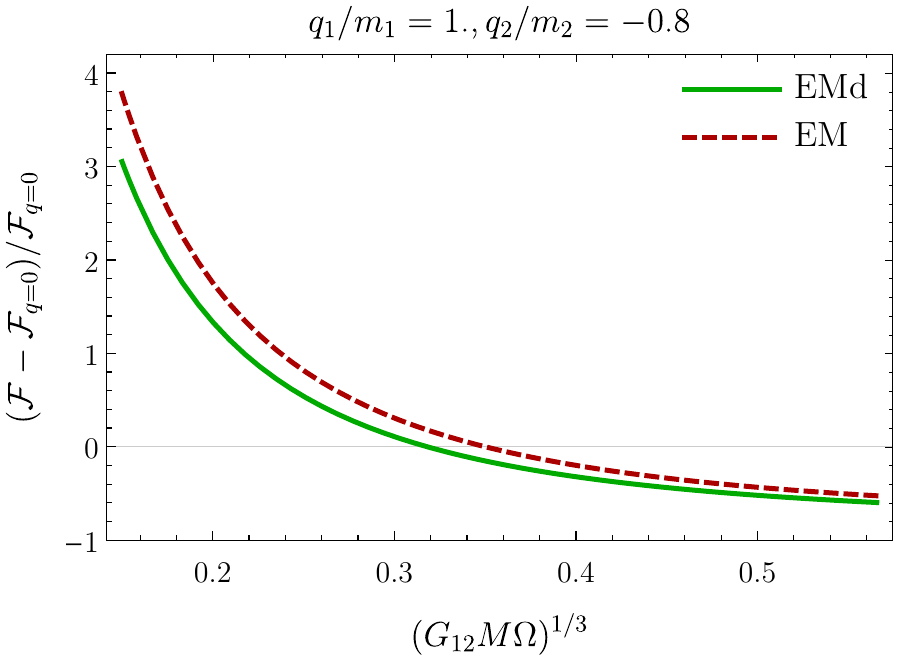}
	\end{minipage}
	\caption{\label{fig:flux} Energy flux in EMd theory and EM relative to the uncharged GR flux plotted versus the gauge-invariant velocity for circular orbits $v \equiv (G_{12} M \Omega)^{1/3}$ for coupling constant $a=1$, for equal masses, and for  charge-to-mass ratio $q_1/m_1=1, q_2/m_2=0.8$ (left) and $q_1/m_1=1, q_2/m_2=-0.8$ (Right).
	}	
\end{figure*}

From the 1PN expansion, we computed the next-to-leading order scalar, vector, and tensor energy fluxes for general orbits (see Appendix \ref{app:flux} for the derivation).
In a $1/c$ expansion, the leading terms are the scalar and vector dipole  fluxes, which are of order $1/c^3$, while the leading order tensor flux is of order $1/c^5$, which is the same as the next-to-leading order scalar and vector fluxes.
We computed the next-to-leading order tensor flux, which is of order $1/c^7$, because that is the maximum level of approximation accessible by use of the 1PN near-field equations.
The scalar and vector dipole fluxes depend on the difference between the charges of the two bodies. The scalar flux also includes a monopole term that vanishes for circular orbits. 

The total energy flux is the sum of the scalar, vector, and tensor fluxes
\begin{equation}
\mathcal{F} = \mathcal{F}_S + \mathcal{F}_V + \mathcal{F}_T,
\end{equation}
where the expressions for the fluxes through next-to-leading order for general orbits are given in Appendix \ref{app:flux}. 
The fluxes for circular orbits are given by
\begin{widetext}
\begin{subequations}
\begin{align} \label{NLOflux}
\mathcal{F}_S&= \frac{\nu^2x^4}{3G_{12}^2}(\alpha_1-\alpha_2)^2
+\frac{\nu^2 x^5}{15G_{12}^2} \left[20f_\gamma(\alpha_1-\alpha_2)^2 +5\left(f_{v^2}^S+f_{1/r}^S\right)  +16\left(X_1\alpha_2 +X_2\alpha_1\right)^2\right]+\Order\left(\frac{1}{c^7}\right), \\
\mathcal{F}_V&=  \frac{2\nu^2x^4}{3G_{12}^2} \left(\frac{q_1}{m_1}-\frac{q_2}{m_2}\right)^2 +\frac{2\nu^2 x^5}{15G_{12}^2}  \left[20f_\gamma\left(\frac{q_1}{m_1}-\frac{q_2}{m_2}\right)^2 +8\left(X_2\frac{q_1}{m_1} +X_1\frac{q_2}{m_2}\right)^2 + 5\left(f_{v^2}^V +f_{1/r}^V\right) \right] +\Order\left(\frac{1}{c^7}\right), \\
\mathcal{F}_T &=  \frac{32\nu^2x^5}{5G_{12}^2}  +\frac{2\nu^2x^6}{105G_{12}^2}\left(f_{v^4}^T + f_{v^2/r}^T + f_{1/r^2}^T + 672 f_\gamma\right)+\Order\left(\frac{1}{c^9}\right),
\end{align}
\end{subequations}
\end{widetext}
where the coefficients $f$ are given by Eqs.~\eqref{coeffSdip}, \eqref{coeffEMdip}, \eqref{coeffGW}, and \eqref{gammax}.
The energy flux is expressed in terms of the parameter $x$ defined by
\begin{equation}
x\equiv \left(G_{12}M\Omega\right)^{2/3},
\end{equation}
where $\Omega$ is the orbital frequency, which is ``perturbatively gauge-invariant'' in the sense that it remains fixed under coordinate transformations to arbitrary PN order.

\begin{figure}
	\includegraphics[width=\linewidth]{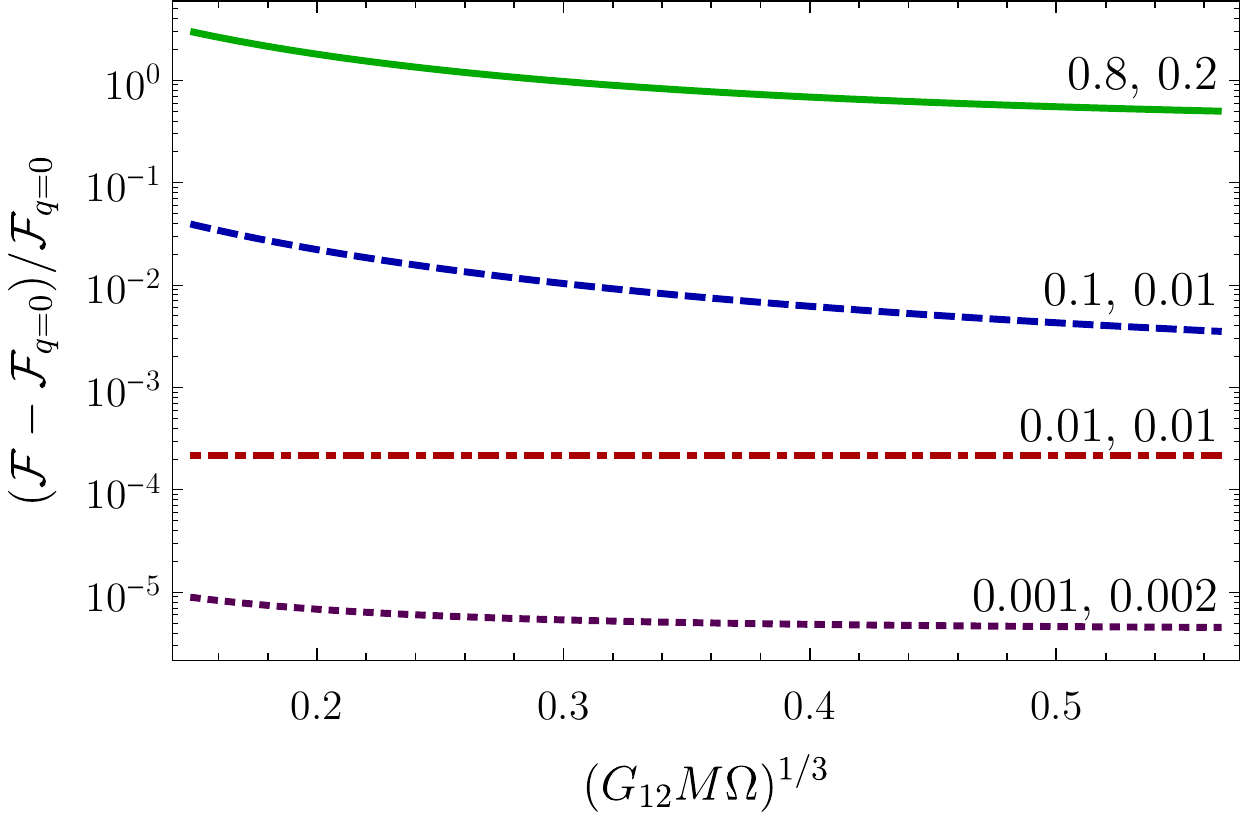} 
	\caption{\label{fig:flux2} Energy flux in EMd theory relative to the uncharged GR flux for coupling constant $a=1$ plotted versus $v=(G_{12}M\Omega)^{1/3}$, for equal masses, and for various charge-to-mass ratios. 
	}
\end{figure}

In Figs.~\ref{fig:flux} and \ref{fig:flux2}, we plot the  total energy flux in EMd theory with $a=1$ relative to the flux when all charges are zero versus the binary's gauge-invariant velocity $v = (G_{12} M \Omega)^{1/3}$, i.e., we plot $(\mathcal{F}-\mathcal{F}_{q=0})/\mathcal{F}_{q=0}$. For comparison, Fig.~\ref{fig:flux} also includes the energy flux in EM, when scalar charges are zero but not the electric charges.
The plots start at $v = (G_{12} M \Omega
)^{1/3}=0.15$ which corresponds to a total mass $M=20M_\odot$, and a lower GW frequency in the detector of 10 Hz.  In the plots, we used the next-to-leading order scalar and vector fluxes, but only used the leading Newtonian order tensor flux,  because 
the 1PN energy flux in GR is given by $\mathcal{F}_\text{GR}\sim x^5-\text{const.}~x^6$; the minus sign of the second term causes the flux to become negative at large frequencies.

From the plots, we see that at small frequencies (large separations), the difference with GR is greater than at larger frequencies because the dipole scalar and vector fluxes dominate  ($\mathcal{F}_S\sim x^4$ while $\mathcal{F}_T\sim x^5$).
For equal charges, the scalar and vector dipole fluxes are both zero, which means the total energy flux is the tensor flux that is proportional to $x^5$. Hence, the next-to-leading order flux in EMd theory becomes a constant shift to the GR flux, and the relative flux plotted in the figures becomes a straight line, as can be seen in Fig.~\ref{fig:flux2}.

In the two panels of Fig.~\ref{fig:flux}, we use charge-to-mass ratios $q_1/m_1=1,~ q_2/m_2=0.8$ (left) and $q_1/m_1=1,~ q_2/m_2=-0.8$ (right). 
For same-sign charges, at a fixed frequency, there is a greater difference from GR than for opposite-sign charges and also a greater difference between EMd and EM. 
This is because the energy flux is inversely proportional to  $G_{12}^2=(1+\alpha_1\alpha_2-q_1q_2/m_1m_2)^2$, which is larger when the electric charges have opposite signs than when they have the same sign.
In the right panel, the plotted curves become negative when $\mathcal{F}<\mathcal{F}_{q=0}$, which occurs because $G_{12}>1$ for opposite-sign charges, which makes the EMd flux smaller than the GR flux at some frequency.

\begin{figure}
	\includegraphics[width=\linewidth]{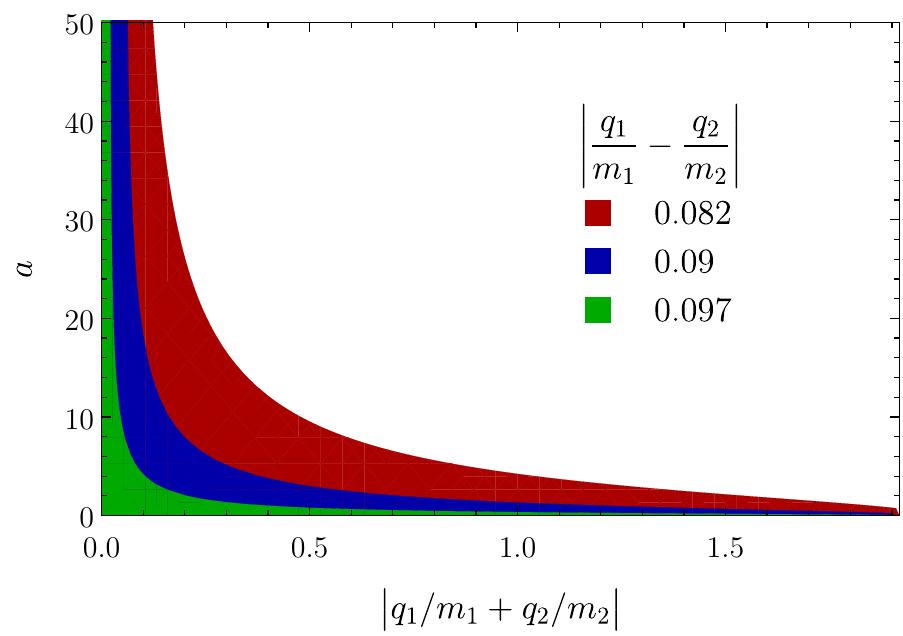} 
	\caption{\label{fig:bounds}  Allowed values of EMd coupling $a$ consistent with a dipole flux constraint of $|B| \leq 10^{-3}$ as a function of mass-weighted total electric charge.  Colors indicate various possible electric dipoles consistent with the bound on $B$.}
\end{figure}

In Fig.~\ref{fig:flux2}, we plot the energy flux for several charge-to-mass ratios. In that figure, we do not plot the flux in  EM theory, because it is almost the same as the EMd flux for charges $q_i/m_i\lesssim 0.5$ since $\mathcal{F}_S \propto \alpha_i^2 \propto q_i^4/m_i^4$, which is much smaller than $\mathcal{F}_V \propto q_i^2/m_i^2$ for small charges.
The plot shows the flux for same-sign charges in a log plot; for small charges $\lesssim 0.01$, the EMd flux decreases significantly and becomes very close to the  GR flux.

The most salient feature that differentiates EMd theory from GR from the perspective of GW observations is the presence of dipole radiation.
At leading order, the energy flux can be written as
\begin{equation}
\mathcal{F}=\mathcal{F}_\text{GR}\left(1 + B x^{-1}\right),
\end{equation}
where $\mathcal{F}_\text{GR}$ is the GR quadrupole flux, and $B$ parameterizes the strength of dipolar emission, which is given by
\begin{equation}
B = \frac{5}{96}\left[(\alpha_1-\alpha_2)^2 + 2\left(\frac{q_1}{m_1} -\frac{q_2}{m_2}\right)^2\right].\label{eq:FluxRatio}
\end{equation}

The presence of dipole flux has been constrained in several types of binary systems.
The best constraints on the $B$ come from radio observations of pulsar--white-dwarf binaries, which lead to the bound $|B|\lesssim 10^{-9}$ \cite{Yunes:2010qb}.
For binaries containing a single BH, the strongest bound comes from low-mass X-ray binaries, in which the companion is a main-sequence star: $|B|\lesssim 2\times 10^{-3}$ \cite{Barausse:2016eii}.
To date, no bound has been set from GW observations of binary BHs, but at design sensitivity, LIGO could set a bound of $|B| \lesssim 8\times 10^{-4}$ for a GW150914-like event, and LISA could lower that bound to $10^{-8}$ \cite{Barausse:2016eii}.

We wish to understand how well such a bound on dipole radiation in binary BHs can constrain EMd theory.
Given the discussion above, we consider a hypothetical binary BH observation that constrains the dipole flux to $|B| \lesssim 10^{-3}$.
The coupling $a$ that characterizes EMd theory enters the prediction of $B$ through the dimensionless scalar charges of the two bodies.
Equation~\eqref{eq:FluxRatio} demonstrates that for a given value of $B$, the scalar and electric dipoles are degenerate, and thus no constraint can be set on $a$ directly with only a bound on the dipole flux.
However, if an independent measurement of the total charges could be made --- e.g., through measurements of the ringdown spectrum of the final remnant---one can potentially break this degeneracy and constrain EMd theory.

In Fig.~\ref{fig:bounds}, we show the values of $a$ consistent with $|B|\leq 10^{-3}$ as a function of mass-weighted total charge $|q_1/m_1+q_2/m_2|$ for various possible values of the electric dipole $|q_1/m_1 - q_2/m_2|$.
The maximum allowed electric dipole is achieved in the limit that $a=0$, wherein the scalar charges of the BHs vanish and our bound on the dipole flux translates directly to the bound on the electric dipole $|q_1/m_1-q_2/m_2|\lesssim 0.098$.
Unsurprisingly, we find that the constraint that can be set on $a$ depends primarily on the magnitude of the electric charges in the binary: for equal-mass systems, the strongest constraints can be set when the BHs have large, nearly-equal charges, and the weakest constraints when the BHs have small, opposite charges.
We see that for any realistic constraint on dipole flux, the parameter $a$ is completely unbounded without an independent measurement of the electric charges.

\subsection{Gravitational-wave phase in the stationary-phase approximation}

Equipped with PN descriptions of the conservative and dissipative sectors of binary dynamics in EMd theory, we compute 
a key observable for GW detections: the Fourier-domain gravitational waveform.
We utilize the stationary-phase approximation to perform this calculation, relying on the fact the GW phase evolves much more rapidly than its amplitude during the adiabatic inspiral along quasi-circular orbits.

We consider a GW detector a distance ${R\gg\lambda_\text{GR}\sim r/v}$ from a binary BH.
In the vicinity of the detector, the metric takes the form
\begin{align}
g_{\mu \nu}=\eta_{\mu \nu} + h_{\mu \nu},
\end{align}
where $\eta_{\mu \nu}$ is the Minkowski metric and $h_{\mu \nu}$ contains two propagating, transverse-traceless polarizations $h_+$ and $h_\times$, which comprise the GW produced by the binary.\footnote{A GW detector also responds to the scalar field through the coupling given in Eq.~\eqref{EMdaction}.
These scalar waves represent a transverse breathing polarization of perturbations to the Jordan-Fierz metric.
Because standard search techniques are targeted at the transverse-traceless polarizations, we consider only those gravitational modes in this work.
Differentiating between the various polarizations of GWs requires a network of detectors; our ability to identify additional GW polarizations will improve as more ground-based detectors come online.} 
At the fixed distance $R$, the GW can be decomposed into spin-weighted spherical harmonics
\begin{align}
h_+ - i h_\times & = \sum_{\ell \geq 2} \sum_{m=-\ell}^{\ell} {}_{-2}Y_{\ell m}(\Theta,\Phi) h_{\ell m}(t),
\end{align}
where $\Phi, \Theta$ are angular coordinates that define the propagation direction from the source to the detector~\cite{Blanchet:2013haa}.
We further decompose each mode into an amplitude and complex phase
\begin{align}
h_{\ell m}(t) = A_{\ell m}(t) e^{i m \phi(t)},
\end{align}
where $\phi(t)$ is the orbital phase of the binary.

We compute the Fourier-transform of the GW using
\begin{align}
\tilde{h}_{\ell m}(f) = \int_{- \infty}^{\infty} dt \, h_{\ell m}(t) e^{-2 i \pi f t}. \label{eq:FourierTransform}
\end{align}
During the adiabatic inspiral, the amplitude and orbital frequency evolve much more slowly than the orbital phase, i.e., $|\dot{A}_{\ell m} / A_{\ell m}| \ll \Omega$ and $|\dot{\Omega}| \ll \Omega^2$ for $m\neq 0$ modes.
Thus, the integral in Eq.~\eqref{eq:FourierTransform} is highly oscillatory and can be approximated by expanding the integrand about the time at which the complex phase is stationary.
Using the stationary-phase approximation, the Fourier-domain waveform is then given by
\begin{align}
\tilde{h}^\text{SPA}_{\ell m}(f) &= \mathcal{A}_{\ell m}(f)e^{-i \psi_{\ell m}(f)-i\pi/4}, \\
\psi_{\ell m}(f) &= 2 \pi f t_f^{(m)} - m \phi(t_f^{(m)}),\label{eq:SPAphase}\\
\mathcal{A}_{\ell m}(f) &=A_{\ell m}(t_f^{(m)})\sqrt{\frac{2 \pi}{m \dot{\Omega}(t_f^{(m)})}},\label{eq:SPAamp}
\end{align}
where $t_f^{m}$ is defined implicitly as the time at which $m \Omega (t_f^{(m)}) = 2 \pi f$.
Following the notation common in the literature, we employ the binary's gauge-invariant velocity for circular orbits $v\equiv x^{1/2} = (G_{12} M \Omega)^{1/3}$ and introduce a similar notation for the GW frequency $f$ as $v_f \equiv (\pi G_{12} M f)^{1/3}$.
Then, by construction, one finds  $v(t_f^{(m)})= (2/m)^{1/3} v_f$ and can rewrite Eq.~\eqref{eq:SPAphase} as
\begin{align}
\psi_{\ell m}(f) = m \left(\frac{1}{G_{12} M} v^3 t(v) - \phi(v) \right)\bigg|_{v=(2/m)^{1/3} v_f}. \label{eq:SPAphaseReduced}
\end{align}
From here onwards, we focus only on the dominant ${\ell = |m| = 2}$ modes and drop the explicit mode numbers for notational simplicity; because we restrict our attention to non-spinning systems, the modes obey the symmetry relation
\begin{align}
h_{\ell m}=(-1)^\ell h_{\ell, -m}^*,
\end{align}
and thus we can consider only the $m= 2$ mode without loss of generality.

The orbital phase and frequency are computed using the balance equation
\begin{align}
\frac{d E}{d t} = - \mathcal{F}.
\end{align}
From this equation, we deduce
\begin{align}
\phi(v)= & \phi_\text{ref} - \frac{1 }{G_{12} M} \int_{v_\text{ref}}^{v} d \hat{v} \hat{v}^3 \frac{E'(\hat{v})}{\mathcal{F}(\hat{v})},\\
t(v) = & t_\text{ref} - \int_{v_\text{ref}}^{v} d \hat{v} \frac{dE / d \hat{v}}{\mathcal{F}(\hat{v})},
\end{align}
where $\phi_\text{ref}$ and $t_\text{ref}$ refer to an arbitrary reference point in the evolution of the binary.
Inserting these results into Eq.~\eqref{eq:SPAphaseReduced}, the Fourier-domain phase is given by
\begin{align} 
\psi(f)=2\pi f t_\text{ref} -\phi_\text{ref} +\frac{2}{G_{12}M}\int_{v_f}^{v_\text{ref}}\left(v_f^3-v^3\right)\frac{E'(v)}{\mathcal{F}(v)}dv. \label{eq:SPAphaseFinal}
\end{align}

The energy flux in terms of $x$ is given by Eq.~\eqref{NLOflux}.
The energy $E$ is given by Eq.~\eqref{energy}, and it can be expressed in terms of $x$ using Eqs.~\eqref{kepler3rd} and \eqref{gammax}, which leads to
\begin{align}
E &= -\frac{\mu}{2}x \left[ 1 + f_E x +\Order\left(1/c^4\right) \right],
\end{align}
where the coefficient $f_E$ is given by
\begin{align}
f_E&=\frac{-1}{3G_{12}^2} \bigg[G_{12}^2\left(\frac{1+\nu}{4}+\frac{3-\alpha_1\alpha_2}{1+\alpha_1\alpha_2-\frac{q_1q_2}{M\mu}}\right) \nonumber\\
&\quad -(1+\alpha_1\alpha_2)^2 -X_2\alpha_2^2\beta_1 
-X_1\alpha_1^2\beta_2 \nonumber\\
&\quad -X_1\frac{q_2^2}{M\mu}(1+a\alpha_1) 
-X_2\frac{q_1^2}{M\mu}(1+a\alpha_2) \nonumber\\
&\quad +2\frac{q_1q_2}{M\mu} \left(1+aX_1\alpha_1 +aX_2\alpha_2\right) 
\bigg].
\end{align}

To evaluate the integral in Eq.~\eqref{eq:SPAphaseFinal}, we need to distinguish between two regimes, similarly to what was done in Ref.~\cite{Sennett:2016klh}. 
In one regime, the  electric charges are small and the inspiral is driven by the tensor quadrupole flux. In the other regime, the electric charges are large and the inspiral is driven by the dipole flux.

For the quadrupole-driven (QD) case, we approximate the integrand  in Eq.~\eqref{eq:SPAphaseFinal} by
\begin{equation} 
\label{approxQD}
\frac{E'(v)}{\mathcal{F}(v)}\simeq\frac{E'(v)}{\mathcal{F}_T(v)}\left[1 -\frac{\mathcal{F}_S(v)+\mathcal{F}_V(v)}{\mathcal{F}_T(v)}\right].
\end{equation}
Then, we expand the integrand using the next-to-leading order fluxes. Evaluating the integral leads to the phase
\begin{widetext}
\begin{equation} \label{phaseQD}
\psi^\text{QD}(f)=2\pi f t_\text{ref} -\phi_\text{ref}  +\frac{1}{v^5}\left[\rho_0^\text{QD} +\frac{\rho_{-2}^\text{QD}}{v^2} + \rho_2^{QD}v^2 +  \Order\left(v^4\right)\right],
\end{equation} 
with the coefficients
\begin{subequations}
\begin{align}
\rho_0^\text{QD} &=-\frac{G_{12}}{4096\nu}\Bigg\lbrace
\frac{5}{168}\left(336f_E-672f_\gamma -f_{1/r^2}^T -f_{v^2/r}^T -f_{v^4}^T\right) \left[2\left(\frac{q_1}{m_1}-\frac{q_2}{m_2}\right)^2 +(\alpha_1-\alpha_2)^2\right] \nonumber\\
&\quad -96+ 5\left(f_{1/r}^S+f_{v^2}^S\right)
+10\left(f_{1/r}^V+f_{v^2}^V\right)+ 40f_\gamma \left(\frac{q_1}{m_1}-\frac{q_2}{m_2}\right)^2
+16\left(X_2\frac{q_1}{m_1} +X_1\frac{q_2}{m_2}\right)^2 \nonumber\\
&\quad +20f_\gamma (\alpha_1-\alpha_2)^2 
+16\left(X_2\alpha_1 +X_1\alpha_2\right)^2 
\Bigg\rbrace, \\
\rho_{-2}^\text{QD} &=- \frac{5G_{12}}{7168\nu}\left[2\left(\frac{q_1}{m_1}-\frac{q_2}{m_2}\right)^2 +(\alpha_1-\alpha_2)^2 \right], \\
\rho_{2}^\text{QD} &= -\frac{5G_{12}}{1548288\nu} \Bigg\lbrace
-32256 f_E + \left[48 -20f_E\left(\frac{q_1}{m_1}-\frac{q_2}{m_2}\right)^2 -10f_E\left(\alpha_1-\alpha_2\right)^2\right] \left(672f_\gamma +f_{1/r^2}^T +f_{v^2/r}^T +f_{v^4}^T\right) \nonumber\\
&\quad  +\frac{5}{224} \left[2\left(\frac{q_1}{m_1}-\frac{q_2}{m_2}\right)^2 +(\alpha_1-\alpha_2)^2\right] \left(672f_\gamma +f_{1/r^2}^T +f_{v^2/r}^T +f_{v^4}^T\right)^2 - \left(672f_\gamma +f_{1/r^2}^T +f_{v^2/r}^T +f_{v^4}^T-336f_E\right) \nonumber\\
&\quad  \times \bigg[5\left(f_{1/r}^S+f_{v^2}^S\right)
+10\left(f_{1/r}^V+f_{v^2}^V\right) +20f_\gamma (\alpha_1-\alpha_2)^2 + 40f_\gamma \left(\frac{q_1}{m_1}-\frac{q_2}{m_2}\right)^2
+16\left(X_2\frac{q_1}{m_1} +X_1\frac{q_2}{m_2}\right)^2 
 \nonumber\\
&\quad\qquad  
+16\left(X_2\alpha_1 +X_1\alpha_2\right)^2  \bigg]
\Bigg\rbrace,
\end{align}
\end{subequations}
\end{widetext}
where the coefficients $f$ are given by Eqs.~\eqref{coeffSdip}, \eqref{coeffEMdip}, \eqref{coeffGW}, and \eqref{gammax}.
When the charges are zero, this phase reduces to the next-to-leading order GR result, i.e., $\rho_0^\text{QD} \to 3/128\nu$, $\rho_2^\text{QD} \to 5(743+924\nu)/32256\nu$, and $\rho_{-2}^\text{QD}\to 0$.

For the dipole-driven (DD) case, we take the tensor flux at the same order as the scalar and vector fluxes, i.e., to $\Order(x^5)$. Evaluating the integral in \eqref{eq:SPAphaseFinal} leads to
\begin{widetext}
\begin{equation}
\psi^\text{DD}(f)=2\pi f t_\text{ref} -\phi_\text{ref} + \frac{\rho_0^\text{DD}}{v^3} \left[1+\rho_2^\text{DD} v^2 +\Order(v^4)\right],
\end{equation} 
where the coefficients are given by
\begin{subequations}
\begin{align}
\rho_0^\text{DD} &= \frac{G_{12}}{\nu}\left[2\left(\frac{q_1}{m_1}-\frac{q_2}{m_2}\right)^2 +(\alpha_1-\alpha_2)^2\right]^{-1}, \\
\rho_2^\text{DD} &=\frac{-9}{10}\left[2\left(\frac{q_1}{m_1}-\frac{q_2}{m_2}\right)^2 +(\alpha_1-\alpha_2)^2\right]^{-1}  \bigg[ 
96 + 10\left(f_{1/r}^V + f_{v^2}^V\right) 
+ 5\left(f_{1/r}^S + f_{v^2}^S\right)
\nonumber\\
&\quad 
-10(f_E-2f_\gamma)(\alpha_1-\alpha_2)^2 
+ 16\left(X_2\alpha_1 +X_1\alpha_2\right)^2 
-20(f_E-2f_\gamma)\left(\frac{q_1}{m_1}+\frac{q_2}{m_2}\right)^2 \nonumber\\
&\quad 
+ 80(f_E-2f_\gamma)\frac{q_1q_1}{m_1m_2}
+16\left(X_2\frac{q_1}{m_1} +X_1\frac{q_2}{m_2}\right)^2
\bigg].
\end{align}
\end{subequations}
\end{widetext}
When we set the electric charges to zero, but keep the scalar charges nonzero, this result agrees with the (ST) result derived in Ref.~\cite{Sennett:2016klh}.

We wish to understand how well a GW signal produced in EMd theory [e.g. Eq.~\eqref{phaseQD}] can be distinguished observationally from a signal in GR.
Answering this question definitively falls beyond the scope of this paper.
To perform such a study, one would need to perform a Bayesian hypothesis test on injections of EMd signals into detectors with realistic noise, comparing the relative evidence that the signal matches template waveforms in either EMd theory or GR; for examples of such analyses for other modifications to GR, see Refs.~\cite{DelPozzo:2011pg,Cornish:2011ys,Li:2011cg,Gossan:2011ha,Sampson:2014qqa}.
Instead of this detailed study, we compute two comparatively simple measures of distinguishability: the difference in total phase, and, in Sec.~\ref{sec:usecycles}, the number of ``useful'' GW cycles.

To compare the phase calculated in EMd theory with that in GR, we need to align the waveforms and  then compute dephasing from this alignment point. We choose to do the alignment around the ``merger frequency," which for simplicity we choose to be the innermost-stable circular orbit (ISCO) frequency $f_\text{ISCO}=6^{-3/2}/\pi M$ for a Schwarzschild BH. Next, we determine $t_\text{ref}$ and $\phi_\text{ref}$ such that the waveform reaches a local maximum at this point and the phase reaches some fixed value, e.g., zero. 
To satisfy these two conditions, one can choose $t_\text{ref}$ and $\phi_\text{ref}$ such that at $f_\text{ISCO}$, $d\psi(f)/df=0$ and $\psi(f)=0$.
For the QD case, this leads to
\begin{align}
t^\text{QD}_\text{ref}&=108MG_{12}^{-10/3}\left(10G_{12}^{2/3}\rho_0^\text{QD}+G_{12}^{4/3}\rho_2^\text{QD}+84\rho_{-2}^\text{QD}\right),\nonumber\\
\phi^\text{QD}_\text{ref}& = 12\sqrt{6}G_{12}^{-7/3}\left(8G_{12}^{2/3}\rho_0^\text{QD}+G_{12}^{4/3}\rho_2^\text{QD}+60\rho_{-2}^\text{QD}\right).
\end{align}
Similarly, for the DD case, we get
\begin{align}
t^\text{DD}_\text{ref}&=6MG_{12}^{-2}\rho_0^\text{DD}\left(18+G_{12}^{2/3}\rho_2^\text{DD}\right),\nonumber\\
\phi^\text{DD}_\text{ref}& = 4\sqrt{\frac{2}{3}}G_{12}^{-1}\rho_0^\text{DD}\left(9+G_{12}^{2/3}\rho_2^\text{DD}\right).
\end{align}

\begin{figure}[t]
	\centering
	\includegraphics[width=\linewidth]{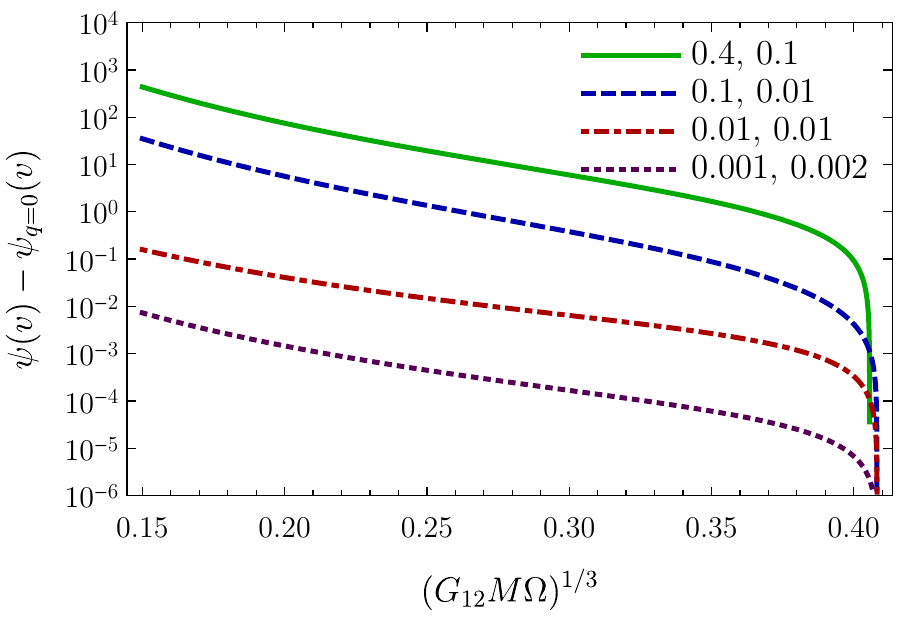} 
	\caption{\label{fig:phase} Phase difference in radians between EMd theory and GR as a function of $v$, computed in the quadrupole-driven regime, for various charge-to-mass ratios, and for equal masses ($\nu=1/4$).}
\end{figure}

In Fig.~\ref{fig:phase}, we plot the difference between the phase calculated in EMd theory with $a=1$ and the phase when all charges are zero, which is the phase in GR up to 1PN order.
For the configurations considered here, $v=0.15$ corresponds to approximately 10 Hz for a $20 M_\odot$ system.
Because the charges are relatively small, we compute the phase using Eq.~\eqref{phaseQD}.
For systems whose component's charge-to-mass ratio $q_i/m_i\lesssim 0.01$, the two waveforms differ by less than one radian over the frequency range of a ground-based GW detector.
The phase difference does not depend strongly on the value of $a$; for values of $a\sim 1000$ and charge-to-mass ratios $q_i / m_i \lesssim 10^{-3}$ analogous to those considered in Ref.~\cite{Hirschmann:2017psw}, the phase difference agrees with that shown in Fig.~\ref{fig:phase} within $10\%$.

\subsection{Number of useful gravitational-wave cycles}
\label{sec:usecycles}

The total number of GW cycles between frequencies $f_\text{min}$ and $f_\text{max}$ is given by
\begin{equation}
N_\text{tot}=\int_{f_\text{min}}^{f_\text{max}} \frac{df}{2\pi}\frac{d\phi}{df},
\end{equation}
where $\phi$ is the gravitational wave phase. The instantaneous number of cycles spent near some frequency $f$ is defined by multiplying the above integrand by $f$
\begin{equation}
\label{Ninst}
N(f)\equiv\frac{f}{2\pi}\frac{d\phi}{df}.
\end{equation}
However, GW detectors are not equally sensitive to all parts of the waveform because the noise spectral density of 
the detector is frequency dependent. A better proxy for how observationally different two waveforms are is to compare the number of ``useful" cycles in each. This measure was originally introduced in Ref.~\cite{Damour:2000gg}.
One computes the total phase accumulated in each frequency bin and then weights this estimate by the sensitivity of a detector at that frequency. Because the strain sensitivity of the detector is concentrated in just a window of frequency space, the result would also depend on the mass of the system. The number of useful cycles is defined by~\cite{Damour:2000gg}
\begin{equation}
\label{Nuseful}
N_\text{useful}(f)\equiv\left[\int_{f_\text{min}}^{f_\text{max}} \frac{df}{f} w(f)N(f)\right] \left[\int_{f_\text{min}}^{f_\text{max}}\frac{df}{f}w(f)\right]^{-1},
\end{equation}
where the weight  $w(f)\equiv A^2(f)/fS_n(f)$, while $A(f)$ is the GW amplitude, and $S_n(f)$ is the noise spectral 
density of the detector.
We use the zero-detuned high-power noise spectral density of Advanced LIGO at design sensitivity \cite{LIGOsensitivity}.

Using the balance equation $dE/dt=-\mathcal{F}$, and the relation
between the GW phase and orbital frequency $\dot{\phi}=\Omega$, the
instantaneous number of cycles in Eq.~\eqref{Ninst} can be
reformulated as
\begin{equation}
N(f)=-\frac{v^4}{3\pi MG_{12}}\frac{E'(v)}{\mathcal{F}(v)}\,,
\end{equation}
which can be computed in the quadrupole-driven regime using Eq.~\eqref{approxQD}.
For the GW amplitude, we used the Newtonian order approximation for the transverse-traceless polarizations $A(f)\propto v^2$, since the effect from the amplitude on the number of cycles is small compared to the phase.
We can then calculate numerically the number of useful cycles using Eq.~\eqref{Nuseful}. 

\begin{figure}[t]
	\centering
	\includegraphics[width=\linewidth]{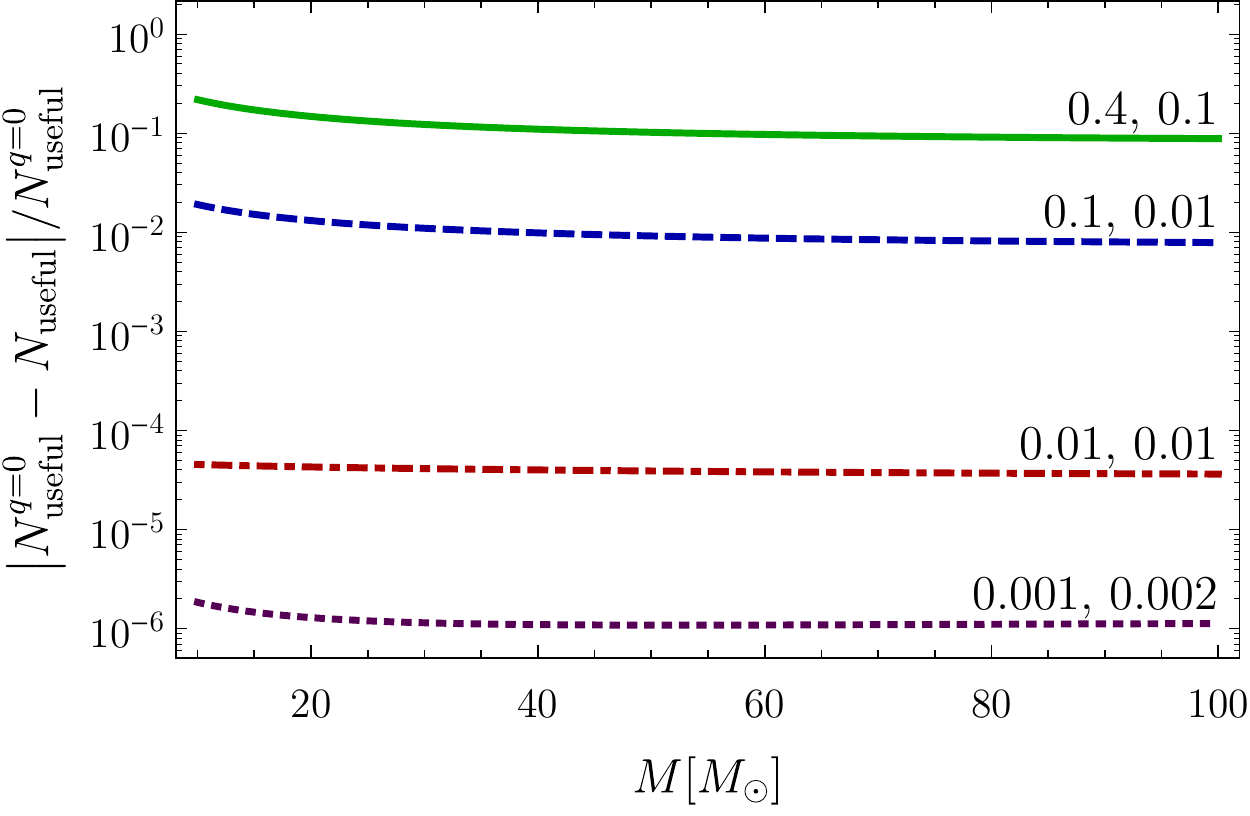}
	\caption{\label{fig:Nuseful} Number of useful cycles versus the total mass for various charge-to-mass ratios, and for equal masses ($\nu=1/4$). The number of cycles in EMd theory is less than in GR except for equal charges. }
\end{figure}

In Fig.~\ref{fig:Nuseful}, we show the relative difference between $N_\text{useful}$ in EMd theory with $a=1$ and the same quantity when all charges are zero (GR to 1PN order).
The number of cycles in EMd theory is less than in GR except for equal charges, because the leading dipole radiation dominates the Newtonian order corrections to the binding energy. 
We find that for systems with $q_i/m_i \sim 0.1$, the number of useful cycles in GR and EMd differs by $\mathcal{O}(1)$.

The quantity plotted in Fig.~\ref{fig:Nuseful} provides a rough estimate of the observable size of deviations from GR relative to the overall GW signal strength.
We recast this quantity in terms of the optimal signal-to-noise ratio (SNR) of the waveforms, defined by
\begin{align}\label{eq:SNRdef}
\text{SNR}^2 = 4 \int_{f_\text{min}}^{f_\text{max}} df \frac{|\mathcal{A}(f)|^2}{S_n(f)}.
\end{align}
Using Eq.~\eqref{eq:SPAamp}, this relation can be rewritten as
\begin{align}
\text{SNR}^2= 4 \int_{f_\text{min}}^{f_\text{max}} \frac{df}{f} w(f) N(f),
\end{align}
and thus
\begin{align}
\frac{|N^{q=0}_\text{useful}-N_\text{useful}|}{N^{q=0}_\text{useful}} =& \frac{|\left(\text{SNR}^2\right)^{q=0} - \left(\text{SNR}^2\right)|}{\left(\text{SNR}^2\right)^{q=0}} \\
& = \frac{2|\Delta\text{SNR}|}{\text{SNR}}+\mathcal{O}\left(\left(\frac{\Delta \text{SNR}}{\text{SNR}}\right)^2\right),
\end{align}
where, $\Delta \text{SNR} = \left(\text{SNR}^{q=0}-\text{SNR}\right)$ is the difference in SNR between signals in GR and EMd theory.
Thus, Fig.~\ref{fig:Nuseful} indicates that corrections arising from the presence of electric and scalar charges in EMd theory can account for only a few percent of the total SNR for systems with electric dipole $\sim 0.1$.

\section{Effective-one-body  framework}
\label{sec:EOB}

In this section, we construct two EOB Hamiltonians: one based on the EMd metric in Eq.~\eqref{EMdmetric}, in which the potential $C(r)\neq 1$, which we call the GHS gauge; the other is based on an approximation to the EMd metric by making a transformation to a gauge were the potential $C(r)=1$, which we call the Schwarzschild gauge.

The EOB Hamiltonian in the GHS gauge is more physical in the strong-gravity regime since it exactly reproduces the test-body limit of the two-body dynamics.
That is, it belongs to a class of Hamiltonians implementing exact solutions to the field equations for isolated objects/BHs.
However, this class of Hamiltonians is very theory specific --- for example the analytic ST vacuum metric in Refs.~\cite{Coquereaux:1990qs,Julie:2017ucp} is distinct from the analytic EMd metric when we set the electromagnetic fields to zero.
In addition, many BH solutions in alternative theories do not even have an analytic solution that can be used.
The advantage of using a Hamiltonian based on the approximate metric in the Schwarzschild gauge, is that it is easier to implement in data-analysis studies of GWs observed by LIGO and Virgo.
One would take the existing EOB Hamiltonians in GR as a starting point and add EMd corrections in the same way as, e.g., tidal corrections are added~\cite{Damour:2009wj}.
Within the regime of small deviations from GR, the two EOB Hamiltonians in EMd theory are expected to closely agree.

In Refs.~\cite{Julie:2017pkb, Julie:2017ucp}, the EOB framework was extended to ST theories.
In Ref.~\cite{Julie:2017ucp}, the motion of a binary BH was mapped to the motion of a test body, such that the effective metric is a $\nu$-deformation of the ST metric.  
This approach is similar to our EOB Hamiltonian in the GHS gauge, but we find a different mapping for the scalar charge.
In Ref.~\cite{Julie:2017pkb}, the motion of the binary in ST theory was mapped to the motion of a test body around an effective BH in GR, but the effective metric does not reproduce exactly the test-body limit of ST theory.
In contrast, whereas our EOB Hamiltonian in the Schwarzschild gauge is also not exact in the test-body limit, it still maps the real problem to an effective one in EMd theory (not in GR).

\subsection{Effective-one-body Hamiltonian in Garfinkle-Horowitz-Strominger gauge}
\label{sec:GHSEOB}

In the EOB framework, the motion of a binary is mapped to the motion of a test body in the background of an effective metric.
In the effective problem in EMd theory, we assume that a test body, with mass $\mu$ and electric charge $q$, is moving in the background of a charged BH with mass $M$ and electric charge $Q$.
To relate the real two-body problem to the effective one, we impose the following conditions:
(a) $M$ and $\mu$ are the total mass and reduced mass of the real description, i.e., $M=m_1+m_2$ and $\mu=m_1m_2/M$;
(b) the effective charges $Q$ and $q$ are related to the real charges by $Qq = q_1q_2$, but we do not assume that $Q$ is the total charge;
and (c) the mapping between the real and effective Hamiltonians takes the form
\begin{equation}
\label{HrealHeff}
\frac{H_\text{eff}^\text{NR}(\bm{R},\bm{P})}{\mu} = \frac{H^\text{NR}(\bm{r},\bm{p})}{\mu}\left[1+\frac{\nu}{2}\frac{H^\text{NR}(\bm{r},\bm{p})}{\mu}\right],
\end{equation}
where the superscript NR means non-relativistic, i.e., $H^\text{NR}=H-M$, and the real Hamiltonian $H$ is given by Eq.~\eqref{Hreal}.
The form (\ref{HrealHeff}) for the ``EOB energy map'' \cite{Buonanno:1998gg} has proven useful in GR up to 4PN order  \cite{Damour:2015isa}, in classical electrodynamics to 2PN order \cite{Buonanno:2000qq}, and in ST gravity to 2PN order \cite{Damour:2016gwp, Julie:2017pkb}.  In the first post-Minkowskian approximation, i.e., to all orders in $v/c$ at linear order in $G$, it can be shown to exactly resum the dynamics, producing the arbitrary-mass-ratio two-body Hamiltonian from the test-body Hamiltonian  \cite{Damour:2016gwp,Vines:2017hyw}.
For the coordinates in the effective problem, we use uppercase letters, such as $R$ and $P$, while for the real problem, we keep using lowercase letters, such as $r$ and $p$.

The effective action for the test body  is given by
\begin{equation}
S_\text{eff}=\int\left[-\mathfrak{m}(\varphi)\,d\tau_\text{eff}+qA_\mu dX^\mu\right],
\end{equation}
where $\tau_\text{eff}$ is the proper time of the BH and the effective test-mass $\mathfrak{m}(\varphi)$ depends on the scalar field $\varphi$ generated by the  BH, and has the expansion in terms of the parameters $\alpha$ and $\beta$  as
\begin{equation}\label{mphi}
\mathfrak{m}(\varphi) = \mu \left[1 + \alpha \varphi
+ \frac{1}{2} (\alpha^2 + \beta) \varphi^2  +\Order\left(1/c^6\right) \right].
\end{equation}
Since we do not know, a priori, how the parameters $\alpha$ and $\beta$ of the effective test body are related to the real problem, we 
expand the mass in a $1/R$ expansion 
\begin{equation}\label{mRf}
\mathfrak{m}(R)=\mu\left[1+\frac{f_1}{R}+\frac{f_2}{R^2}+\Order\left(1/c^6\right)\right]
\end{equation}
and solve for the unknown coefficients $f_1$ and $f_2$. 

We take the effective metric of the background to be a deformation of the EMd metric in the GHS gauge
\begin{equation}
\label{metricansatz}
ds_\text{eff}^2=-d\tau_\text{eff}^2=-A(R)dT^2+B(R)dR^2+R^2 C(R) d\Omega^2 \,,
\end{equation}
with 
\begin{subequations}
\begin{align}
A(R)&= \left(1-\frac{R_+}{R}\right)\left(1-\frac{R_-}{R}\right)^{\frac{1-a^2}{1+a^2}}, \\
B(R)&= \frac{1}{A(R)} \left(1+\frac{b_1}{R}\right), \\
C(R)&= \left(1-\frac{R_-}{R}\right)^{\frac{2a^2}{1+a^2}},
\end{align}
\end{subequations}
where $R_-$ and $R_+$ are the radii of the inner and outer horizons of the effective BH, which are given by Eqs.~\eqref{rminus} and \eqref{rplus}, i.e., 
\begin{equation}\label{Rpm}
R_- = \frac{1+a^2}{a} D\,,  \qquad
R_+ =2M - \frac{1-a^2}{a}D\,.
\end{equation}
We choose to define $R_-$ and $R_+$ by these relations in terms of $D$, but not in terms of $Q$, because the relation between $Q$ and $D$ is deformed by the mapping. We note that in the above metric's ansatz, we have added a deformation to $B(R)$ only because, in EMd theory at 1PN order, the mapping leads to three equations in $f_1$, $f_2$, and any deformation to the metric.
Thus, we can only determine uniquely one unknown coefficient in the effective metric. So we choose to take that coefficient to be $b_1$, and assume the possible deformations to $A(R)$ or $C(R)$ to be zero at 1PN order.

The scalar field for a single BH is given by Eq.~\eqref{phiBH}; we add a PN deformation $g_2/R^2$ such that the effective scalar field is given by
\begin{equation} 
\label{phiGHS}
\varphi(R)=\frac{a}{1+a^2} \ln\left(1-\frac{R_-}{R} +\frac{1+a^2}{a}\frac{g_2}{R^2}\right).
\end{equation}
The electric potential is given by
\begin{equation}
A_0(R)= -\frac{Q}{R}\,.
\end{equation}
We do not add PN corrections to $A_0$ because those corrections can be absorbed in the PN corrections to the scalar field or to the relation between $D$ and $Q$.
The coefficient $g_2$ is not independent of $f_1$ and $f_2$, because the mass expansion can also be expanded directly in $\varphi$ [see Eq.~\eqref{Mphi}]
\begin{align}\label{mRs}
\mathfrak{m}(R) &= \mu\bigg[1 -\frac{D\alpha}{R} +\frac{1}{R^2}\bigg(g_2\alpha -\frac{D^2\alpha}{2a} -\frac{a}{2}D^2\alpha \nonumber\\
&\qquad\quad +\frac{1}{2}D^2\alpha^2+\frac{1}{2}D^2\beta\bigg) + \Order\left(1/c^6\right) \bigg].
\end{align}
In what follows, we uniquely solve for the coefficients $b_1$, $f_1$, and $f_2$ by matching the real Hamiltonian to the effective one by a canonical transformation. 
Matching the two mass expansions in Eqs.~\eqref{mRf} and \eqref{mRs} allows us to determine the mapping for the parameters $\alpha$ and $\beta$, and for the coefficient $g_2$.
The mapping for $\alpha$ is unique, but the mapping for $\beta$ and $g_2$ is not unique at 1PN order.

To find the effective Hamiltonian, we first find the effective Lagrangian, in the equatorial plane $\Theta = \pi/2$,
\begin{align}
L_\text{eff}&=qA_0 - \mathfrak{m}(\varphi)\sqrt{-g_{\mu\nu}\frac{dX^\mu}{dT}\frac{dX^\nu}{dT}}, \nonumber\\
&=qA_0 - \mathfrak{m}(\varphi)\sqrt{A(R)-B(R)\dot{R}^2-C(R)R^2\dot{\Phi}^2}.
\end{align}
Then, applying the Legendre transformation $H_\text{eff}=P_R\dot{R}+P_\Phi\dot{\Phi}-L_\text{eff}$ yields the effective Hamiltonian
\begin{equation}
\label{Heff}
H_{\text{eff}} = -qA_0 + \sqrt{A(R)\left[\mathfrak{m}^2(\varphi)+\frac{P_\Phi^2}{C(R)R^2}+\frac{P_R^2}{B(R)}\right]},
\end{equation}
where $P_\Phi=\partial L_\text{eff}/\partial \dot{\Phi}$ is the angular momentum, and $P_R=\partial L_\text{eff}/\partial \dot{R}$ is the radial momentum.

Before matching the Hamiltonians, we need to apply a  canonical transformation from the real variables,  $\bm{r}$ and $\bm{p}$, to the  effective ones, $\bm{R}$ and $\bm{P}$. At 1PN order, this transformation is given  by \cite{Buonanno:1998gg}
\begin{equation}
\label{cantrans}
R^i=r^i+\frac{\partial G_\text{1PN}}{\partial p_i}\,, \qquad
P_i=p_i-\frac{\partial G_\text{1PN}}{\partial r^i} \,, 
\end{equation}
with the generating function 
\begin{equation}\label{genfunc}
G_\text{1PN}(\bm{r},\bm{p})=(\bm{r}\cdot \bm{p})\left(c_1 \bm{p}^2+\frac{c_2}{r}\right),
\end{equation}
where the coefficients $c_1$ and $c_2$ are to be determined by the mapping.

Inserting the expansions of the real and effective Hamiltonians into Eq.~\eqref{HrealHeff}, and applying the canonical transformation, we obtain the five equations:
\begin{subequations}
\begin{align}
&2c_1\mu^2+\nu=0\,, \\
&f_1+M\alpha_1\alpha_2=0\,, \\
&M-c_2+\mu+\mu\alpha_1\alpha_2-\frac{qQ}{M}+aD+c_1M\mu^2 \nonumber\\
&\qquad -\mu c_1qQ -\mu c_1 f_1\mu =0\,, \\
&b_1+\frac{qQ}{M} +2M +2aD +4c_1\mu qQ +4c_1\mu^2 f_1 -2c_2  \nonumber\\
&\qquad -\mu-\mu\alpha_1\alpha_2 -4c_1 M\mu^2=0\,, 
\end{align}
\begin{widetext}
\begin{align}
&\frac{q^2Q^2}{M^2} +q_2^2X_1\left(1+a\alpha_1\right) +q_1^2X_2\left(1+a\alpha_2\right) 
-2\mu c_2+4\mu f_1 +2\nu c_2f_1 -\nu f_1^2 -2\nu f_2 
+2M\mu -\frac{2D\mu}{a} +2aD\mu  \nonumber\\
&\quad\quad 
-\nu D^2 +\nu\frac{D^2}{a^2} +4M\mu \alpha_1\alpha_2 +2M\mu\alpha_1^2\alpha_2^2 +X_2\nu \beta_1\alpha_2^2 + X_1\nu\beta_2\alpha_1^2 +\mu^2 
+2\mu^2\alpha_1\alpha_2 +\mu^2\alpha_1^2\alpha_2^2 \nonumber\\
&\quad\quad 
+qQ\left(-2 +2\frac{c_2}{M} -2\frac{f_1}{M} -2a\alpha_1X_1 -2a\alpha_2X_2 -2\alpha_1\alpha_2 -2\nu -2\nu\alpha_1\alpha_2\right)=0\,.
\end{align}
\end{widetext}
\end{subequations}
Solving these equations respectively for the coefficients $c_1$, $f_1$, $c_2$, $b_1$, and $f_2$  yields
\begin{subequations}
\begin{align}
c_1 &=-\frac{\nu}{2\mu^2}\,,  \\
f_1 &= -M\alpha_1\alpha_2\,,  \\
c_2 &=M+\frac{M\nu}{2}+\frac{1}{2}M\nu\alpha_1\alpha_2-\frac{qQ\nu}{2\mu}+aD\,,  \label{c2GHS}\\
b_1 &=0\,, \\
f_2 &= \frac{D^2}{2a^2} -\frac{D^2}{2} -\frac{MD}{a} -aMD\alpha_1\alpha_2 + \frac{aqQD}{\mu} \nonumber\\
&\quad
-M^2\left[\alpha_1\alpha_2 -\frac{1}{2}(\alpha_1\alpha_2)^2 -\frac{1}{2}\left(X_2\alpha_2^2\beta_1 +X_1\alpha_1^2\beta_2\right)  \right] \nonumber\\
&\quad +\frac{M}{2}\left[\frac{q_2^2}{m_2}(1+a\alpha_1) +\frac{q_1^2}{m_1}(1+a\alpha_2)\right]
 \nonumber\\
&\quad -aM \frac{q_1q_2}{\mu}\left(X_1\alpha_1+X_2\alpha_2\right).
\label{f2GHS}
\end{align}
\end{subequations}
	
To find the mapping of the scalar charge, we identify the mass expansion in Eq.~\eqref{mRf} with the expansion in Eq.~\eqref{mRs} to give
\begin{subequations}
\begin{align}
&-D\alpha = f_1\,, \\
&g_2\alpha -\frac{D^2\alpha}{2a} -\frac{a}{2}D^2\alpha +\frac{1}{2}D^2\alpha^2+\frac{1}{2}D^2\beta = f_2\,.
\end{align}
\end{subequations}
Inserting the solution for $f_1$ and $f_2$ gives a unique mapping for $\alpha$ 
\begin{equation}
\alpha = \frac{M}{D}\alpha_1\alpha_2\,, 
\end{equation}
and suggests the following mapping for $\beta$
\begin{equation}
\beta = \frac{M^2}{D^2} \left(X_2\alpha_2^2\beta_1 +X_1 \alpha_1^2\beta_2\right).
\end{equation}
Further, we take the mapping of the dilaton charge $D$ of the effective BH to be the sum of the asymptotic value of the scalar charges of the two bodies, i.e.,
\begin{equation} 
D = m_1 \alpha_1 + m_2 \alpha_2 \,.
\end{equation}
The mapping for $\alpha$ and $\beta$ agrees with what was found in Ref.~\cite{Julie:2017ucp}, but the mapping for $D$ is different.
The reason we choose this mapping for $D$ is that it leads to a simple deformation to the scalar field 
\begin{equation}
\label{g2GHS}
g_2=-\frac{1-a^2}{2a^2}\frac{(\alpha_1-\alpha_2)^2}{\alpha_1\alpha_2}DM\nu\,.
\end{equation} 
This deformation vanishes in the test-mass-limit $\nu\to0$, and also when $a=1$ or $\alpha_1=\alpha_2$. Other choices for $D$ lead to complicated expressions for $g_2$.
In obtaining this result for $g_2$, we used the expression for the electric charge in terms of the scalar charge, which is valid for BHs only,
\begin{equation}
\frac{q_i^2}{m_i^2}  = \frac{2}{a} \alpha_i - \frac{1-a^2}{a^2} \alpha_i^2\,.
\end{equation}
This relation follows from Eq.~\eqref{alphaphi} after solving for $q_i$ in terms of $\alpha_i$ and setting the scalar field to its asymptotic value.

A convenient mapping for the electric charge is 
\begin{equation}
Q^2 = M\left(\frac{q_1^2}{m_1} + \frac{q_2^2}{m_2}\right).
\end{equation}
The reasoning behind this choice is that it is symmetric under the exchange of the two bodies; it has the correct test-body limit, $Q\to q_1$ when $m_2/m_1\rightarrow 0$ with $q_2/m_2$ held constant; and it appears naturally in EM theory as we show in the next subsection. 
With that mapping for $Q$ and $D$, the relation between them is given by
\begin{equation}\label{QDEOB}
Q^2 = \frac{2M}{a} D - \frac{1-a^2}{a^2} D^2 - \frac{1-a^2}{a^2}(\alpha_1-\alpha_2)^2 M^2\nu\,.
\end{equation}
One could choose to enforce Eq.~\eqref{QD} for generic masses by making a different choice for $D$ or $Q$, but this seems to lead to very complicated expressions for them.

\subsection{Effective-one-body Hamiltonian in Schwarzschild gauge}
\label{sec:RNEOB}

In the EMd metric, the potential $C(r)\neq 1$, but the standard EOB gauge is the Schwarzschild gauge $C(r)=1$. 
This is the gauge that was used to derive the original EOB Hamiltonian~\cite{Buonanno:1998gg}, which was then improved by calibrating it 
to numerical-relativity simulations \cite{Pan:2009wj}.
Therefore, to profit from the best available EOB Hamiltonian in GR, we need to construct an EMd-EOB 
Hamiltonian that is also in the Schwarzschild gauge.

The EMd metric can be transformed to the Schwarzschild gauge by the coordinate transformation $\bar{r}^2=r^2C(r)$. However, for arbitrary values of the coupling constant $a$, the metric cannot be analytically transformed. 
Instead, we expand the EMd metric \eqref{EMdmetric} and transform it to get an approximate EMd metric in the Schwarzschild gauge.
We make the coordinate transformation, valid to 1PN order, 
\begin{align}
\label{transf}
\bar{r}^2  &= r^2\left[1-\frac{2a^2r_-}{(1+a^2)r}\right], \nonumber\\
\Rightarrow  \quad r&= \bar{r}+\frac{a^2}{1+a^2}r_-=\bar{r}+aD\,.
\end{align}
With that transformation, and inserting the expressions for $r_-$ and $r_+$ in terms of $M$ and $Q$ [Eqs.~\eqref{rminus} and \eqref{rplus}], we get 
\begin{equation}
\label{approxmetric}
ds^2=-\left(1-\frac{2M}{\bar{r}}+\frac{Q^2}{\bar{r}^2}\right) dt^2 
+\left(1+\frac{2M}{\bar{r}}\right)d\bar{r}^2+\bar{r}^2d\Omega^2,
\end{equation}
which is the same as the Reissner--Nordstr\"{o}m metric to 1PN order.

As an ansatz for the effective metric, we assume a metric based on the approximate metric (\ref{approxmetric})
\begin{equation}
ds_\text{eff}^2=-A(R)dt^2+B(R)dR^2+R^2 d\Omega^2 \,,
\end{equation}
with 
\begin{subequations}
\begin{align}
A(R)&=\, 1+\frac{a_1}{R}+\frac{a_2}{R^2}+\dots\,, \\
B(R)&=\, 1+\frac{b_1}{R}+\dots\,,
\end{align}
\end{subequations}
and we write the mass expansion as
\begin{equation}
\label{mR1}
\mathfrak{m}(R)=\mu\left[1+\frac{f_1}{R}+\frac{f_2}{R^2} +\Order\left(1/c^6\right)\right],
\end{equation}
where the unknown coefficients $a_1, a_2, b_1, f_1,$ and $f_2$ are to be determined by the mapping.
However, the mapping leads to three equations in those five coefficients, making two of them arbitrary. 
We choose to take $a_1=-2M$ and $a_2=Q^2$ so that the effective metric would agree with the EMd metric in the Schwarzschild gauge to 1PN order. 
When we solve for $b_1$, we get $b_1=2M$, in agreement with the EMd approximate metric.

For the effective electric potential, we apply the coordinate transformation \eqref{transf} with $\bar{r}=R$ to get
\begin{equation}
A_0(R) = -\frac{Q}{R+aD}\,.
\end{equation}
Applying the same transformation to the scalar field, and adding a PN deformation $g_2/R^2$, we obtain
\begin{equation} 
\label{phiSchw}
\varphi(R)=\frac{a}{1+a^2} \ln\left[1-\frac{1+a^2}{a}\frac{D}{R + aD} +\frac{1+a^2}{a}\frac{g_2}{R^2}\right].
\end{equation}
The mass expansion in terms of $\varphi$, Eq.~\eqref{mphi}, can now be written as an expansion in $1/R$ by
\begin{align}
\label{mR2}
\mathfrak{m}(R) &= \mu\bigg[1 -\frac{D\alpha}{R} +\frac{1}{R^2}\bigg(g_2\alpha -\frac{D^2\alpha}{2a} +\frac{a}{2}D^2\alpha \nonumber\\
&\qquad +\frac{1}{2}D^2\alpha^2+\frac{1}{2}D^2\beta\bigg) + \Order\left(1/c^6\right) \bigg].
\end{align}

Following the same method used in the previous subsection, the effective Hamiltonian is given by Eq.~\eqref{Heff} with the potential $C(R)=1$. The relation between the real and effective Hamiltonians is given by Eq.~\eqref{HrealHeff}, and the canonical transformation that relates the real and effective variables is given by Eq.~\eqref{cantrans}. 
Matching the real and effective Hamiltonians, we obtain the five equations:
\begin{subequations}
\begin{align}
&2c_1\mu^2+\nu=0\,,  \\ &a_1+2f_1+2M(1+\alpha_1\alpha_2)=0\,, \\
&2c_2 -a_1 -4M + c_1a_1\mu^2 +2f_1c_1\mu^2 +2\frac{qQ}{M} \nonumber\\
&\qquad -2\mu(1+\alpha_1\alpha_2 -c_1qQ) =0\,, \\
&b_1 -2c_2 +4c_1\mu qQ +2a_1c_1\mu^2 +4c_1f_1\mu^2-\mu \nonumber\\
&\qquad -\mu\alpha_1\alpha_2+\frac{qQ}{M} =0\,, \\
&M^2\alpha_1^2\alpha_2^2 +M^2X_2\alpha_2\beta_1 +M^2X_1\alpha_1^2\beta_2 -a_2 -2M^2\alpha_1\alpha_2 \nonumber\\
&\qquad  +\frac{2aqQD}{\mu} +\frac{m_1q_2^2}{\mu}(1+a\alpha_1) +\frac{m_2q_1^2}{\mu}(1+a\alpha_2)  \nonumber\\
&\qquad  -2a\frac{qQ}{\mu}(m_1\alpha_1 +m_2\alpha_2)-2f_2=0\,.
\end{align}
\end{subequations}
Solving these equations respectively for the coefficients $c_1$, $f_1$, $c_2$, $b_1$, and $f_2$  yields
\begin{subequations}
\begin{align}
c_1&=-\frac{\nu}{2\mu^2}\,, \\ f_1&=-M\alpha_1\alpha_2\,,  \\ c_2&=M+\frac{M\nu}{2}+\frac{1}{2}M\nu\alpha_1\alpha_2-\frac{qQ\nu}{2\mu}\,, \label{c2Schw}\\
b_1&=2M\,,  \\
f_2&=-\frac{a_2}{2} + \frac{aq_1q_2D}{\mu} -a \frac{q_1q_2}{\mu}\left(m_1\alpha_1 +m_2\alpha_2\right) \nonumber\\
&\quad -M^2\left[\alpha_1\alpha_2 -\frac{1}{2}(\alpha_1\alpha_2)^2 -\frac{1}{2}\left(X_2\alpha_2^2\beta_1 +X_1\alpha_1^2\beta_2\right)  \right] 
\nonumber\\
&\quad  +\frac{M}{2}\left[\frac{q_2^2}{m_2}(1+a\alpha_1) +\frac{q_1^2}{m_1}(1+a\alpha_2)\right]. \label{f2Schw}
\end{align}
\end{subequations}

Choosing $a_2=Q^2$, so that the effective metric agrees with the EMd metric to 1PN order, the above solution for $f_2$ leads to the mapping
\begin{equation}\label{Qmap}
Q^2 = M\left(\frac{q_1^2}{m_1} + \frac{q_2^2}{m_2}\right).
\end{equation}
This is because, for the case of EM theory, when we take the parameters $\alpha$ and $\beta$ in the solution for $f_2$ to be zero, we get $f_2=-a_2/2 + M(q_1^2/m_1+q_2^2/m_2)/2$. Hence, requiring that $f_2=0$ in EM theory and that $a_2=Q^2$, naturally leads to the charge map \eqref{Qmap}.

Identifying the mass expansion in Eq.~\eqref{mR1} with that in Eq.~\eqref{mR2}, leads to the following mapping for $\alpha$ and $\beta$
\begin{align}
\alpha &=\frac{M}{D}\alpha_1\alpha_2\,, \\
\beta &= \frac{M^2}{D^2} \left(X_2\alpha_2^2\beta_1 +X_1\alpha_1^2\beta_2\right), \label{effbeta}
\end{align}
which is the same mapping that was found in the previous subsection.
Further, taking the mapping of the dilaton charge to also be given as in the previous subsection
\begin{equation}
\label{simpleD}
D=m_1\alpha_1 + m_2\alpha_2,
\end{equation}
leads to the astonishingly simple result
\begin{equation}
g_2 = 0.
\end{equation}
With that mapping for $D$ and $Q$, the relation between them is given by Eq.~\eqref{QDEOB}.

Interestingly, the above mappings also lead to a ST EOB Hamiltonian  in Schwarzschild gauge at 1PN order. A 2PN EOB Hamiltonian based on an exact analytic solution  for the metric and scalar field can be found in Ref.~\cite{Julie:2017ucp}.
The metric in that work also includes a potential $C(R)\neq 1$, Eq.~(II.3) in Ref.~\cite{Julie:2017ucp}, and that metric is unrelated to the EMd metric when the electric charges are zero. 
The scalar field is given by
\begin{equation}
\varphi_\text{ST} = \frac{D}{a_*} \log \left[ 1 - \frac{a_*}{r} + \frac{a_*^2 - 2 M a_*}{2 r^2} \right] ,
\end{equation}
where $a_*^2 = 4 (M^2 + D^2)$. The author of Ref.~\cite{Julie:2017ucp} found the same mapping for $\alpha$ and $\beta$ that we got, but used a different mapping for $D$ (at 2PN order).
When we approximately transform the metric and the scalar field to the Schwarzschild gauge, in which the potential $C(R)=1$, and repeat the same analysis in this section, we get an EOB Hamiltonian with the same mapping for the scalar charge given in Eq.~\eqref{simpleD}, and with no deformation to the metric or the scalar field to 1PN order.
The point is that the mapping of the scalar charge would be the same in EMd theory and ST theory, which is another hint that Eq.~\eqref{simpleD} is a good choice at 1PN order.

\subsection{Comparison of two effective-one-body Hamiltonians in Einstein-Maxwell-dilaton theory}

\begin{figure*}[t]
	\centering
	\begin{minipage}[b]{0.49\linewidth}
		\includegraphics[width=\linewidth]{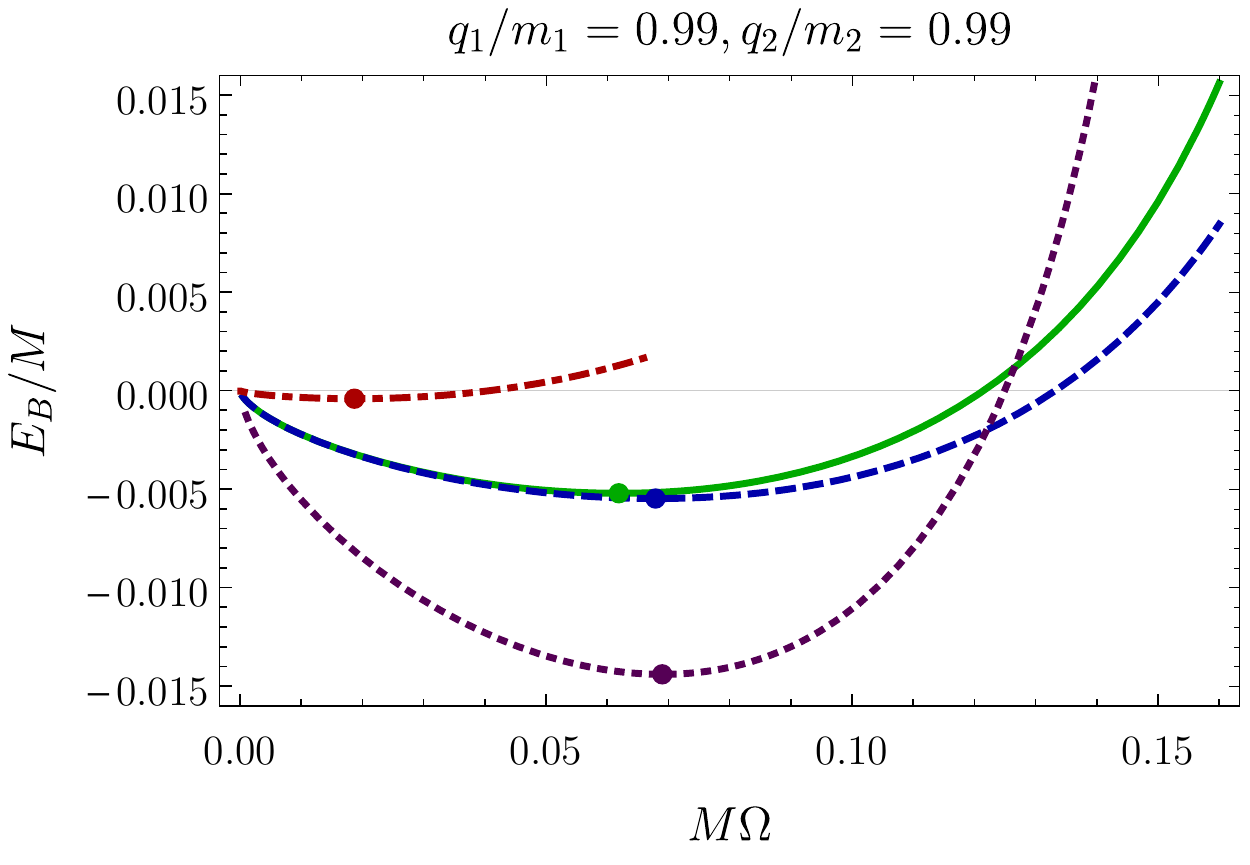}
		\includegraphics[width=\linewidth]{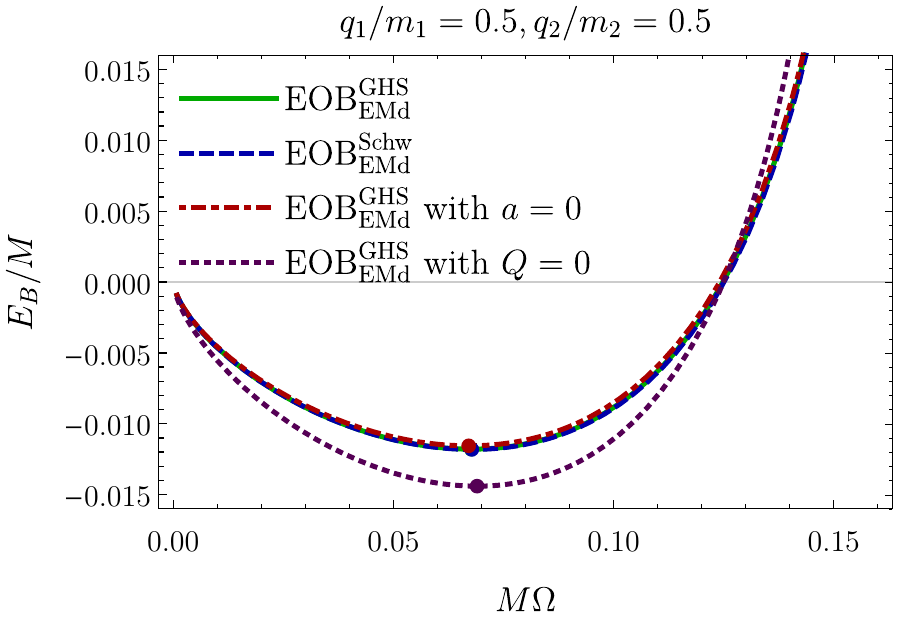}
	\end{minipage}
	\begin{minipage}[b]{0.49\linewidth}
		\includegraphics[width=\linewidth]{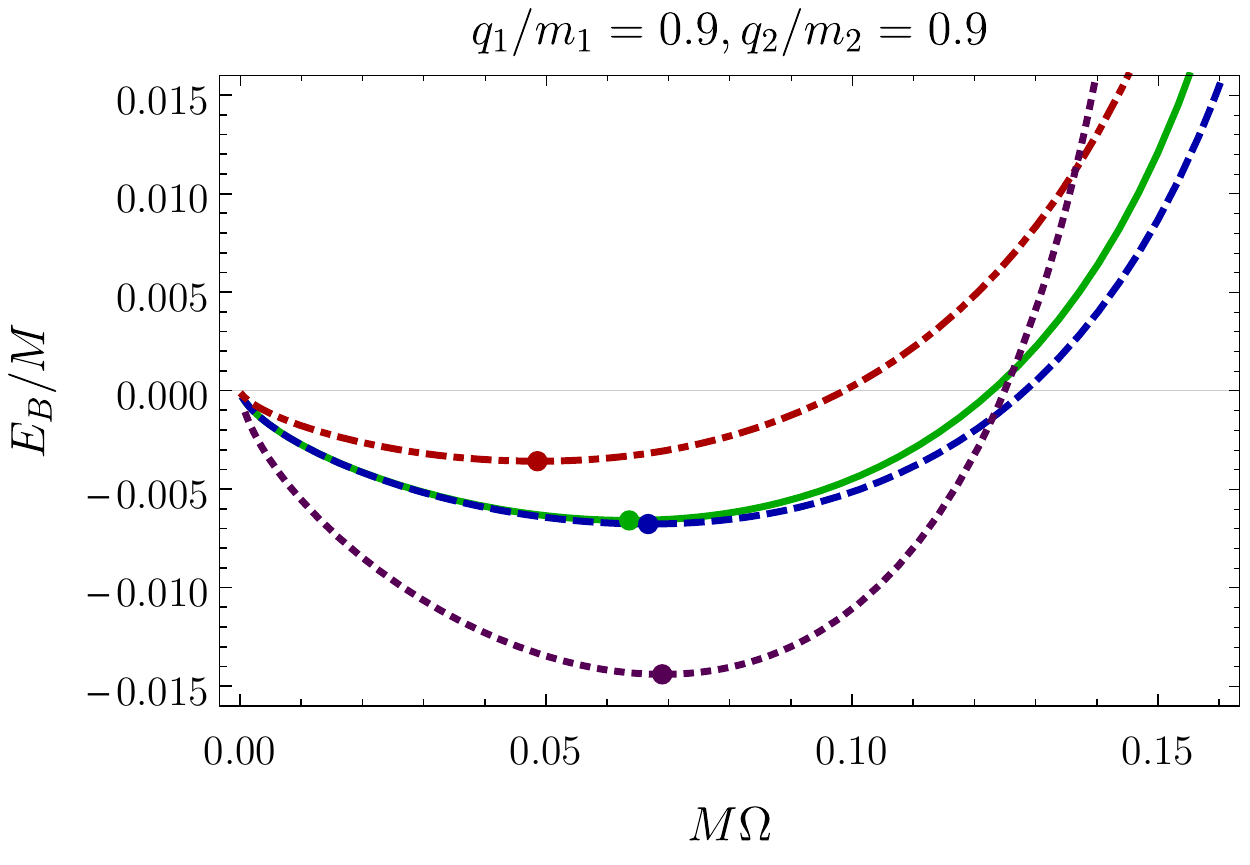}
		\includegraphics[width=\linewidth]{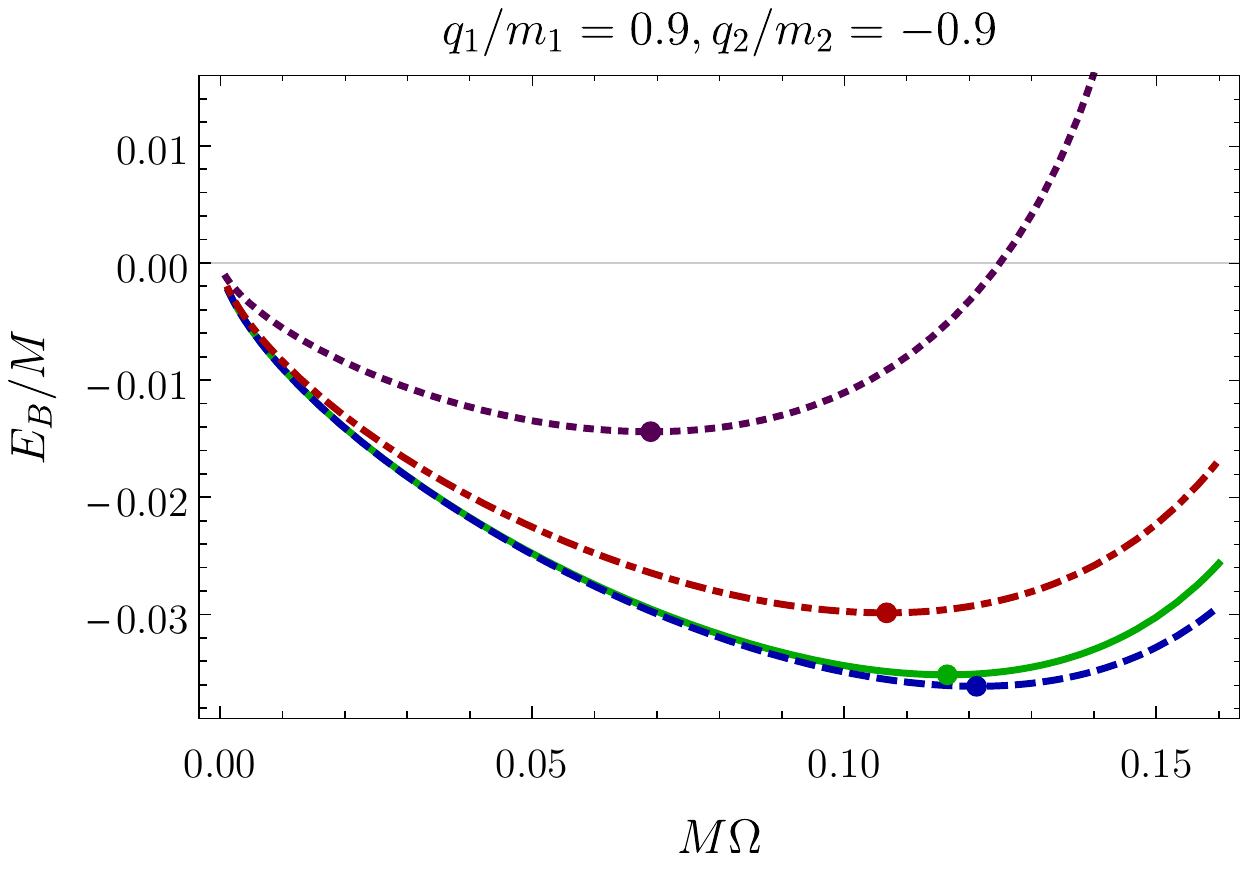}
	\end{minipage}
	\caption{\label{fig:ex}  Binding energy $E_B$ normalized by the total mass $M$ as a function of $M\Omega$ for equal masses, $\nu=1/4$, and for charge-to mass-ratios $q_1/m_1=q_2/m_2=0.99,~ 0.9,~ 0.5$, and $q_1/m_1=-q_2/m_2=0.9$. To improve readability, we show the plots only up to the frequency corresponding to $R = 1.05R_\text{LR}$ or to energy $E_B/M=0.015$.
	The point on each curve indicates the location of the ISCO.} 
	
\end{figure*}

In this subsection, we compare the two EMd-EOB Hamiltonians with each
other, and also with the EOB Hamiltonian in GR, by calculating the binding energy and the ISCO. The goal is to
investigate the range of parameter space where the two EMd-EOB Hamiltonians agree.

The mappings of the electric charge, scalar charge, and the parameters $\alpha$ and $\beta$ are the same for the two EMd-EOB Hamiltonians, i.e.,
\begin{align}
Q^2&=M\left(\frac{q_1^2}{m_1}+\frac{q_2^2}{m_2}\right), \quad 
D=m_1\alpha_1+m_2\alpha_2\,,\nonumber\\
\alpha&=\frac{M}{D}\alpha_1\alpha_2\,,
\quad 
\beta = \frac{M^2}{D^2}\left(X_2\alpha_2^2\beta_1 + X_1\alpha_1^2\beta_2\right).
\end{align}
For the EOB Hamiltonian in the GHS gauge, the effective metric is the GHS metric for $\nu=0$ [Eqs.~\eqref{metricansatz}--\eqref{Rpm} with $b_1=0$]. For the EOB Hamiltonian in the Schwarzschild gauge, the effective metric agrees with the Reissner-Nordstr\"{o}m metric for $\nu=0$ [Eq.~\eqref{approxmetric}].
Other differences between the two Hamiltonians are in the parameters of the mass expansion \eqref{mRf}, the canonical transformation \eqref{genfunc}, and the correction to the scalar field [Eqs.~\eqref{phiGHS} and \eqref{phiSchw}]. The parameters in those equations are shown in Table \ref{tab:EOBparms}.

\begin{table}[h]
\caption{Difference between the two EOB Hamiltonians in terms of the effective metric and the parameters of the mass expansion, the canonical transformation, and the scalar field.}
\begin{ruledtabular}
\begin{tabular}{ccc}
        &  EOB in GHS gauge & EOB in Schw gauge \\ 
	\hline 
effective metric & Eqs.~\eqref{metricansatz}--\eqref{Rpm}  & Eq.~\eqref{approxmetric} \\
$c_1$	&  \multicolumn{2}{c}{$c_1=-\nu/2\mu^2$} \\ 
$c_2$	& Eq.~\eqref{c2GHS} & Eq.~\eqref{c2Schw} \\ 
$f_1$	&  \multicolumn{2}{c}{$f_1=-M\alpha_1\alpha_2$} \\ 
$f_2$	& Eq.~\eqref{f2GHS} &  Eq.~\eqref{f2Schw}\\ 
$g_2$	& Eq.~\eqref{g2GHS} & $g_2=0$  \\ 
\end{tabular} 
\end{ruledtabular}
\label{tab:EOBparms}
\end{table}

\begin{figure*}[th]
	\centering
	\begin{minipage}[b]{0.49\linewidth}
		\includegraphics[width=\linewidth]{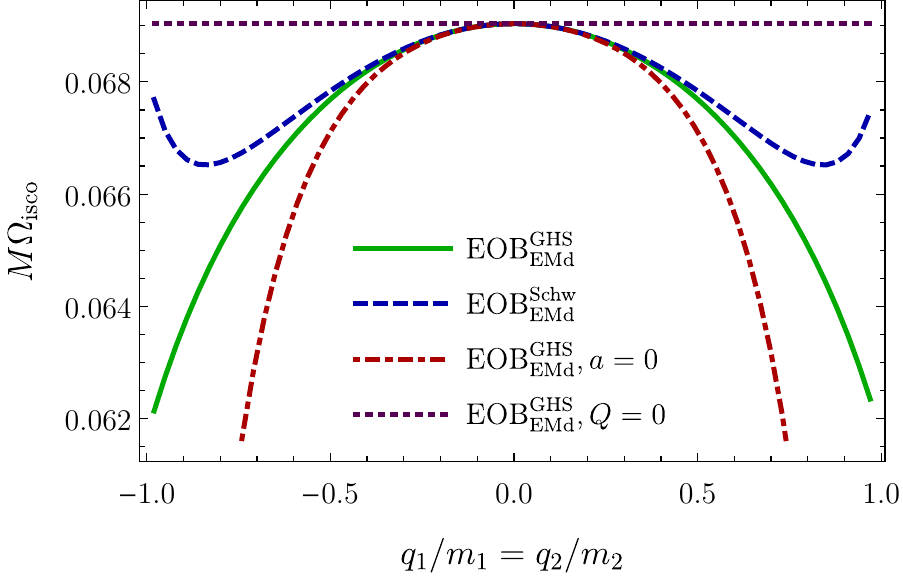} 
	\end{minipage}
	\begin{minipage}[b]{0.49\linewidth}
		\includegraphics[width=\linewidth]{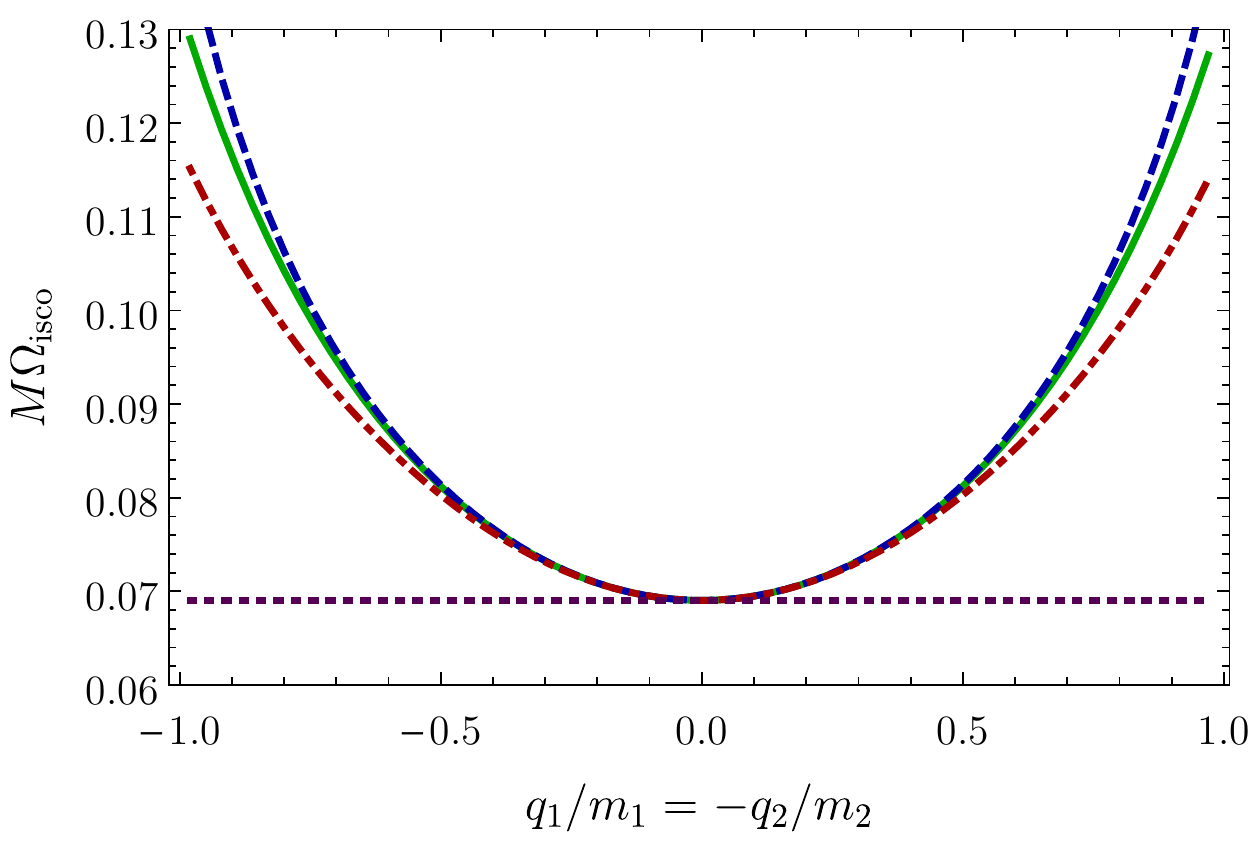}
	\end{minipage}
	\caption{\label{fig:Risco} Angular frequency at ISCO as a function of the charge-to-mass ratio $q_1/m_1$ from -0.99 to 0.99. In the left panel, $q_2/m_2=q_1/m_1$, while in the right, $q_2/m_2=-q_1/m_1$. An ISCO frequency of $0.062$ corresponds to an ISCO radius $\sim 6.4M$, and a frequency of $0.13$ corresponds to radius $\sim 3.9M$.}		
\end{figure*}

To find the binding energy from the two EOB Hamiltonians, we start with the energy map in Eq.~\eqref{HrealHeff}, which gives the relation between the effective Hamiltonian and the real  Hamiltonian.
Inverting that relation, we obtain the resummed EOB Hamiltonian
\begin{equation}
H_{\text{EOB}}^{\text{NR}}=M\sqrt{1+2\nu\left(\frac{H_{\text{eff}}}{\mu}-1\right)}-M\,.
\end{equation}
To obtain the binding energy for circular orbits, we set $P_R=0$, and solve $\dot{P}_R=-\partial H_{\text{eff}}/\partial R=0$ for the angular momentum $P_\Phi$. However, that equation cannot be solved analytically because of the non-linearity of the Hamiltonian. Hence, we solve the equation numerically for  $P_\Phi$ at specific values of $R$.
Since we want to plot the binding energy as a function of the orbital frequency $\Omega$, we need to calculate the orbital frequency via
\begin{equation}
\label{orbfreq}
\Omega=\frac{\partial H_\text{EOB}}{\partial P_\Phi}=\frac{\partial H_\text{EOB}}{\partial H_\text{eff}}\frac{\partial H_\text{eff}}{\partial P_\Phi}.
\end{equation}
Then, we calculate the binding energy and orbital frequency as $R$ goes from $100M$ to the radius of the light ring.
The light ring (or photon orbit) of a (charged) BH metric in GR is defined as the circular-orbit
  solution to the geodesic equation of massless particles. This geodesic
  equation is actually encoded by our effective Hamiltonian if we set
  $q=0$ (geodesic motion) and $\mu=0$ (massless particle).
To obtain the light-ring solution in EMd theory, we hence take the effective Hamiltonian for the case $\mu = 0 = q$, and impose the conditions for circular orbits $P_R = 0$ and $\dot{P}_R = 0$. The latter condition means that we look for an extremum of the effective Hamiltonian,
\begin{equation}
\begin{split}
  0 = \dot{P}_R &= - \left. \frac{\partial H_\text{eff}}{\partial R} \right|_{\mu = q = P_R = 0},
\end{split}
\end{equation}
which is actually a maximum, $\partial^2 H_\text{eff} / \partial R^2 < 0$, and the light-ring solution is therefore unstable.
For the Schwarzschild metric in GR, solving this equation for $R$ gives the known value $R_\text{LR}=3M$.
For the EMd metric in the GHS gauge
\begin{align}
R_\text{LR} &=\frac{3}{2}M +\frac{aD}{2} +\frac{1}{2a}\big[9a^2M^2 -16aMD +6a^3MD \nonumber\\
&\qquad\qquad  +8D^2 -8a^2D^2+a^4D^2\big]^{1/2},
\end{align}
while for the approximate metric in the Schwarzschild gauge
\begin{equation}
R_\text{LR}=\frac{1}{2}\left[3M+\sqrt{9M^2-8Q^2}\right],
\end{equation}
which is the same as the Reissner-Nordstr\"{o}m metric since the potential $A(R)$ is the same in both cases.

In Fig.~\ref{fig:ex}, we plot the binding energy scaled by the total mass, $E_B/M$, versus the orbital frequency $M\Omega$ for equal masses, $\nu=1/4$, and for charge-to mass-ratios $q_1/m_1=q_2/m_2=0.99,~ 0.9,~ 0.5$ and $q_1/m_1=-q_2/m_2=0.9$. 
The binding energy diverges at the light ring; to improve readability, we show the plots only up to the frequency corresponding to $R = 1.05R_\text{LR}$ or to energy $E_B/M=0.015$.
We plot the binding energy for four cases: (a) EMd-EOB Hamiltonian in the GHS gauge; (b) EMd-EOB Hamiltonian in the Schwarzschild gauge; (c) EMd-GHS Hamiltonian with $a=0$, which is EM theory; and (d) EMd-GHS Hamiltonian in the limit where all charges are zero $Q=0$, which is the standard uncharged GR case.
[The effective Hamiltonian for case (c) is that of a charge moving in the Reissner-Nordstr{\"o}m spacetime, and for (d) it is that of a reduced mass in Schwarzschild spacetime.]
The difference between the EM case ($a=0$ curve) and the standard astrophysical scenario of uncharged BHs ($Q=0$ curve)
quantifies the effect of the electric charges, while the difference between the EMd Hamiltonian(s) and the
EM case quantifies the effect of the scalar charges. 

We see from Fig.~\ref{fig:ex} that
the electric charges have a larger effect on the binding energy than the additional scalar charges
in EMd theory (except for almost extreme charges).
For small electric charges $\lesssim 0.5$ (lower left panel of Fig.~\ref{fig:ex}), the difference in binding energy between EMd theory and EM theory at the ISCO is 
only $9 \%$ of the difference between EMd theory and GR with no charges, i.e., the scalar charge has a very small effect.
The difference between the two EMd-EOB Hamiltonians increases with increasing electric charge and frequency, but they still agree well. 
The binding energy of the two Hamiltonians at the ISCO differs by $\sim 6\%$ for charge-to-mass ratio 0.99 and by $\sim 0.1 \%$ for charge-to-mass ratio 0.5.
For charge-to-mass ratios larger than one, a naked singularity appears in the effective metric in the Schwarzschild gauge; this is an unphysical feature arising from the choice of gauge, and thus the EOB Hamiltonian should not be used for small separations (high frequencies) approaching this singularity.
Note that, if one is only interested in the
inspiral, then the comparison of the Hamiltonians via the binding energy can be stopped already at the ISCO frequency instead of the LR frequency.

The ISCO marks the end of the inspiral phase of the binary coalescence and the beginning of the plunge.
To find the value of the ISCO, we set both the first and second derivatives of the effective Hamiltonian to zero $\partial H_{\text{eff}}/\partial R=0=\partial^2H_{\text{eff}}/\partial R^2$ and set $P_R=0$.
Then, we solve the two equations numerically for the ISCO radius and angular momentum. The orbital frequency at ISCO can then be calculated from Eq.~\eqref{orbfreq}.

In Fig.~\ref{fig:ex}, the location of the ISCO is indicated by the point on each curve.
In Fig.~\ref{fig:Risco}, we plot the orbital frequency at ISCO, scaled by the mass, i.e., $M\Omega_\text{ISCO}$, versus the charge-to-mass ratio $q_1/m_1$ with $q_2/m_2=q_1/m_1$ in the left panel, and $q_2/m_2=-q_1/m_1$ in the right. 
From the left panel, we see that for high charge-to-mass ratios, the two EOB Hamiltonians do not agree well at this high frequency.
For same-sign charges, the ISCO orbital frequency is lower than the uncharged case, which means the ISCO radius is greater than the Schwarzschild value of $6M$. This is because the binding energy of charged BHs is higher (less bound) than the energy of uncharged BHs, as can be seen from the binding energy in Fig.~\ref{fig:ex}. 
For opposite-sign charges, the ISCO orbital frequency is higher than the uncharged case because the binding energy is lower than the energy of uncharged BHs.

\section{Conclusions}
\label{sec:conclusions}

In this paper, we analytically modeled the dynamics of binary BHs in EMd theory.
In this theory, electrically charged BHs also carry a scalar charge, whereas in GR (and many modified theories of gravity) the scalar charge is zero.
Thus, the identification of a BH with scalar charge through GW observations could point to modifications of gravity in the strong-field regime and violations of the strong equivalence principle.
Observation of a large electric charge on BHs could be a trace of minicharged dark matter and/or dark photons.

We began by considering the case of a test BH in the background of a more massive companion in EMd theory, wherein the scalar charge of the test BH decreases as it moves radially inwards.
Consistent with the results of Ref.~\cite{Julie:2017rpw}, we found that the dimensionless charge $\alpha(\varphi)$ exhibits a sharp transition [see Figs.~\ref{fig:alpha} and \ref{fig:alphar}].
However, we showed that in a binary system, the scalar charge of the test BH will change dramatically  only very close to the horizon of the background BH and only if both BHs are nearly-extremally charged. Thus, these features can be observationally 
  relevant only in minicharged dark matter and dark photons models, but not in the Standard Model of particle physics. Our study also showed that binary BHs in EMd theory will not exhibit non-perturbative phenomena akin to induced or dynamical scalarization that are found in certain ST theories [see Fig.~\ref{fig:sensitivityRatio}].

We then used the PN approximation in EMd theory to study the dynamics of a two-body system with an arbitrary mass ratio.
We derived the two-body 1PN Lagrangian and Hamiltonian, and investigated how the bodies' scalar charges decrease with their separation at next-to-leading PN order.
As in the test-BH case, we expect that dramatic changes could occur only for nearly-extremal charged BHs on very compact orbits; this is a regime most easily probed by systems with extreme mass ratios and/or rapidly spinning BHs. 
We derived the scalar, vector, and tensor energy fluxes at next-to-leading PN order.
From the energy flux and binding energy, we calculated the Fourier-domain gravitational waveform for binaries on quasi-circular orbits using the stationary-phase approximation.

Using our  PN result, we discussed the possibility of constraining EMd theory with GWs.
Given current and projected constraints on dipole radiation, we examined how the degeneracies between electric and scalar charges limit the bounds that can be set on the EMd parameter $a$ --- constraining this parameter requires one to measure the electric charges of each BH independently, and the strength of this bound improves for larger total electric charge [see Fig.~\ref{fig:bounds}].
We also estimated the observational deviations from GR predicted in EMd theory with two measures: the dephasing between PN waveforms in the stationary-phase approximation [Fig.~\ref{fig:phase}], and the difference in the number of useful GW cycles [Fig.~\ref{fig:Nuseful}].
For ground-based GW detectors, we found that the presence of electric and scalar charges contributes $\lesssim 1$ radian to the phase provided the black holes have charge-to-mass ratios of $q_i/m_i\lesssim 0.01$ for coupling constant $a=1$.
We showed that the relative difference in useful cycles between EMd theory and GR provides an estimate of the fractional correction to SNR by non-GR corrections; for systems with $q_i/m_i\lesssim 0.1$, the deviations from GR affect the total SNR by a few percent.

Finally, we constructed two EOB Hamiltonians for binary BHs in EMd theory: an EOB Hamiltonian in the GHS gauge, which is based on the exact BH solution, and an EOB Hamiltonian in the Schwarzschild gauge, which is based on an approximation to that solution. 
The EOB Hamiltonian in the GHS gauge is more physical in the strong-gravity regime, since it exactly reproduces the dynamics of a test body, and hence will be more accurate for systems with a very asymmetric mass ratio.
The EOB Hamiltonian in Schwarzschild gauge is easier to implement by taking the existing EOB Hamiltonians in GR as a starting point and adding to it corrections due to EMd theory.
We compared the two Hamiltonians by calculating the binding energy and the innermost stable circular orbit, and found that they agree well, except for nearly-extremal charges at high frequencies [see Figs.~\ref{fig:ex} and \ref{fig:Risco}]. 
The binding energy of the two Hamiltonians at the ISCO differs by $\sim 6\%$ for charge-to-mass ratio 0.99 and by $\sim 0.1 \%$ for charge-to-mass ratio 0.5.

An important goal in future continuations of our work would be the construction of a full (inspiral-merger-ringdown) EOB waveform model in EMd theory.
For accurate predictions in the late inspiral, one likely needs PN results for the Hamiltonian, fluxes, and modes to the same order as they are available in GR, next to a calibration of the model to NR simulations in EMd theory.
Modeling the merger and ringdown requires predictions for the parameters of the final black hole and its quasi-normal modes as a function of the EMd coupling constant $a$ (see, e.g., Refs.~\cite{Jai-akson:2017ldo, Ferrari:2000ep} for partial results).
Since EOB waveform models in existing data-analysis infrastructure are formulated in the Schwarzschild gauge, this gauge is probably the best compromise for the purpose of GW data analysis.
This gauge is also better suited for creating a single EOB waveform model covering various alternative theories; for example, we demonstrated that our EOB Hamiltonian in the Schwarzschild gauge can describe both ST and EMd theories.
Ultimately, one could aim to construct a generalized EOB framework that uses a physically motivated parameterization to encode a range of possible deviations from GR.

\section*{Acknowledgments}

M.K. thanks the graduate school and the physics department at the University of Maryland for the International Graduate Research Fellowship. He also thanks the Max Planck Institute for Gravitational Physics for its hospitality and for co-funding his stay during the completion of this work.

\appendix

\section{The 1PN two-body Lagrangian in Einstein-Maxwell-dilaton theory}
\label{app:lag}
In this appendix, we derive the 1PN two-body Lagrangian in EMd theory using the Fokker action method \cite{Fokker:1929} (see also Refs.~\cite{Damour:1995kt, Bernard:2015njp, Bernard:2018hta}).
To derive the Lagrangian, we expand the EMd action in Eq.~\eqref{EMdaction}, together with the matter action for point particles in Eq.~\eqref{Smatter} and the mass expansion from Eq.~\eqref{Mphi}. After that, we obtain the field equations for the potentials, solve them, and plug the solutions back into the action to get the Lagrangian. 
Throughout, we work in the harmonic gauge $g^{\mu\nu}\Gamma^\lambda_{\mu\nu}=0$ and the Lorenz gauge $\partial_\mu A^\mu=0$.
Also, in this appendix and the next, we explicitly write $c$ and $G$ for bookkeeping.

\subsection{Expanding the metric and connection coefficients}

Before expanding the EMd action, we start by expanding the metric in powers of $v/c$ \cite{Damour:1990pi},
\begin{align}
g_{00}&= -1+2V-2V^2+\dots , \nonumber\\ 
g_{0i}&=-4V_i+\dots , \nonumber\\ 
g_{ij}&=\delta_{ij}+2V\delta_{ij}+\dots ,
\end{align}
where the potentials $V\sim \mathcal{O}(1/c^2)$, and $V_i\sim \mathcal{O}(1/c^3)$. The  inverse metric satisfies $g^{\mu\lambda}g_{\lambda\nu}=\delta^\mu_{~\nu}$.

The connection coefficients in terms of the metric are given by
\begin{equation}
\Gamma{^\mu}_{\nu\lambda}=\frac{1}{2}g^{\mu\rho}\left(\partial_\lambda g_{\rho\nu}+\partial_\nu g_{\rho\lambda}-\partial_\rho g_{\nu\lambda}\right).
\end{equation}
Plugging the metric expansion in terms of the potentials yields the connection coefficients to $\Order(1/c^4)$
\begin{align}
&\Gamma{^0}_{00}=-\partial_0 V\,, \nonumber\\ 
&\Gamma{^0}_{0i}=-\partial_i V\,, \nonumber\\
&\Gamma{^i}_{00}=-\partial_i V+2\partial_i V^2-4\partial_0 V_i\,, \nonumber\\
&\Gamma{^0}_{ij}= 2\left(\partial_j V_i-\partial_i V_j\right)+\delta_{ij}\partial_0 V\,,  \nonumber\\
&\Gamma{^i}_{0j}=2\left(\partial_i V_j-\partial_j V_i\right)+\delta_{ij}\partial_0 V\,, \nonumber\\
&\Gamma{^i}_{jk}=-(1+2 V)\left(\delta_{ij}\partial_k V+\delta_{ik}\partial_j V-\delta_{jk}\partial_i V\right).
\label{Gamma}
\end{align}

\subsection{Expanding the action}
The action of EMd theory is given by Eq.~\eqref{EMdaction}.
We can divide that action into four pieces
\begin{equation}
S=S_g+S_\varphi+S_{\text{em}}+S_m\,,
\end{equation}
where $S_g$ is the gravitational action,  $S_\varphi$ is the dilaton action, $S_{\text{em}}$ is the electromagnetic action with the dilaton coupling, and $S_m$ is the matter action.

In the Einstein frame, the gravitational action is the same as in GR. The Einstein-Hilbert gravitational action can be written in the Landau-Lifshitz form 
\begin{align}\label{Sg}
S_{g} &= \frac{c^4}{16\pi G} \int dtd^{3}\bm{x} \, \sqrt{-g} 
g^{\mu\nu} \left( \Gamma^{\rho}_{\mu\lambda}
\Gamma^{\lambda}_{\nu\rho} - \Gamma^{\rho}_{\mu\nu}
\Gamma^{\lambda}_{\rho\lambda} \right).
\end{align}
Substituting the connection coefficients from Eq.~\eqref{Gamma} in terms of the potentials leads to
\begin{align}
S_{g} &= \frac{c^4}{16\pi G} \int dtd^{3}\bm{x}\Big[-2\partial_i V\partial_i V -16\partial_i V\partial_0 V_i \nonumber\\
&\quad\qquad  -6\partial_0 V\partial_0 V +8\partial_i V_j\partial_i V_j -8\partial_i V_j\partial_j V_i\Big].
\end{align}
Imposing the harmonic gauge condition $g^{\mu\nu}\Gamma^\lambda_{\mu\nu}=0$  gives $\partial_0 V+\partial_i V_i=0$. Applying that condition in the action and integrating by parts yields 
\begin{align}
\label{Sgravity}
S_{g} &= \frac{c^4}{16\pi G} \int dtd^{3}\bm{x}\Big[-2\partial_i V\partial_i V +2\partial_0 V\partial_0 V \nonumber\\
&\qquad\qquad\qquad\qquad  +8\partial_i V_j\partial_i V_j\Big].
\end{align}

The dilaton action is given by
\begin{equation}
S_\varphi=-\frac{c^4}{8\pi G}\int dtd^{3}\bm{x}\sqrt{-g}g^{\mu\nu}\partial_\mu\varphi\partial_\nu\varphi\,.
\end{equation}
Since  $\varphi$ is of order $1/c^2$, then to ${\cal O}(1/c^2)$
\begin{equation}
S_\varphi =-\frac{c^4}{8\pi G}\int dtd^{3}\bm{x}\left(-\partial_0\varphi\partial_0\varphi+\partial_i\varphi\partial_i\varphi\right).
\end{equation}

The electromagnetic action including the dilaton coupling is given by
\begin{equation}
S_{\text{em}}=\frac{-1}{16\pi}\int dtd^{3}\bm{x} \sqrt{-g}e^{-2a\varphi}F_{\mu\nu}F^{\mu\nu}\,,
\end{equation}
with the electromagnetic field $F_{\mu\nu}=\nabla_{\mu}A_{\nu}-\nabla_{\nu}A_{\mu}= \partial_{\mu}A_{\nu}-\partial_{\nu}A_{\mu}$ and the vector potential $A_\mu=(A_0, A_i)$.
The component $A_0={\cal{O}}(1)+{\cal{O}}(1/c^2)+\dots$, while the components $A_i={\cal{O}}(1/c)+\dots$. Therefore, expanding $F_{\mu\nu}F^{\mu\nu}$ to ${\cal{O}}(1/c^2)$ leads to
\begin{align}
F_{\mu\nu}F^{\mu\nu}=&-2\partial_i A_0\partial_i A_0+2\partial_j A_i\partial_j A_i+4\partial_0 A_i\partial_i A_0 \nonumber\\
&-2\partial_i A_j\partial_j A_i \,.
\label{FF}
\end{align}
Because the last two terms in Eq.~\eqref{FF} are of order $1/c^2$, we can use integration by parts and the Lorentz gauge condition ($\partial_{\mu}A^\mu=0$) to replace these last two terms  by $2\partial_0A_0\partial_0A_0$. 
Since $\sqrt{-g}=1+2 V$, and $e^{-2a\varphi}\simeq 1-2a\varphi+\dots$, the action becomes
\begin{align}
\label{Sem}
S_{\text{em}}&=\frac{1}{8\pi} \int dtd^{3}\bm{x} \big[(1+2V-2a\varphi)\partial_i A_0\partial_i A_0 \nonumber\\
&\quad\qquad-\partial_j A_i\partial_j A_i-\partial_0A_0\partial_0A_0\big].
\end{align}

The matter action $S_{m}$ for point particles at monopolar order
(dipole/spin and higher multipoles neglected) is given by
\begin{align}
S_{m} &= - \sum_A \int dt \bigg[ \mathfrak{m}_A(\varphi)c^2 \sqrt{-g_{\mu\nu}\,v_A^\mu v_A^\nu/c^2} \nonumber\\
&\quad - \frac{1}{c} q_A A_\mu \frac{dx^\mu}{dt} \bigg],
\end{align}
where the field-dependent mass of each body has the expansion given by Eq.~\eqref{Mphi}
\begin{equation}
\mathfrak{m}(\varphi)=m\left[1+\alpha\varphi+\frac{1}{2}(\alpha^2+\beta)\varphi^2+ \Order\left(1/c^6\right)\right] .
\end{equation}
Defining the mass density $\rho_g$  in terms of the constant masses
\begin{equation}
\rho_g \equiv\sum_A m_A\delta^3(\bm{x}-\bm{x}_A),
\end{equation}
and defining the electric charge density by
\begin{equation}
\rho_e \equiv\sum_A q_A\delta^3(\bm{x-\bm{x}}_A),
\end{equation}
then the matter action to ${\cal O}(1/c^2)$ can be written as
\begin{align}
\label{Sparticle}
S_m &= \int dtd^{3}\bm{x}\,\bigg[\rho_g \bigg(-c^2+\frac{1}{2}v^2 + Vc^2 +\frac{1}{8}\frac{v^4}{c^2} +\frac{3}{2} V v^2 \nonumber\\
&\qquad-\frac{1}{2} V^2c^2 -4 V_i v^ic\bigg) 
+\rho_g \alpha\varphi\left(-c^2+\frac{1}{2}v^2+ Vc^2 \right) \nonumber\\
&\qquad -\frac{1}{2}c^2\rho_g(\alpha^2+\beta)\varphi^2 
+\rho_e\left( A_0+ \frac{1}{c}A_iv^i\right)
\bigg].
\end{align}
The parameters $\alpha$ and $\beta$ will be assigned a subscript when multiplied by the delta functions in $\rho_g$.

\subsection{The field equations}

Combining the expansion of the action from the previous subsection, the total action at 1PN order is given by
\begin{widetext}
\begin{align}
S=\int dtd^{3}\bm{x}\,\bigg\{&\frac{c^4}{16\pi  G}\left[-2\partial_i V\partial_i V +2\partial_0 V\partial_0 V +8\partial_i V_j\partial_i V_j\right] \nonumber\\
&+\rho_g \left[-c^2+\frac{1}{2}v^2+ Vc^2 +\frac{1}{8}\frac{v^4}{c^2}+\frac{3}{2} V v^2-\frac{1}{2} V^2c^2 -4 V_i v^ic \right] \nonumber\\
&-\frac{c^4}{8\pi G} \left(-\partial_0\varphi\partial_0\varphi+\partial_i\varphi\partial_i\varphi\right) 
+\rho_g\alpha\varphi\left(-c^2+\frac{1}{2}v^2+ Vc^2 \right)-\frac{1}{2}c^2\rho_g(\alpha^2+\beta)\varphi^2 \nonumber\\
&+\frac{1}{8\pi}  \left[(1+2V-2a\varphi)\partial_i A_0\partial_i A_0-\partial_j A_i\partial_j A_i-\partial_0A_0\partial_0A_0\right]
+\rho_e\left( A_0+ \frac{1}{c^2}A_iv^i\right)
\bigg\}.
\end{align}
\end{widetext}
	
Varying the action with respect to the potentials  $V_i$, $A_i$, $V$, $\varphi$, and $A_0$ respectively yields the field equations
\begin{align}
\nabla^2 V_i&=-\frac{4\pi G}{c^3}\rho_g v^i\,, \label{eqVi} \\
\nabla^2A_i&=-\frac{4\pi}{c} \rho_ev^i\,, \label{eqAi} \\
\square V&=-\frac{4\pi G}{c^2}\rho_g - \frac{4\pi G}{c^4}\rho_g\left(\frac{3}{2}v^2- V c^2\right)  \nonumber\\
&\quad -\frac{4\pi G}{c^2}\rho_g\alpha\varphi -\frac{G}{c^4} \partial_i A_0\partial_i A_0 \,, \label{eqpsi} \\
\square\varphi&=-\frac{4\pi G}{c^4}\rho_g\left[-\alpha+\frac{1}{2}\alpha v^2+\alpha V-(\alpha^2+\beta)\varphi\right] \nonumber\\
&\quad +\frac{Ga}{c^4} \partial_iA_0\partial_iA_0\,, \label{eqphi}\\
\square A_0&=4\pi  \rho_e -2 V\nabla^2A_0 -2\partial_i V\partial_i A_0+2a\varphi\nabla^2 A_0 \nonumber\\
&\quad+2a\partial_i\varphi\partial_i A_0\,,  \label{eqA0}
\end{align}
where $\square=-\partial_0^2+\nabla^2$ is the flat d'Alembertian.

The first two equations can be solved directly for $V_i$, and $A_i$
\begin{align}
V_i&=\frac{G}{c^3} \left(\frac{m_1 v_1^i}{|\bm{x}-\bm{x}_1|}+\frac{m_2 v_2^i}{|\bm{x}-\bm{x}_2|}\right), \\
A_i&= \frac{1}{c} \left(\frac{q_1v_1^i}{|\bm{x}-\bm{x}_1|}+\frac{q_2v_2^i}{|\bm{x}-\bm{x}_2|}\right) \,.
\end{align}
To solve the other three equations, we first rewrite the terms $\partial_iA_0\partial_iA_0$, $\partial_i\varphi\partial_i A_0$, and $\partial_i V\partial_i A_0$ using the identity
\begin{equation}
\nabla^2(\chi\xi)= \chi\nabla^2\xi+\xi\nabla^2 \chi+2\partial_i \chi\partial_i \xi\,,
\end{equation}
where $\chi$ and $\xi$ are any scalar functions. Using that identity, Eqs. \eqref{eqpsi}, \eqref{eqphi}, and \eqref{eqA0} can be written as
\begin{align}
\square V&= -\frac{4\pi G}{c^2}\rho_g -\frac{4\pi G}{c^4}\rho_g\left(\frac{3}{2}v^2 - Vc^2\right) -\frac{4\pi G}{c^2}\rho_g\alpha\varphi \nonumber\\
&\quad -\frac{G}{2c^4} \nabla^2(A_0)^2  +\frac{G}{c^4}A_0\nabla^2A_0\,, \label{eqV2}\\
\square\varphi&=-\frac{4\pi G}{c^4}\rho_g\left[-c^2\alpha+\frac{1}{2}\alpha v^2 + c^2\alpha V-c^2(\alpha^2+\beta)\varphi\right] \nonumber\\
&\quad- \frac{Ga}{c^4}A_0\nabla^2A_0 +\frac{Ga}{2c^4} \nabla^2(A_0)^2\,,  \label{eqphi2}\\
\square A_0&=4\pi  \rho_e -\nabla^2(VA_0) - V\nabla^2A_0 +A_0\nabla^2V \nonumber\\
&\quad+a\nabla^2(\varphi A_0)+a\varphi\nabla^2 A_0 -aA_0\nabla^2\varphi\,.  \label{eqA02} 
\end{align}

At this point, one could split the fields into separate PN orders, followed by further simplifications
  of the action through partial integrations and use of the field equations. Eventually one would
  only need an explicit expression for the leading order solution to the field equations here
  in order to obtain the 1PN Fokker action. This is essentially the ``n+2'' method from
  Ref.~\cite{Bernard:2015njp}. However, at this order this does overall not provide a big
  simplification, and we need a solution for the 1PN scalar field for
  Figs.~\ref{fig:Drnu} and \ref{fig:Drqm}. We therefore proceed by solving the 1PN field equations
  and straightforwardly insert the solution into the complete action.

To solve those three equations, we first solve for the leading order terms of $V$, $\varphi$, and $A_0$, and then insert that solution back into the right hand side of the equations. Equation \eqref{eqV2} yields
\begin{widetext}
\begin{align}
V&= \frac{G}{c^2} \left(\frac{m_1}{|\bm{x}-\bm{x}_1|}+\frac{m_2}{|\bm{x}-\bm{x}_2|}\right)
+\frac{G}{2c^4}\left(m_1\frac{\partial^2}{\partial t^2}|\bm{x}-\bm{x}_1|+m_2\frac{\partial^2}{\partial t^2}|\bm{x}-\bm{x}_2|\right)
+\frac{3G}{2c^4}\left(\frac{m_1v_1^2}{|\bm{x}-\bm{x}_1|}+\frac{m_2v_2^2}{|\bm{x}-\bm{x}_2|}\right)  \nonumber\\
&\quad -\frac{G^2}{c^4}m_1m_2\left(\frac{1}{r|\bm{x}-\bm{x}_1|} +\frac{1}{r|\bm{x}-\bm{x}_2|}\right)
-\frac{G^2}{c^4}\alpha_1\alpha_2m_1m_2\left(\frac{1}{r|\bm{x}-\bm{x}_1|}+\frac{1}{r|\bm{x}-\bm{x}_2|}\right) 
 \nonumber\\
&\quad -\frac{G}{2c^4} \left(\frac{q_1}{|\bm{x}-\bm{x}_1|}+\frac{q_2}{|\bm{x}-\bm{x}_2|}\right)^2
+\frac{G}{c^4}q_1q_2\left(\frac{1}{r|\bm{x}-\bm{x}_1|}+\frac{1}{r|\bm{x}-\bm{x}_2|}\right) ,
\end{align}
where $r\equiv|\bm{x}_1-\bm{x}_2|$ and
\begin{equation}
\label{partialt}
\frac{\partial^2}{\partial t^2}|\bm{x}-\bm{x}_1|=\frac{v_1^2}{|\bm{x}-\bm{x}_1|}-\bm{n}_1\cdot\bm{a}_1-\frac{(\bm{n}_1\cdot \bm{v}_1)^2}{|\bm{x}-\bm{x}_1|}\,,
\end{equation}
with $\bm{n}_1\equiv (\bm{x}-\bm{x}_1)/|\bm{x}-\bm{x}_1|$, and $\bm{a}_1=d\bm{v}_1/dt$ is the acceleration.

Solving Eq.~\eqref{eqphi2} and using Eq.~\eqref{partialt}, we get
\begin{align}
\label{1PNphi}
\varphi=&-\frac{G}{c^2}\left(\frac{\alpha_1m_1}{|\bm{x}-\bm{x}_1|}+\frac{\alpha_2m_2}{|\bm{x}-\bm{x}_2|}\right)
+\frac{G}{2c^4}\alpha_1m_1\left(\bm{n}_1\cdot\bm{a}_1+\frac{(\bm{n}_1\cdot \bm{v}_1)^2}{|\bm{x}-\bm{x}_1|}\right) +\frac{G}{2c^4}\alpha_2m_2\left(\bm{n}_2\cdot\bm{a}_2+\frac{(\bm{n}_2\cdot \bm{v}_2)^2}{|\bm{x}-\bm{x}_2|}\right)
\nonumber\\
&+\frac{G^2}{c^4}m_1m_2\left(\frac{\alpha_1+\alpha_2(\alpha_1^2+\beta_1)}{r|\bm{x}-\bm{x}_1|}+\frac{\alpha_2+\alpha_1(\alpha_2^2+\beta_2)}{r|\bm{x}-\bm{x}_2|}\right) \nonumber\\
&-\frac{Ga}{c^4} q_1q_2\left(\frac{1}{r|\bm{x}-\bm{x}_1|}+\frac{1}{r|\bm{x}-\bm{x}_2|}\right) +\frac{a}{2} \left(\frac{q_1}{|\bm{x}-\bm{x}_1|}+\frac{q_2}{|\bm{x}-\bm{x}_2|}\right)^2.
\end{align}

The solution of Eq.~\eqref{eqA02} for $A_0$ is given by
\begin{align}
A_0=&- \left(\frac{q_1}{|\bm{x}-\bm{x}_1|}+\frac{q_2}{|\bm{x}-\bm{x}_2|}\right)
-\frac{q_1}{2c^2}  \left(\frac{v_1^2}{|\bm{x}-\bm{x}_1|}-\bm{n}_1\cdot\bm{a}_1-\frac{(\bm{n}_1\cdot \bm{v}_1)^2}{|\bm{x}-\bm{x}_1|}\right) 
-\frac{q_2}{2c^2}  \left(\frac{v_2^2}{|\bm{x}-\bm{x}_2|}-\bm{n}_2\cdot\bm{a}_2-\frac{(\bm{n}_2\cdot \bm{v}_2)^2}{|\bm{x}-\bm{x}_2|}\right) \nonumber\\
& +\frac{G}{c^2}\left((1+a\alpha_1)\frac{m_1}{|\bm{x}-\bm{x}_1|}+(1+a\alpha_2)\frac{m_2}{|\bm{x}-\bm{x}_2|}\right)\left(\frac{q_1}{|\bm{x}-\bm{x}_1|}+\frac{q_2}{|\bm{x}-\bm{x}_2|}\right) \nonumber\\
&+ \frac{G}{c^2}\left((1+a\alpha_2)\frac{q_1m_2}{r|\bm{x}-\bm{x}_1|}+ (1+a\alpha_1)\frac{q_2m_1}{r|\bm{x}-\bm{x}_2|}\right) 
- \frac{G}{c^2}\left((1+a\alpha_1)\frac{m_1q_2}{r|\bm{x}-\bm{x}_1|}+ (1+a\alpha_2)\frac{m_2q_1}{r|\bm{x}-\bm{x}_2|}\right).
\end{align}

\subsection{The 1PN Lagrangian}

The total action, after using the field equations and integrating by parts, can be written as 
\begin{align}
S=\int dtd^{3}\bm{x}& \bigg[\rho_g\left(-c^2+\frac{1}{2}v^2+\frac{v^4}{8c^2}+\frac{1}{2} Vc^2 +\frac{3}{4} V v^2 -2 V_iv^ic\right)   +\rho_e\left(\frac{1}{2}A_0+\frac{1}{2c}A_iv^i\right) \nonumber\\
&+\frac{1}{2}\rho_g\alpha\varphi\left(-c^2+\frac{1}{2}v^2\right) 
+\frac{1}{2}\rho_eA_0V-\frac{1}{2}a\rho_eA_0\varphi
+\frac{G}{4c^2} \rho_g(1+a\alpha)A_0^2\bigg].
\end{align}
Substituting the potentials gives acceleration terms that can be eliminated using integration by parts in the action
\begin{equation}
\int dt\, (\bm{n}\cdot \bm{a}_1)= \int dt\left(-\frac{v_1^2}{r} +\frac{(\bm{n}\cdot\bm{v}_1)^2}{r} -\frac{(\bm{n}\cdot\bm{v}_1)(\bm{n}\cdot\bm{v}_2)}{r} +\frac{\bm{v}_1\cdot\bm{v}_2}{r}\right),
\end{equation}
where $\bm{n}\equiv (\bm{x}_1-\bm{x}_2)/|\bm{x}_1-\bm{x}_2|$, and $\bm{a}_1 = \dot{\bm{v}}_1$.
Finally, integrating over space term by term and simplifying leads to the 1PN Lagrangian
\begin{equation}
L=-m_1c^2-m_2c^2+L_0+\frac{1}{c^2}L_1\,,
\end{equation}
with
\begin{align}
\label{lagrangian}
L_0 =&\, \frac{1}{2}m_1v_1^2+\frac{1}{2}m_2v_2^2+ G(1+\alpha_1\alpha_2)\frac{m_1m_2}{r}- \frac{q_1q_2}{r}, \nonumber\\
L_1 =&\, \frac{1}{8}m_1v_1^4+\frac{1}{8}m_2v_2^4+ \frac{q_1q_2}{2r}\left[\bm{v_1\cdot v_2}+(\bm{n}\cdot\bm{v_1})(\bm{n}\cdot\bm{v_2})\right] \nonumber\\
& +\frac{Gm_1m_2}{2r}\left[(3-\alpha_1\alpha_2)(v_1^2+v_2^2)- (7-\alpha_1\alpha_2)(\bm{v_1}\cdot\bm{v_2})-(1+\alpha_1\alpha_2)(\bm{n}\cdot\bm{v_1})(\bm{n}\cdot\bm{v_2})\right]\nonumber\\
&-\frac{G^2m_1m_2}{2r^2}\left[(1+2\alpha_1\alpha_2)(m_1+m_2)+m_1\alpha_1^2(\alpha_2^2+\beta_2)+m_2\alpha_2^2(\alpha_1^2+\beta_1)\right] \nonumber\\
& +\frac{Gq_1q_2}{r^2} \left[m_1(1+a\alpha_1)+m_2(1+a\alpha_2)\right]
- \frac{G}{2r^2} \left[m_1q_2^2(1+a\alpha_1)+m_2q_1^2(1+a\alpha_2)\right].
\end{align}
\end{widetext}

\section{Energy flux  to next-to-leading PN order in Einstein-Maxwell-dilaton theory}
\label{app:flux}
In this appendix, we derive the next-to-leading order scalar, vector, and tensor energy fluxes for general orbits. The derivation follows the one used  in Ref.~\cite{Damour:1992we} in the context of ST theory.

\subsection{Scalar energy flux}

The scalar field in a radiative coordinate system can be written as
\begin{equation}
\varphi(\bm{X}^\mu) = \varphi_0 +\frac{1}{R}\psi(U,\bm{N})+ \Order\left(\frac{1}{R^2}\right),
\end{equation}
where $R\equiv |\bm{X}|$, $U\equiv T-R/c$, $\bm{N}\equiv \bm{X}/R$, and the Einstein-frame radiative scalar multipole moments are defined by
\begin{equation}
\psi(U,\bm{N}) = G \sum_{\ell\geq 0} \frac{1}{\ell!c^{\ell+2}} N^L\Psi_L^{(\ell)}(U).
\end{equation} 
In this notation, an uppercase index denotes a multi-index, such as $N^L=N^{i_1}N^{i_2}\dots N^{i_\ell}$. A superscript in parentheses denotes derivative, such as $\Psi^{(\ell)}(U)=d^\ell\Psi/dU^\ell$.

Next, to relate the radiative moments to the source moments, one defines `algorithmic' moments that serve as functional parameters for a general external metric. Based on the arguments in Refs.~\cite{Blanchet:1989ki,Damour:1992we}, the radiative moments coincide with the algorithmic ones to  $\Order(1/c^3)$, and the algorithmic moments agree with the source moments $K_L$ to order $\Order(1/c^4)$ 
\begin{align}
\Psi_L&=\Psi_L^{\text{(alg)}} + \Order(1/c^3), \\
\Psi_L^{\text{(alg)}} &= K_L + \Order(1/c^4).
\end{align}
The source moments are defined by
\begin{equation}
K_L=\int d^3x \, \left[\widehat{x}_LS +\frac{1}{2(2\ell+3)c^2}\bm{x}^2\widehat{x}_L\frac{\partial^2S}{\partial t^2}\right],
\end{equation}
where the hat on $x_L$ denotes a symmetric trace-free projection on the $\ell$ indices.
The source function $S$ is defined by the field equation for $\varphi$  as
\begin{equation}
\square \varphi=-\frac{4\pi G}{c^2} S.
\end{equation}

The scalar energy flux
\begin{equation}
\mathcal{F}_S=-cR^2 \oint T_{0i}^SN^i d\Omega ,
\end{equation}
where the scalar part of the  stress-energy tensor is given by 
\begin{equation}
T_{\mu\nu}^S=\frac{c^4}{4\pi G }\left[\nabla_\mu\varphi \nabla_\nu\varphi -\frac{1}{2} g_{\mu\nu}(\nabla\varphi)^2\right].
\end{equation}
In the far zone, 
\begin{equation}
T_{0i}^S\simeq\frac{c^4}{4\pi G} \partial_0\varphi\partial_i\varphi \simeq-\frac{c^4}{4\pi G} N_i (\partial_0\varphi)^2,
\end{equation}
where, in the last step, we used the relation
\begin{equation}
\partial_i\varphi =-N_i\partial_0\varphi + \Order(r/R^2).
\end{equation}
The scalar flux becomes
\begin{align}
\mathcal{F}_S& = \frac{c^3}{4\pi G} \int d\Omega \left(\frac{\partial\psi}{\partial U}\right)^2 \nonumber\\
&= G\sum_{\ell\geq 0}  \frac{1}{c^{2\ell+1}(\ell!)^2 } \int\frac{d\Omega}{4\pi} N^LN^P\Psi^{(\ell+1)}_L(U)\Psi^{(\ell+1)}_P(U).
\end{align}
To integrate over the solid angle, we use the integration formula given by Eq.~(A 29a) in Ref.  \cite{blanchet1986radiative}, which yields
\begin{align}
\mathcal{F}_S& =G\sum_{\ell\geq 0} \frac{1}{c^{2\ell+1}\ell! (2\ell+1)!!}\Psi^{(\ell+1)}_L(U)\Psi^{(\ell+1)}_L(U) \nonumber\\
&=G\left[\frac{\Psi^{(1)}\Psi^{(1)}}{c} + \frac{\Psi_i^{(2)}\Psi_i^{(2)}}{c^3} + \frac{\Psi_{ij}^{(3)}\Psi_{ij}^{(3)}}{c^5} +\dots\right],
\end{align}
where the first term is the monopole flux, the second is the dipole flux, and the third is the quadrupole flux. In terms of the source function $S$, those multipole moments needed for the calculation of the next-to-leading order flux are given by
\begin{align}
\Psi &=\int d^3x \left[S +\frac{1}{6c^2} \frac{d}{dt}(x^2S)\right], \\
\Psi_i &= \int d^3x \left[x^iS +\frac{1}{10c^2} \frac{d}{dt}\left(x^2x^iS\right)\right], \\
\Psi_{ij} &= \int d^3x \left(x^ix^j -\frac{1}{3}x^2\delta_{ij}\right)S.
\end{align}

The 1PN field equation for $\varphi$ is given by Eq.~\eqref{eqphi2}
\begin{align}
\square\varphi=&-\frac{4\pi G}{c^2}\rho_g\left[-\alpha+\frac{1}{2c^2}\alpha v^2 +\alpha V-(\alpha^2+\beta) \varphi\right] \nonumber\\
&- \frac{Ga}{c^4}A_0\nabla^2A_0  +\frac{Ga}{2c^4} \nabla^2(A_0)^2\,. 
\end{align}
The last term in that equation can be moved to the left hand side by a redefinition of the field, and since $A^2_0\sim 1/R^2$, we can neglect that term to $\mathcal{O}(1/R)$. 
The other terms are expressed in terms of delta functions. Hence, we can  write the source function $S$ as
\begin{equation}
S(\bm{x},t)= \sum_A \sigma_A\delta^3(\bm{x}-\bm{x}_A),
\end{equation}
with
\begin{align}
\sigma_1=& -m_1\alpha_1 \left(1-\frac{v_1^2}{2c^2}\right) +\frac{m_1m_2}{c^2r}\left(\alpha_1+\alpha_1^2\alpha_2 +\beta_1\alpha_2\right) \nonumber\\
&-\frac{aq_1q_2}{c^2r}\,,
\end{align} 
and similarly for $\sigma_2$, where $\bm{r}\equiv \bm{x}_1-\bm{x}_2$. In the center-of-mass coordinates, we define
\begin{align}
\label{x1x2CM}
&\bm{v}\equiv \frac{d\bm{r}}{dt}\,, \qquad \bm{a}\equiv \frac{d\bm{v}}{dt}\,, \nonumber\\
& \bm{x}_1=\frac{m_2}{M}\bm{r} +\mathcal{O}\left(\frac{1}{c^2}\right), \nonumber\\
& \bm{x}_2=-\frac{m_1}{M}\bm{r} +\mathcal{O}\left(\frac{1}{c^2}\right).
\end{align}
Thus, $\sigma_1$ can be written as
\begin{align}
\sigma_1=& -m_1\alpha_1 +\frac{\nu}{2c^2}m_1\alpha_1v^2 \nonumber\\
&+\frac{M^2\nu}{c^2r}\left[\alpha_1+\alpha_1^2\alpha_2 +\beta_1\alpha_2-\frac{aq_1q_2}{M\mu}\right].
\end{align}
The multipole moments can now be written in terms of $\sigma$ after integrating the delta functions
\begin{align}
\Psi^{(1)}&=\frac{d\sigma_1}{dt}- \frac{m_1\alpha_1}{6c^2} \frac{d^2}{dt^2}x_1^2 + 1\leftrightarrow 2 , \\
\Psi_i^{(2)}&=\frac{d^2}{dt^2}(x_1^i\sigma_1) - \frac{m_1\alpha_1}{10c^2} \frac{d^4}{dt^4}x_1^2x_1^i + 1\leftrightarrow 2, \\
\Psi_{ij}^{(3)}&=-m_1\alpha_1\frac{d^3}{dt^3}\left(x_1^ix_1^j -\frac{1}{3}x_1^2\delta_{ij}\right) + 1\leftrightarrow 2,
\end{align}
where, in the higher order terms, we used $\sigma_1=-m_1\alpha_1$.

For the monopole and quadrupole fluxes, the multipole moments in the center-of-mass coordinates can be written as
\begin{align}
\Psi^{(1)} &=\frac{d}{dt}(\sigma_1+\sigma_2) -\frac{\nu}{6c^2}\left(m_2\alpha_1+m_1\alpha_2\right)\frac{d^3r^2}{dt^3}\,, \\
\Psi_{ij}^{(3)} &= -\nu(m_2\alpha_1+m_1\alpha_2)\frac{d^3}{dt^3}\left(r^ir^j-\frac{1}{3}r^2\delta_{ij}\right).
\end{align}
Differentiating, and using the relations
\begin{align}
&\frac{d^2\bm{r}}{dt^2}=-\frac{G_{12}M}{r^2}\bm{n} + \mathcal{O}\left(\frac{1}{c^2}\right),  \nonumber\\
&\frac{d\bm{n}}{dt}=\frac{\bm{v}-\dot{r}\bm{n}}{r}\,,
\end{align}
where
\begin{equation}
G_{12}\equiv G\left(1+\alpha_1\alpha_2-\frac{q_1q_2}{M\mu}\right),
\end{equation}
we get
\begin{widetext}
\begin{align}
\Psi^{(1)} &= -\frac{2}{3}\frac{G_{12}M\mu}{c^2r^2}\dot{r}\left(X_2\alpha_1 +X_1\alpha_2\right) 
-\frac{M\mu}{c^2r^2}\dot{r}\left[\alpha_1+\alpha_2 +\alpha_1^2\alpha_2 +\alpha_2^2\alpha_1 +\beta_1\alpha_2 +\beta_2\alpha_1 -\frac{2aq_1q_2}{M\mu}\right], \nonumber \\
\Psi_{ij}^{(3)} &=-\frac{G_{12}M\mu}{r^2}\left(X_1\alpha_2+X_2\alpha_1\right)\left[6\dot{r}n^in^j -4(n^iv^j+n^jv^i) +\frac{2}{3}\dot{r}\delta_{ij} \right].
\end{align}
Squaring leads to the monopole and quadrupole scalar fluxes
\begin{align}
\mathcal{F}_S^\text{Mon}&= G\frac{\Psi^{(1)}\Psi^{(1)}}{c} =\frac{G}{c^5}\left(\frac{G_{12}M\mu}{r^2}\right)^2\dot{r}^2 \bigg[
\frac{1}{1+\alpha_1\alpha_2-\frac{q_1q_2}{M\mu}}\left(\alpha_1+\alpha_2 +\alpha_1^2\alpha_2 +\alpha_2^2\alpha_1 +\beta_1\alpha_2 +\beta_2\alpha_1 -\frac{2aq_1q_2}{M\mu}\right) \nonumber\\
&\qquad\qquad\qquad\qquad\qquad\qquad\qquad\qquad +\frac{2}{3}\left(X_2\alpha_1 +X_1\alpha_2\right)  \bigg]^2. \\
\mathcal{F}_S^\text{Quad}&= G \frac{\Psi_{ij}^{(3)}\Psi_{ij}^{(3)}}{c^5} =\frac{G}{30c^5}\left(\frac{G_{12}M\mu}{r^2}\right)^2\left(X_1\alpha_2+X_2\alpha_1\right)^2\left(32v^2-\frac{88}{3}\dot{r}^2\right).
\end{align}
\end{widetext}

For the dipole flux, we need to write $\bm{x}_1$ and $\bm{x}_2$ in the center-of-mass coordinates to 1PN order. From the boost invariance of the Lagrangian, we obtain \cite{Damour:1992we}
\begin{align}
&\bm{x}_1=\frac{\mu_2}{\mu_1+\mu_2}\bm{r} + \mathcal{O}\left(\frac{1}{c^4}\right), \nonumber\\ &\bm{x}_2=-\frac{\mu_1}{\mu_1+\mu_2}\bm{r} + \mathcal{O}\left(\frac{1}{c^4}\right),
\end{align}
where
\begin{align}
\mu_1 &\equiv m_1\left(1+\frac{v_1^2}{2c^2}-\frac{G_{12}m_2}{2c^2r}\right) +\Order\left(\frac{1}{c^4}\right)\nonumber\\
&=M\left(X_1+X_2\frac{\nu v^2}{2c^2}-\frac{G_{12}M\nu}{2c^2r}\right) +\Order\left(\frac{1}{c^4}\right),
\end{align}
and similarly for $\mu_2$.
This leads to the dipole moment $\Psi_i$ in the center-of-mass coordinates
\begin{align}
\Psi_i^{(2)} =& \frac{d^2}{dt^2}\left(\frac{\mu_2}{\mu_1+\mu_2} r^i\sigma_1\right) - \frac{d^2}{dt^2}\left(\frac{\mu_1}{\mu_1+\mu_2} r^i\sigma_2\right) \nonumber\\
& +\frac{\mu}{10c^2}\left(X_1^2\alpha_2 -X_2^2\alpha_1\right)\frac{d^4}{dt^4}(r^2r^i).
\end{align}
To calculate the dipole flux, we also need the 1PN acceleration, which can be derived from the 1PN Lagrangian, and we obtain
\begin{widetext}
\begin{align}
\label{acc1pn}
\frac{d^2\bm{r}}{dt^2} =& -\frac{G_{12}M}{r^2}\bm{n} \Bigg\lbrace 1+\frac{v^2}{c^2}\left[3\nu + \frac{1-\alpha_1\alpha_2 +q_1q_2/2M\mu}{1+\alpha_1\alpha_2-q_1q_2/M\mu}\right] -\frac{3}{2c^2}\nu \dot{r}^2 -2\nu\frac{G_{12}M}{c^2r} \nonumber\\
&- \frac{G_{12}M}{c^2r} \frac{1}{\left(1+\alpha_1\alpha_2-\frac{q_1q_2}{M\mu}\right)^2} \Bigg[ 2\nu \left(1+\alpha_1\alpha_2-\frac{q_1q_2}{M\mu}\right)^2 + 
X_2\frac{q_1^2}{M\mu}(1+a\alpha_2) +X_1\frac{q_2^2}{M\mu}(1+a\alpha_1)
\nonumber\\
&
- 2a\frac{q_1q_2}{M\mu}\left(X_1\alpha_1 +X_2\alpha_2\right)
- \frac{q_1q_2}{M\mu} (5-\alpha_1\alpha_2) +4(1+\alpha_1\alpha_2)
+ X_2\beta_1\alpha_2^2+ X_1\beta_2\alpha_1^2
\Bigg]
\Bigg\rbrace
\nonumber\\
&-\frac{G_{12}M}{c^2r^2}\dot{r}\bm{v} \left[ 2\nu - \frac{4-q_1q_2/M\mu}{1+\alpha_1\alpha_2-q_1q_2/M\mu}\right] + \Order\left(\frac{1}{c^4}\right).
\end{align}

Finally, we obtain the dipole scalar flux
\begin{align} 
\mathcal{F}_S^\text{dip} =& \frac{G}{3c^3}\left(\frac{G_{12}M\mu}{r^2}\right)^2 \Bigg\lbrace (\alpha_1-\alpha_2)^2
+f_{\dot{r}^2}^S \frac{\dot{r}^2}{c^2} 
+ f_{v^2}^S \frac{v^2}{c^2} 
+f_{1/r}^S \frac{G_{12}M}{c^2r}
\Bigg\rbrace,
\end{align}
with the coefficients
\begin{subequations}
\label{coeffSdip}
\begin{align} 
f_{\dot{r}^2}^S &= \frac{-1}{1+\alpha_1\alpha_2-\frac{q_1q_2}{M\mu}} 
\bigg\lbrace 8(\alpha_1-\alpha_2)^2 -2\nu(\alpha_1-\alpha_2)^2(1+\alpha_1\alpha_2)
\nonumber\\
&\quad 
+(\alpha_1-\alpha_2)(1+\alpha_1\alpha_2) \left[X_1(\alpha_1+3\alpha_2) -X_2(\alpha_2+3\alpha_1)\right] 
+2(\alpha_1-\alpha_2)\left(X_1\alpha_1\beta_2 -X_2\alpha_2\beta_1\right)
\nonumber\\
&\quad
-\frac{q_1q_2}{M\mu}\left[2(1-\nu)(\alpha_1-\alpha_2)^2 +\left(X_1-X_2\right)(\alpha_1-\alpha_2)(2a+\alpha_1+\alpha_2)\right] 
\bigg\rbrace ,
\\
f_{v^2}^S&= \frac{2}{5\left(1+\alpha_1\alpha_2-\frac{q_1q_2}{M\mu}\right)} \bigg\lbrace
5(\alpha_1-\alpha_2)^2(1-\alpha_1\alpha_2) 
+5(\alpha_1-\alpha_2)\left(X_1\alpha_1\beta_2 -X_2\alpha_2\beta_1\right) 
\nonumber\\
&\quad
+(\alpha_1-\alpha_2)(1+\alpha_1\alpha_2)\left[-\frac{25}{2}(1-\nu)(\alpha_1-\alpha_2) +\frac{11}{2}\left(X_2^2\alpha_1 -X_1^2\alpha_2\right) +\frac{35}{2}\left(X_1\alpha_1-X_2\alpha_2\right)\right] \nonumber\\
&\quad
+\frac{q_1q_2}{2M\mu}(\alpha_1-\alpha_2) \left[5(1-5\nu)(\alpha_1-\alpha_2) -10a\left(X_1-X_2\right)
-11\left(X_2^2\alpha_1 -X_1^2\alpha_2\right)\right]
\nonumber\\
&\quad
-\frac{5q_1q_2}{2M\mu}(\alpha_1-\alpha_2) \left[2\left(X_1\alpha_1 -X_2\alpha_2\right) 
+3\left(X_1\alpha_2 -X_2\alpha_1\right)\right]
\bigg\rbrace, \\
f_{1/r}^S &= -\frac{2}{5} \bigg\lbrace
5\nu(\alpha_1-\alpha_2)^2
+5(\alpha_1^2-\alpha_2^2)\left(X_1-X_2\right) 
+6(\alpha_1-\alpha_2)\left(X_2^2\alpha_1 -X_1^2\alpha_2\right) 
\nonumber\\
&\quad
+\frac{5(\alpha_1-\alpha_2)^2}{\left(1+\alpha_1\alpha_2-\frac{q_1q_2}{M\mu}\right)^2}
\left[-\frac{q_1q_2}{M\mu}\left(5-\alpha_1\alpha_2 +2aX_1\alpha_1 +2aX_2\alpha_2\right) +X_2\frac{q_1^2}{M\mu}(1+a\alpha_2) +X_1\frac{q_2^2}{M\mu}(1+a\alpha_1)\right]
\nonumber\\
&\quad
+\frac{5(\alpha_1-\alpha_2)^2}{\left(1+\alpha_1\alpha_2-\frac{q_1q_2}{M\mu}\right)^2} \left[4(1+\alpha_1\alpha_2) +X_2\alpha_2^2\beta_1 +X_1\alpha_1^2\beta_2\right]
\bigg\rbrace.
\end{align}
\end{subequations}
This flux reduces to the ST dipole flux derived in Ref.~\cite{Damour:1992we} in the limit where the electric charges are zero.
\end{widetext}

\subsection{Vector energy flux}

The calculation of the vector flux is similar to that of the scalar flux. The vector potential can be written in terms of radiative multipole moments as \cite{Damour:1990gj}
\begin{align}
A_0(\bm{X}, T) &= \frac{1}{R}\sum_{\ell\geq 0}\ \frac{1}{\ell!c^\ell}N^L Q_L^{(\ell)}(U), \nonumber\\
A_i(\bm{X}, T) &= \frac{1}{R}\sum_{\ell\geq 1} \frac{1}{\ell! c^\ell} \bigg[N_{L-1}Q_{iL-1}^{(\ell)}(U) \nonumber\\
&\qquad -\frac{\ell}{(\ell+1)c}\varepsilon_{iab}N_{aL-1}M_{bL-1}^{(\ell)}\bigg].
\end{align}
As was done in the previous subsection, the radiative moments can be related to the source moments using algorithmic moments. At leading order, the three agree, and we can express the electric and magnetic multipole moments directly in terms of the source moments 
\begin{align}
Q_L(U) =& \int d^3x \, \bigg[\widehat{x}_L\rho +\frac{1}{2(2\ell+3)c^2}x^2\widehat{x}_L \frac{d^2\rho}{dt^2} \nonumber\\
& -\frac{2\ell+1}{(\ell+1)(2\ell+1)c^2}\widehat{x}_{aL}\frac{dJ_a}{dt}\bigg], \quad \ell\geq 0\,,\\
M_L(U) =& \int d^3x \, \bigg[\widehat{x}_{\langle L-1}m_{i_\ell \rangle} + \nonumber\\
& \frac{1}{2(2\ell+3)c^2}x^2\widehat{x}_{\langle L-1}\frac{d}{dt^2}m_{i_\ell\rangle}\bigg] , \quad \ell\geq 1\,,
\end{align}
where the magnetization density $\bm{m}=\bm{x}\times\bm{J}$.
The source functions $\rho$ and $J_i$ are defined by
\begin{equation}
\square A_0 = 4\pi\rho\,, \qquad \square A_i =-\frac{4\pi}{c} J_i.
\end{equation}

The vector flux
\begin{equation}
\mathcal{F}_V = -cR^2\oint N^iT_{0i}^\text{EM}   d\Omega,
\end{equation}
where the electromagnetic part of the stress-energy tensor is given by 
\begin{equation}
T^{\text{EM}}_{\mu\nu}=\frac{1}{8\pi}e^{-2a\varphi}\left(2F_{\mu\alpha}{F_\nu}^\alpha -\frac{1}{2}g_{\mu\nu}F^2\right).
\end{equation}
In the far zone, 
\begin{align}
T^{\text{EM}}_{0i} &=\frac{1}{4\pi}F_{0j}{F_i}^j \nonumber\\
&=\frac{1}{4\pi}\left(\partial_0A_j-\partial_jA_0\right) \left(\partial_iA_j -\partial_jA_i\right).
\end{align}
The vector flux becomes
\begin{equation}
\mathcal{F}_V = \frac{R^2}{4\pi c}\int d\Omega \left[\frac{\partial A_i}{\partial U}\frac{\partial A_i}{\partial U}-N^iN^j\frac{\partial A_i}{\partial U}\frac{\partial A_j}{\partial U}\right].
\end{equation}

The vector potential $A_i$, to the required order, has the multipole expansion
\begin{equation}
A_i = \frac{1}{R}\left[\frac{1}{c}Q_i^{(1)} -\frac{1}{2c^2}\varepsilon_{ijk}N^jM_k^{(1)} +\frac{1}{2c^2}N^jQ_{ij}^{(2)} \right],
\end{equation}
which leads to
\begin{align}
&\frac{R^2}{4\pi c}\int d\Omega \left(\frac{\partial A_i}{\partial U}\right)^2 \nonumber\\
&\quad= \frac{Q_i^{(2)}Q_i^{(2)}}{c^3} + \frac{M_i^{(2)}M_i^{(2)}}{6c^5} + \frac{Q_{ij}^{(3)}Q_{ij}^{(3)}}{12c^5}+\Order\left(\frac{1}{c^7}\right) \nonumber\\
&\quad=\sum_{\ell\geq 1} \frac{1}{c^{2\ell+1}\ell!(2\ell+1)!!}\bigg[\frac{2\ell+1}{\ell} Q_L^{(\ell+1)}Q_L^{(\ell+1)} \nonumber\\
&\qquad\qquad +\frac{\ell}{c^2(\ell+1)}M_L^{(\ell+1)}M_L^{(\ell+1)}\bigg],
\end{align}
and
\begin{align}
&\frac{R^2}{4\pi c}\int d\Omega N^iN^j\frac{\partial A_i}{\partial U}\frac{\partial A_j}{\partial U} \nonumber\\
&\quad= \frac{Q_i^{(2)}Q_i^{(2)}}{3c^3}  + \frac{Q_{ij}^{(3)}Q_{ij}^{(3)}}{30c^5} +\Order\left(\frac{1}{c^7}\right) \nonumber\\
&\quad=\sum_{\ell\geq 1} \frac{1}{c^{2\ell+1}\ell!(2\ell+1)!!} Q_L^{(\ell+1)}Q_L^{(\ell+1)}.
\end{align}
Hence, the vector flux
\begin{align}
\mathcal{F}_V &=\sum_{\ell\geq 1} \frac{1}{c^{2\ell+1}\ell!(2\ell+1)!!}\bigg[\frac{\ell+1}{\ell} Q_L^{(\ell+1)}Q_L^{(\ell+1)} \nonumber\\
&\qquad\qquad +\frac{\ell}{c^2(\ell+1)}M_L^{(\ell+1)}M_L^{(\ell+1)}\bigg] \\
& =\frac{2Q_i^{(2)}Q_i^{(2)}}{3c^3}  +\frac{M_i^{(2)}M_i^{(2)}}{6c^5} +\frac{Q_{ij}^{(3)}Q_{ij}^{(3)}}{20c^5} + \dots .
\end{align}
The first two terms give the dipole flux, and the third term is the quadrupole flux. There is no monopole flux because of the conservation of the total electric charge.

The 1PN field equations are given by Eqs.~\eqref{eqAi} and \eqref{eqA02}, which are
\begin{align}
\square A_i&=-\frac{4\pi}{c} \rho_ev^i\,, \\
\square A_0 &=\, 4\pi  \rho_e- V\nabla^2A_0 +A_0\nabla^2V +a\varphi\nabla^2 A_0 \nonumber\\
&\quad -aA_0\nabla^2\varphi -\nabla^2(VA_0) +a\nabla^2(\varphi A_0).
\end{align} 
The last two terms in the above equation are of order $1/R^2$, and hence do not contribute to the next-to-leading order flux.
The source functions $\rho$ and $J^i$ are then given by
\begin{align}
\rho &= \rho_e = q_1\delta^3(\bm{x}-\bm{x}_1) + q_2\delta^3(\bm{x}-\bm{x}_2), \\
J^i &= \rho_e v^i = q_1 v_1^i\delta^3(\bm{x}-\bm{x}_1) + q_2 v_2^i\delta^3(\bm{x}-\bm{x}_2).
\end{align}
The function $\rho$ is simply the electric charge density because the higher order terms from the field equation cancel when summed over the two bodies.

For the dipole flux, we need $Q_i$ and $M_i$ to $\Order(1/c^2)$
\begin{align}
Q_i =&\int d^3x \left[x_i\rho_e +\frac{1}{10c^2}\rho_e\frac{d^2}{dt^2}\left(x^2x^i\right) -\frac{3}{10c^2}\frac{d}{dt}\left(\widehat{x}_{ij}J_j\right) \right] \nonumber\\
=&\left(q_1\frac{\mu_2}{\mu_1+\mu_2} -q_2\frac{\mu_1}{\mu_1+\mu_2}\right)r^i \nonumber\\
& +\frac{1}{10c^2}\left(q_1\frac{m_2^3}{M^3}-q_2\frac{m_1^3}{M^3}\right)\frac{d^2}{dt^2}\left(r^ir^2\right) \nonumber\\ &-\frac{3}{10c^2}\left(q_1\frac{m_2^3}{M^3}-q_2\frac{m_1^3}{M^3}\right) \frac{d}{dt}\left(r^ir^j-\frac{1}{3}r^2\delta^{ij}\right)v^j\,, \\
M_i=&\, q_1\varepsilon_{ijk}x_1^jv_1^k + q_2\varepsilon_{ijk}x_2^jv_2^k \nonumber\\
=& \left(q_1\frac{m_2^2}{M^2}+q_2\frac{m_1^2}{M^2}\right)\varepsilon_{ijk}r^jv^k\,.
\end{align}
Differentiating and using the 1PN acceleration from Eq.~\eqref{acc1pn}, we obtain the next-to-leading order vector dipole flux
\begin{widetext}
\begin{align}
\mathcal{F}_V^\text{Dip} &=
\frac{2}{3c^3}\left(\frac{G_{12}M\mu}{r^2}\right)^2 
\Bigg[
\left(\frac{q_1}{m_1}-\frac{q_2}{m_2}\right)^2 
+ f_{v^2}^V \frac{v^2}{c^2} 
+ f_{\dot{r}^2}^V \frac{\dot{r}^2}{c^2} 
+ f_{1/r}^V\frac{G_{12}M}{c^2r}
\Bigg],
\end{align}
with the coefficients
\begin{subequations}
\label{coeffEMdip}
\begin{align}
f_{v^2}^V &= \frac{2}{5}\left(X_2^2\frac{q_1}{m_1}-X_1^2\frac{q_2}{m_2}\right)\left(\frac{q_1}{m_1}-\frac{q_2}{m_2}\right)
+\frac{2}{M}(X_1-X_2)\left(q_1+q_2\right)\left(\frac{q_1}{m_1}-\frac{q_2}{m_2}\right)
\nonumber \\
&\quad +\frac{1}{1+\alpha_1\alpha_2-\frac{q_1q_2}{M\mu}}\left(\frac{q_1}{m_1}-\frac{q_2}{m_2}\right)^2\left(2+6\nu +2\alpha_1\alpha_2(3\nu-1) +\frac{q_1q_2}{M\mu}(1-6\nu) \right), 
\\
f_{\dot{r}^2}^V &= -\left(\frac{q_1}{m_1}-\frac{q_2}{m_2}\right)^2 \left[
3\nu +\frac{(X_1-X_2)(q_1+q_2)}{M \left(\frac{q_1}{m_1}-\frac{q_2}{m_2}\right)} +\frac{ 8 -4\nu(1+\alpha_1\alpha_2) - 2\frac{q_1q_2}{M\mu}(1-2\nu)}{1+\alpha_1\alpha_2 -\frac{q_1q_2}{M\mu}} \right], \\
f_{1/r}^V &= -2 \left(\frac{q_1}{m_1}-\frac{q_2}{m_2}\right)^2 \Bigg[ 
2\nu +\frac{2}{5}\,\frac{X_2^2q_1/m_1-X_1^2q_2/m_2}{q_1/m_1-q_2/m_2} 
-\frac{q_1q_2}{M\mu} \frac{5-\alpha_1\alpha_2 +2a\left(X_1\alpha_1 +X_2\alpha_2\right)}{\left(1+\alpha_1\alpha_2 -\frac{q_1q_2}{M\mu}\right)^2} \nonumber\\
&\quad
+\frac{(X_1-X_2)(q_1+q_2)}{M\left(\frac{q_1}{m_1}-\frac{q_2}{m_2}\right)} 
+\frac{4(1+\alpha_1\alpha_2)+X_2\frac{q_1^2}{M\mu}(1+a\alpha_2) +X_1\frac{q_2^2}{M\mu}(1+a\alpha_1) +X_2\alpha_2^2\beta_1 +X_1\alpha_1^2\beta_2}{\left(1+\alpha_1\alpha_2 -\frac{q_1q_2}{M\mu}\right)^2}
\Bigg] .
\end{align}
\end{subequations}
	
For the quadrupole flux, 
\begin{align}
Q_{ij}^{(3)} &=\int d^3x\left(x_ix_j-\frac{1}{3}x^2\delta_{ij}\right)\rho_e =\left(X_2^2q_1 +X_1^2q_2\right) \frac{d^3}{dt^3}\left(r^ir^j -\frac{1}{3}r^2\delta^{ij}\right),
\end{align}
which leads to 
\begin{align}
\mathcal{F}_V^\text{Quad}&= \frac{Q_{ij}^{(3)}Q_{ij}^{(3)}}{30c^5} =\frac{1}{30c^5}\left(\frac{G_{12}M\mu}{r^2}\right)^2\left(X_2\frac{q_1}{M}+X_1\frac{q_2}{m_2} \right)^2\left(32v^2-\frac{88}{3}\dot{r}^2\right).
\end{align}
\end{widetext}

\subsection{Tensor energy flux}

The metric in radiative coordinates
\begin{equation}
G_{\mu\nu}(\bm{X}^\mu)= \eta_{\mu\nu} +\frac{1}{R} H_{\mu\nu}(U,\bm{N}) + \mathcal{O}\left(\frac{1}{R^2}\right),
\end{equation}
where the radiative multipole moments $M_L$ and $S_L$ are defined by
\begin{align}
H_{ij}^\text{TT}(U,\bm{N}) &=4G\sum_{\ell\geq 2} \frac{1}{\ell! c^{\ell+2}} \bigg[N_{L-2}M_{ijL-2}^{(\ell)}(U) \nonumber\\
&\quad -\frac{2\ell}{(\ell+1)c} N_{hL-2} \varepsilon_{hk(i} S_{j)kL-2}^{(\ell)}\bigg]^\text{TT}.
\end{align}
The radiative multipoles agree with the source multipoles $I_L$ and $J_L$ up to order
\begin{equation}
M_L=I_L +\mathcal{O}(1/c^3), \qquad S_L=J_L+\mathcal{O}(1/c^2) ,
\end{equation}
where \cite{Damour:1990gj,Damour:1992we}
\begin{align}
I_L(t) &=\int d^3x\bigg[\widehat{x}_L\sigma +\frac{1}{2(2\ell+3)c^2}x^2\widehat{x}_L\frac{\partial^2\sigma}{\partial t^2} \nonumber\\
&\qquad\quad -\frac{4(2\ell+1)}{(\ell+1)(2\ell+3)c^2}\widehat{x}_{Ls}\frac{\partial \sigma^s}{\partial t} \bigg], \\
J_L(t) &= \int d^3x \,  \varepsilon_{hk\langle i_\ell}\widehat{x}_{L-1\rangle h}\sigma^k.
\end{align}
In terms of the multipole moments, the tensor flux is given by
\begin{widetext}
\begin{align}
F_g&=\frac{c^3}{32\pi G}\int d\Omega \left(\frac{\partial H_{ij}^\text{TT}}{\partial U}\right)^2 \nonumber\\
&= G\sum_{\ell\geq 2} \frac{1}{c^{2\ell+1}\ell!(2\ell+1)!!} \left[\frac{(\ell+1)(\ell+2)}{\ell(\ell-1)}M_L^{(\ell+1)}(U)M_L^{(\ell+1)}(U) 
+\frac{4\ell(\ell+2)}{c^2(\ell-1)(\ell+1)}S_L^{\ell+1}(U)S_L^{(\ell+1)}(U)\right] \nonumber\\
&= \frac{G}{5c^5}M_{ij}^{(3)}M_{ij}^{(3)} +\frac{G}{189c^7}M_{ijk}^{(4)}M_{ijk}^{(4)}+\frac{16G}{45c^7}S_{ij}^{(3)}S_{ij}^{(3)}+\Order\left(1/c^9\right),
\end{align}
\end{widetext}
where the first term is the mass quadrupole flux, the second is the mass octopole, and the third is the current quadrupole.

The source functions $\sigma$ and $\sigma^i$ are given by
\begin{equation}
\sigma \equiv \frac{T^{00}+T^{ss}}{c^2}, \qquad \sigma^i\equiv\frac{T^{0i}}{c},
\end{equation}
and from the 1PN field equations \eqref{eqV2} and \eqref{eqVi}
\begin{equation}
\square V=-\frac{4\pi G}{c^2}\sigma, \qquad \square V^i=-\frac{4\pi G}{c^3}\sigma^i ,
\end{equation}
with
\begin{align}
\sigma^i &= m_1v_1^i\delta^3(\bm{x}-\bm{x}_1) +m_2v_2^i\delta^3(\bm{x}-\bm{x}_2) , \\
\sigma&= \left[m_1 +\frac{3}{2c^2}m_1v_1^2-\frac{G_{12}m_1m_2}{c^2r}  \right]\delta(\bm{x}-\bm{x}_1) + 1\leftrightarrow 2 .
\end{align}

The multipole moments needed for the next-to-leading order flux are $M_{ij}$, $M_{ijk}$, and $S_{ij}$, which are given by
\begin{align}
M_{ij} &= \left(m_1+\frac{3}{2c^2}m_1v_1^2 -\frac{G_{12}m_1m_2}{c^2r}\right) \widehat{x}_1^{ij} \nonumber\\
&\quad +\frac{m_1}{14c^2}\frac{d^2}{dt^2}x_1^2\widehat{x}_1^{ij}
-\frac{20m_1}{21c^2}\frac{d}{dt}v_1^k\widehat{x}_1^{ijk} + 1\leftrightarrow 2, \\
M_{ijk}&=\, m_1\widehat{x}_1^{ijk} + m_2\widehat{x}_2^{ijk}, \\
S_{ij} &=\, m_1\varepsilon^{hk\langle j}x_1^{i\rangle} x_1^hv_1^k + 1\leftrightarrow 2.
\end{align} 
In the center-of-mass coordinates, this becomes
\begin{align}
M_{ij} &= \mu\left[1+\frac{3}{2c^2}(1-3\nu)v^2 -\frac{G_{12}M}{c^2r}(1-2\nu)\right] \nonumber\\
&\quad +\frac{\mu}{14c^2}(1-3\nu) \frac{d^2}{dt^2}r^2\widehat{r}^{ij} -\frac{20\mu}{21c^2}(1-3\nu) \frac{d}{dt} v^k\widehat{r}^{ijk}, \\
M_{ijk} &=\mu\left[\frac{m_2^2}{M^2}-\frac{m_1^2}{M^2}\right] \widehat{r}^{ijk} , \\
S_{ij} &=\mu \left[\frac{m_2^2}{M^2}-\frac{m_1^2}{M^2}\right]  \varepsilon^{hk\langle j}r^{i\rangle}r^hv^k,
\end{align}
where 
\begin{align}
&\widehat{r}^{ij}= r^ir^j  -\frac{1}{3}r^2\delta_{ij}, \\
&\widehat{r}^{ijk}= r^ir^jr^k -\frac{r^2}{5}\left(r^i\delta^{jk} +r^j\delta^{ik}+r^k\delta^{ij}\right), \\
&\varepsilon^{hk\langle j}r^{i\rangle}r^hv^k = \left[\frac{1}{2}\varepsilon^{hkj}r^i +\frac{1}{2}\varepsilon^{hki}r^j -\frac{1}{3}\varepsilon^{hkm}r^m \right]r^hv^k.
\end{align}
Taking the time derivatives of the multipole moments and squaring, we obtain the tensor flux
\begin{align}
\mathcal{F}_T&=\frac{8G}{15c^5} \left(\frac{G_{12}M\mu}{r^2}\right)^2   \left[12v^2-11 \dot{r}^2\right] \nonumber\\ 
&\quad + \frac{8G}{420c^7} \left(\frac{G_{12}M\mu}{r^2}\right)^2
\bigg[
f_{v^4}^T  v^4  + f_{v^2\dot{r}^2}^T  v^2\dot{r}^2  \nonumber\\
&\quad +  f_{\dot{r}^4}^T \dot{r}^4
 +  f_{v^2/r}^T \frac{G_{12}M v^2}{r} \nonumber\\
&\quad +  f_{\dot{r}^2/r}^T \frac{G_{12}M \dot{r}^2}{r} +  f_{1/r^2}^T\frac{G_{12}^2M^2}{r^2} \bigg],
\end{align}
with the coefficients
\begin{subequations}
\label{coeffGW}
\begin{align}
f_{v^4}^T &= \left[\frac{785+113\alpha_1\alpha_2-281\frac{q_1q_2}{M\mu}}{1+\alpha_1\alpha_2-\frac{q_1q_2}{M\mu}} -852\nu\right], \\
f_{v^2\dot{r}^2}^T &= -2\left[\frac{1487+255\alpha_1\alpha_2-563\frac{q_1q_2}{M\mu}}{1+\alpha_1\alpha_2-\frac{q_1q_2}{M\mu}} -1392\nu\right], \\
f_{\dot{r}^4}^T &= 3 \left[\frac{687+127\alpha_1\alpha_2-267\frac{q_1q_2}{M\mu}}{1+\alpha_1\alpha_2-\frac{q_1q_2}{M\mu}} -620\nu\right], \\
f_{1/r^2}^T &= 16(1-4\nu),
\end{align}
\begin{widetext}
\begin{align} 
f_{v^2/r}^T &= -\frac{8}{\left(1+\alpha_1\alpha_2-\frac{q_1q_2}{M\mu}\right)^2}
\Bigg[
20(1+\alpha_1\alpha_2)(17-\nu) +4\alpha_1\alpha_2(1+\alpha_1\alpha_2)(22-5\nu)
+84\frac{q_1^2}{M\mu}X_2(1+a\alpha_2)
\nonumber\\
&\quad 
+84\frac{q_2^2}{M\mu}X_1(1+a\alpha_1)
+\frac{q_1^2q_2^2}{M^2\mu^2}(67-20\nu)  
-168\frac{aq_1q_2}{M\mu}\left(X_1\alpha_1 +X_2\alpha_2\right) 
-\frac{q_1q_2}{M\mu}(491-40\nu)
\nonumber\\
&\quad 
-\alpha_1\alpha_2\frac{q_1q_2}{M\mu}(71-40\nu)
+84\left(X_1\alpha_1^2\beta_2 +X_2\alpha_2^2\beta_1\right)
\Bigg], \\
f_{\dot{r}^2/r}^T  &= \frac{8}{\left(1+\alpha_1\alpha_2-\frac{q_1q_2}{M\mu}\right)^2}
\Bigg[
(1+\alpha_1\alpha_2)(367-15\nu) +3\alpha_1\alpha_2(1+\alpha_1\alpha_2)(29-5\nu) 
+84\frac{q_1^2}{M\mu}X_2(1+a\alpha_2) \nonumber\\
& \quad
+84\frac{q_2^2}{M\mu}X_1(1+a\alpha_1)
+\frac{q_1^2q_2^2}{M^2\mu^2}(73-15\nu)
-168\frac{aq_1q_2}{M\mu}\left(X_1\alpha_1 +X_2\alpha_2\right)
-\frac{2q_1q_2}{M\mu}(262-15\nu) \nonumber\\
&\quad
-2\frac{q_1q_2}{M\mu}\alpha_1\alpha_2(38-15\nu)
+84\left(X_1\alpha_1^2\beta_2 +X_2\alpha_2^2\beta_1\right)
\Bigg].
\end{align}
\end{widetext}
\end{subequations}
This flux reduces to the one derived in Ref.~\cite{Lang:2014osa}, in the context of ST theory, when the electric charges are zero and after converting the notation to the Jordan-Fierz frame.

\subsection{Energy flux for circular orbits}

In this section, we express the energy flux for circular orbits in terms of the gauge-independent parameter $x$, which is defined by
\begin{equation}
x\equiv \left(\frac{G_{12}M\Omega}{c^3}\right)^{2/3},
\end{equation}
where $\Omega$ is the orbital frequency.
To do that, we need to find the relation between $r$ and $\Omega$ to 1PN order (Kepler's third law). We start by writing the Lagrangian \eqref{1PNlag} in the center-of-mass coordinates  
\begin{align}
L&=-Mc^2 + \frac{1}{2}\mu v^2 +\frac{G_{12}M\mu}{r} \nonumber\\
&\quad +\frac{1}{c^2} \bigg\lbrace
\frac{1}{8}(1-3\nu)\mu v^4 \nonumber\\
&\quad +\frac{G_{12}M\mu}{2r}\left[\left(\frac{3-\alpha_1\alpha_2}{1+\alpha_1\alpha_2-\frac{q_1q_2}{M\mu}} +\nu\right)v^2 +\nu \dot{r}^2\right]  \nonumber\\
&\quad
-\frac{M^2\mu}{2r^2}\bigg[
(1+\alpha_1\alpha_2)^2
+X_2\alpha_2^2\beta_1 +X_1\alpha_1^2\beta_2 \nonumber\\
&\quad -2\frac{q_1q_2}{M\mu}\left(1+a \alpha_1X_1 +a\alpha_2X_2\right) \nonumber\\
&\quad +\frac{q_2^2}{M\mu}X_1(1+a\alpha_1) +\frac{q_1^2}{M\mu}X_2(1+a\alpha_2) 
\bigg] \bigg\rbrace.
\end{align}

Applying the Euler-Lagrange equation and using   $\dot{r}=0$ and $v=r\Omega$ leads to
\begin{equation}
\label{kepler3rd}
\Omega^2 = \frac{G_{12}M}{r^3}\left[1-3 f_\gamma \gamma +\Order\left(\frac{1}{c^4}\right)\right],
\end{equation}
where the parameter $\gamma$ is defined by
\begin{equation}
\gamma \equiv \frac{G_{12}M}{c^2r}\,,
\end{equation}
and the the coefficient $f_\gamma$ is defined by
\begin{align}
f_\gamma &\equiv \frac{1}{6G_{12}^2} \bigg[G_{12}^2(1-2\nu) +G_{12}(3-\alpha_1\alpha_2) \nonumber\\
&\quad +2(1+\alpha_1\alpha_2)^2 +2X_2\alpha_2^2\beta_1 
+2X_1\alpha_1^2\beta_2 \nonumber\\
&\quad +2\frac{q_2^2}{M\mu}X_1(1+a\alpha_1) 
+2\frac{q_1^2}{M\mu}X_2(1+a\alpha_2) \nonumber\\
&\quad -4\frac{q_1q_2}{M\mu} \left(1+aX_1\alpha_1 +aX_2\alpha_2\right) 
\bigg]. 
\end{align}
Substituting $x$ instead of $\Omega$ and inverting Eq.~\eqref{kepler3rd}, we obtain 
\begin{equation}
\label{gammax}
\gamma = x\left[1+ f_\gamma x + \Order\left(1/c^4\right)\right].
\end{equation}

To express the flux for circular orbits in terms of $\gamma$, we set $\dot{r}=0$ and $v=r\Omega$ and then use Eqs.~\eqref{kepler3rd} to obtain 
\begin{widetext}
\begin{subequations}
\begin{align}
\mathcal{F}_S&= \frac{Gc^5}{3G_{12}^2}\nu^2\gamma^4(\alpha_1-\alpha_2)^2
+\frac{Gc^5}{3G_{12}^2}\nu^2 \gamma^5 \left[f_{v^2}^S+f_{1/r}^S +\frac{16}{5}\left(X_1\alpha_2 +X_2\alpha_1\right)^2\right], \\
\mathcal{F}_V&=  \frac{2Gc^5}{3G_{12}^2}\nu^2\gamma^4 \left(\frac{q_1}{m_1}-\frac{q_2}{m_2}\right)^2 +\frac{2Gc^5}{3G_{12}^2} \nu^2 \gamma^5 \left[\frac{8}{5}\left(X_2\frac{q_1}{m_1} +X_1\frac{q_2}{m_2}\right)^2 + f_{v^2}^V +f_{1/r}^V \right], \\
\mathcal{F}_T &=  \frac{32Gc^5}{5G_{12}^2} \nu^2\gamma^5 +\frac{2Gc^5}{105G_{12}^2}\nu^2\gamma^6\left(f_{v^4}^T + f_{v^2/r}^T + f_{1/r^2}^T + 1008 f_\gamma\right).
\end{align}
\end{subequations}
Using Eq.~\eqref{gammax} to express the energy flux in terms of $x$ instead of $\gamma$ leads to Eq.~\eqref{NLOflux}.
\end{widetext}


\begin{thebibliography}{85}%
\makeatletter
\providecommand \@ifxundefined [1]{%
 \@ifx{#1\undefined}
}%
\providecommand \@ifnum [1]{%
 \ifnum #1\expandafter \@firstoftwo
 \else \expandafter \@secondoftwo
 \fi
}%
\providecommand \@ifx [1]{%
 \ifx #1\expandafter \@firstoftwo
 \else \expandafter \@secondoftwo
 \fi
}%
\providecommand \natexlab [1]{#1}%
\providecommand \enquote  [1]{``#1''}%
\providecommand \bibnamefont  [1]{#1}%
\providecommand \bibfnamefont [1]{#1}%
\providecommand \citenamefont [1]{#1}%
\providecommand \href@noop [0]{\@secondoftwo}%
\providecommand \href [0]{\begingroup \@sanitize@url \@href}%
\providecommand \@href[1]{\@@startlink{#1}\@@href}%
\providecommand \@@href[1]{\endgroup#1\@@endlink}%
\providecommand \@sanitize@url [0]{\catcode `\\12\catcode `\$12\catcode
  `\&12\catcode `\#12\catcode `\^12\catcode `\_12\catcode `\%12\relax}%
\providecommand \@@startlink[1]{}%
\providecommand \@@endlink[0]{}%
\providecommand \url  [0]{\begingroup\@sanitize@url \@url }%
\providecommand \@url [1]{\endgroup\@href {#1}{\urlprefix }}%
\providecommand \urlprefix  [0]{URL }%
\providecommand \Eprint [0]{\href }%
\providecommand \doibase [0]{http://dx.doi.org/}%
\providecommand \selectlanguage [0]{\@gobble}%
\providecommand \bibinfo  [0]{\@secondoftwo}%
\providecommand \bibfield  [0]{\@secondoftwo}%
\providecommand \translation [1]{[#1]}%
\providecommand \BibitemOpen [0]{}%
\providecommand \bibitemStop [0]{}%
\providecommand \bibitemNoStop [0]{.\EOS\space}%
\providecommand \EOS [0]{\spacefactor3000\relax}%
\providecommand \BibitemShut  [1]{\csname bibitem#1\endcsname}%
\let\auto@bib@innerbib\@empty
\bibitem [{\citenamefont {Abbott}\ \emph
  {et~al.}(2016{\natexlab{a}})\citenamefont {Abbott} \emph
  {et~al.}}]{Abbott:2016blz}%
  \BibitemOpen
  \bibfield  {author} {\bibinfo {author} {\bibfnamefont {B.~P.}\ \bibnamefont
  {Abbott}} \emph {et~al.} (\bibinfo {collaboration} {Virgo, LIGO
  Scientific}),\ }\bibfield  {title} {\enquote {\bibinfo {title} {{Observation
  of Gravitational Waves from a Binary Black Hole Merger}},}\ }\href {\doibase
  10.1103/PhysRevLett.116.061102} {\bibfield  {journal} {\bibinfo  {journal}
  {Phys. Rev. Lett.}\ }\textbf {\bibinfo {volume} {116}},\ \bibinfo {pages}
  {061102} (\bibinfo {year} {2016}{\natexlab{a}})},\ \Eprint
  {http://arxiv.org/abs/1602.03837} {arXiv:1602.03837 [gr-qc]} \BibitemShut
  {NoStop}%
\bibitem [{\citenamefont {Abbott}\ \emph
  {et~al.}(2016{\natexlab{b}})\citenamefont {Abbott} \emph
  {et~al.}}]{Abbott:2016nmj}%
  \BibitemOpen
  \bibfield  {author} {\bibinfo {author} {\bibfnamefont {B.~P.}\ \bibnamefont
  {Abbott}} \emph {et~al.} (\bibinfo {collaboration} {Virgo, LIGO
  Scientific}),\ }\bibfield  {title} {\enquote {\bibinfo {title} {{GW151226:
  Observation of Gravitational Waves from a 22-Solar-Mass Binary Black Hole
  Coalescence}},}\ }\href {\doibase 10.1103/PhysRevLett.116.241103} {\bibfield
  {journal} {\bibinfo  {journal} {Phys. Rev. Lett.}\ }\textbf {\bibinfo
  {volume} {116}},\ \bibinfo {pages} {241103} (\bibinfo {year}
  {2016}{\natexlab{b}})},\ \Eprint {http://arxiv.org/abs/1606.04855}
  {arXiv:1606.04855 [gr-qc]} \BibitemShut {NoStop}%
\bibitem [{\citenamefont {Abbott}\ \emph
  {et~al.}(2017{\natexlab{a}})\citenamefont {Abbott} \emph
  {et~al.}}]{Abbott:2017vtc}%
  \BibitemOpen
  \bibfield  {author} {\bibinfo {author} {\bibfnamefont {B.~P.}\ \bibnamefont
  {Abbott}} \emph {et~al.} (\bibinfo {collaboration} {VIRGO, LIGO
  Scientific}),\ }\bibfield  {title} {\enquote {\bibinfo {title} {{GW170104:
  Observation of a 50-Solar-Mass Binary Black Hole Coalescence at Redshift
  0.2}},}\ }\href {\doibase 10.1103/PhysRevLett.118.221101} {\bibfield
  {journal} {\bibinfo  {journal} {Phys. Rev. Lett.}\ }\textbf {\bibinfo
  {volume} {118}},\ \bibinfo {pages} {221101} (\bibinfo {year}
  {2017}{\natexlab{a}})},\ \Eprint {http://arxiv.org/abs/1706.01812}
  {arXiv:1706.01812 [gr-qc]} \BibitemShut {NoStop}%
\bibitem [{\citenamefont {Abbott}\ \emph
  {et~al.}(2017{\natexlab{b}})\citenamefont {Abbott} \emph
  {et~al.}}]{Abbott:2017gyy}%
  \BibitemOpen
  \bibfield  {author} {\bibinfo {author} {\bibfnamefont {B.~P.}\ \bibnamefont
  {Abbott}} \emph {et~al.} (\bibinfo {collaboration} {Virgo, LIGO
  Scientific}),\ }\bibfield  {title} {\enquote {\bibinfo {title} {{GW170608:
  Observation of a 19-solar-mass Binary Black Hole Coalescence}},}\ }\href
  {\doibase 10.3847/2041-8213/aa9f0c} {\bibfield  {journal} {\bibinfo
  {journal} {Astrophys. J.}\ }\textbf {\bibinfo {volume} {851}},\ \bibinfo
  {pages} {L35} (\bibinfo {year} {2017}{\natexlab{b}})},\ \Eprint
  {http://arxiv.org/abs/1711.05578} {arXiv:1711.05578 [astro-ph.HE]}
  \BibitemShut {NoStop}%
\bibitem [{\citenamefont {Abbott}\ \emph
  {et~al.}(2017{\natexlab{c}})\citenamefont {Abbott} \emph
  {et~al.}}]{Abbott:2017oio}%
  \BibitemOpen
  \bibfield  {author} {\bibinfo {author} {\bibfnamefont {B.~P.}\ \bibnamefont
  {Abbott}} \emph {et~al.} (\bibinfo {collaboration} {Virgo, LIGO
  Scientific}),\ }\bibfield  {title} {\enquote {\bibinfo {title} {{GW170814: A
  Three-Detector Observation of Gravitational Waves from a Binary Black Hole
  Coalescence}},}\ }\href {\doibase 10.1103/PhysRevLett.119.141101} {\bibfield
  {journal} {\bibinfo  {journal} {Phys. Rev. Lett.}\ }\textbf {\bibinfo
  {volume} {119}},\ \bibinfo {pages} {141101} (\bibinfo {year}
  {2017}{\natexlab{c}})},\ \Eprint {http://arxiv.org/abs/1709.09660}
  {arXiv:1709.09660 [gr-qc]} \BibitemShut {NoStop}%
\bibitem [{\citenamefont {Abbott}\ \emph
  {et~al.}(2017{\natexlab{d}})\citenamefont {Abbott} \emph
  {et~al.}}]{TheLIGOScientific:2017qsa}%
  \BibitemOpen
  \bibfield  {author} {\bibinfo {author} {\bibfnamefont {B.}~\bibnamefont
  {Abbott}} \emph {et~al.} (\bibinfo {collaboration} {Virgo, LIGO
  Scientific}),\ }\bibfield  {title} {\enquote {\bibinfo {title} {{GW170817:
  Observation of Gravitational Waves from a Binary Neutron Star Inspiral}},}\
  }\href {\doibase 10.1103/PhysRevLett.119.161101} {\bibfield  {journal}
  {\bibinfo  {journal} {Phys. Rev. Lett.}\ }\textbf {\bibinfo {volume} {119}},\
  \bibinfo {pages} {161101} (\bibinfo {year} {2017}{\natexlab{d}})},\ \Eprint
  {http://arxiv.org/abs/1710.05832} {arXiv:1710.05832 [gr-qc]} \BibitemShut
  {NoStop}%
\bibitem [{\citenamefont {Will}(2014)}]{Will:2014kxa}%
  \BibitemOpen
  \bibfield  {author} {\bibinfo {author} {\bibfnamefont {C.~M.}\ \bibnamefont
  {Will}},\ }\bibfield  {title} {\enquote {\bibinfo {title} {{The Confrontation
  between General Relativity and Experiment}},}\ }\href {\doibase
  10.12942/lrr-2014-4} {\bibfield  {journal} {\bibinfo  {journal} {Living Rev.
  Rel.}\ }\textbf {\bibinfo {volume} {17}},\ \bibinfo {pages} {4} (\bibinfo
  {year} {2014})},\ \Eprint {http://arxiv.org/abs/1403.7377} {arXiv:1403.7377
  [gr-qc]} \BibitemShut {NoStop}%
\bibitem [{\citenamefont {Berti}\ \emph {et~al.}(2015)\citenamefont {Berti}
  \emph {et~al.}}]{Berti:2015itd}%
  \BibitemOpen
  \bibfield  {author} {\bibinfo {author} {\bibfnamefont {E.}~\bibnamefont
  {Berti}} \emph {et~al.},\ }\bibfield  {title} {\enquote {\bibinfo {title}
  {{Testing General Relativity with Present and Future Astrophysical
  Observations}},}\ }\href {\doibase 10.1088/0264-9381/32/24/243001} {\bibfield
   {journal} {\bibinfo  {journal} {Class. Quant. Grav.}\ }\textbf {\bibinfo
  {volume} {32}},\ \bibinfo {pages} {243001} (\bibinfo {year} {2015})},\
  \Eprint {http://arxiv.org/abs/1501.07274} {arXiv:1501.07274 [gr-qc]}
  \BibitemShut {NoStop}%
\bibitem [{\citenamefont {Yunes}\ \emph {et~al.}(2016)\citenamefont {Yunes},
  \citenamefont {Yagi},\ and\ \citenamefont {Pretorius}}]{Yunes:2016jcc}%
  \BibitemOpen
  \bibfield  {author} {\bibinfo {author} {\bibfnamefont {N.}~\bibnamefont
  {Yunes}}, \bibinfo {author} {\bibfnamefont {K.}~\bibnamefont {Yagi}}, \ and\
  \bibinfo {author} {\bibfnamefont {F.}~\bibnamefont {Pretorius}},\ }\bibfield
  {title} {\enquote {\bibinfo {title} {{Theoretical Physics Implications of the
  Binary Black-Hole Mergers GW150914 and GW151226}},}\ }\href {\doibase
  10.1103/PhysRevD.94.084002} {\bibfield  {journal} {\bibinfo  {journal} {Phys.
  Rev.}\ }\textbf {\bibinfo {volume} {D94}},\ \bibinfo {pages} {084002}
  (\bibinfo {year} {2016})},\ \Eprint {http://arxiv.org/abs/1603.08955}
  {arXiv:1603.08955 [gr-qc]} \BibitemShut {NoStop}%
\bibitem [{\citenamefont {Arun}\ \emph {et~al.}(2006)\citenamefont {Arun},
  \citenamefont {Iyer}, \citenamefont {Qusailah},\ and\ \citenamefont
  {Sathyaprakash}}]{Arun:2006hn}%
  \BibitemOpen
  \bibfield  {author} {\bibinfo {author} {\bibfnamefont {K.~G.}\ \bibnamefont
  {Arun}}, \bibinfo {author} {\bibfnamefont {B.~R.}\ \bibnamefont {Iyer}},
  \bibinfo {author} {\bibfnamefont {M.~S.~S.}\ \bibnamefont {Qusailah}}, \ and\
  \bibinfo {author} {\bibfnamefont {B.~S.}\ \bibnamefont {Sathyaprakash}},\
  }\bibfield  {title} {\enquote {\bibinfo {title} {{Probing the non-linear
  structure of general relativity with black hole binaries}},}\ }\href
  {\doibase 10.1103/PhysRevD.74.024006} {\bibfield  {journal} {\bibinfo
  {journal} {Phys. Rev.}\ }\textbf {\bibinfo {volume} {D74}},\ \bibinfo {pages}
  {024006} (\bibinfo {year} {2006})},\ \Eprint
  {http://arxiv.org/abs/gr-qc/0604067} {arXiv:gr-qc/0604067 [gr-qc]}
  \BibitemShut {NoStop}%
\bibitem [{\citenamefont {Yunes}\ and\ \citenamefont
  {Pretorius}(2009)}]{Yunes:2009ke}%
  \BibitemOpen
  \bibfield  {author} {\bibinfo {author} {\bibfnamefont {N.}~\bibnamefont
  {Yunes}}\ and\ \bibinfo {author} {\bibfnamefont {F.}~\bibnamefont
  {Pretorius}},\ }\bibfield  {title} {\enquote {\bibinfo {title} {{Fundamental
  Theoretical Bias in Gravitational Wave Astrophysics and the Parameterized
  Post-Einsteinian Framework}},}\ }\href {\doibase 10.1103/PhysRevD.80.122003}
  {\bibfield  {journal} {\bibinfo  {journal} {Phys. Rev.}\ }\textbf {\bibinfo
  {volume} {D80}},\ \bibinfo {pages} {122003} (\bibinfo {year} {2009})},\
  \Eprint {http://arxiv.org/abs/0909.3328} {arXiv:0909.3328 [gr-qc]}
  \BibitemShut {NoStop}%
\bibitem [{\citenamefont {Mishra}\ \emph {et~al.}(2010)\citenamefont {Mishra},
  \citenamefont {Arun}, \citenamefont {Iyer},\ and\ \citenamefont
  {Sathyaprakash}}]{Mishra:2010tp}%
  \BibitemOpen
  \bibfield  {author} {\bibinfo {author} {\bibfnamefont {C.~K.}\ \bibnamefont
  {Mishra}}, \bibinfo {author} {\bibfnamefont {K.~G.}\ \bibnamefont {Arun}},
  \bibinfo {author} {\bibfnamefont {B.~R.}\ \bibnamefont {Iyer}}, \ and\
  \bibinfo {author} {\bibfnamefont {B.~S.}\ \bibnamefont {Sathyaprakash}},\
  }\bibfield  {title} {\enquote {\bibinfo {title} {{Parametrized tests of
  post-Newtonian theory using Advanced LIGO and Einstein Telescope}},}\ }\href
  {\doibase 10.1103/PhysRevD.82.064010} {\bibfield  {journal} {\bibinfo
  {journal} {Phys. Rev.}\ }\textbf {\bibinfo {volume} {D82}},\ \bibinfo {pages}
  {064010} (\bibinfo {year} {2010})},\ \Eprint {http://arxiv.org/abs/1005.0304}
  {arXiv:1005.0304 [gr-qc]} \BibitemShut {NoStop}%
\bibitem [{\citenamefont {Li}\ \emph {et~al.}(2012)\citenamefont {Li},
  \citenamefont {Del~Pozzo}, \citenamefont {Vitale}, \citenamefont {Van
  Den~Broeck}, \citenamefont {Agathos}, \citenamefont {Veitch}, \citenamefont
  {Grover}, \citenamefont {Sidery}, \citenamefont {Sturani},\ and\
  \citenamefont {Vecchio}}]{Li:2011cg}%
  \BibitemOpen
  \bibfield  {author} {\bibinfo {author} {\bibfnamefont {T.~G.~F.}\
  \bibnamefont {Li}}, \bibinfo {author} {\bibfnamefont {W.}~\bibnamefont
  {Del~Pozzo}}, \bibinfo {author} {\bibfnamefont {S.}~\bibnamefont {Vitale}},
  \bibinfo {author} {\bibfnamefont {C.}~\bibnamefont {Van Den~Broeck}},
  \bibinfo {author} {\bibfnamefont {M.}~\bibnamefont {Agathos}}, \bibinfo
  {author} {\bibfnamefont {J.}~\bibnamefont {Veitch}}, \bibinfo {author}
  {\bibfnamefont {K.}~\bibnamefont {Grover}}, \bibinfo {author} {\bibfnamefont
  {T.}~\bibnamefont {Sidery}}, \bibinfo {author} {\bibfnamefont
  {R.}~\bibnamefont {Sturani}}, \ and\ \bibinfo {author} {\bibfnamefont
  {A.}~\bibnamefont {Vecchio}},\ }\bibfield  {title} {\enquote {\bibinfo
  {title} {{Towards a generic test of the strong field dynamics of general
  relativity using compact binary coalescence}},}\ }\href {\doibase
  10.1103/PhysRevD.85.082003} {\bibfield  {journal} {\bibinfo  {journal} {Phys.
  Rev.}\ }\textbf {\bibinfo {volume} {D85}},\ \bibinfo {pages} {082003}
  (\bibinfo {year} {2012})},\ \Eprint {http://arxiv.org/abs/1110.0530}
  {arXiv:1110.0530 [gr-qc]} \BibitemShut {NoStop}%
\bibitem [{\citenamefont {Abbott}\ \emph
  {et~al.}(2016{\natexlab{c}})\citenamefont {Abbott} \emph
  {et~al.}}]{TheLIGOScientific:2016src}%
  \BibitemOpen
  \bibfield  {author} {\bibinfo {author} {\bibfnamefont {B.~P.}\ \bibnamefont
  {Abbott}} \emph {et~al.} (\bibinfo {collaboration} {Virgo, LIGO
  Scientific}),\ }\bibfield  {title} {\enquote {\bibinfo {title} {{Tests of
  general relativity with GW150914}},}\ }\href {\doibase
  10.1103/PhysRevLett.116.221101} {\bibfield  {journal} {\bibinfo  {journal}
  {Phys. Rev. Lett.}\ }\textbf {\bibinfo {volume} {116}},\ \bibinfo {pages}
  {221101} (\bibinfo {year} {2016}{\natexlab{c}})},\ \Eprint
  {http://arxiv.org/abs/1602.03841} {arXiv:1602.03841 [gr-qc]} \BibitemShut
  {NoStop}%
\bibitem [{\citenamefont {Abbott}\ \emph
  {et~al.}(2016{\natexlab{d}})\citenamefont {Abbott} \emph
  {et~al.}}]{TheLIGOScientific:2016pea}%
  \BibitemOpen
  \bibfield  {author} {\bibinfo {author} {\bibfnamefont {B.~P.}\ \bibnamefont
  {Abbott}} \emph {et~al.} (\bibinfo {collaboration} {Virgo, LIGO
  Scientific}),\ }\bibfield  {title} {\enquote {\bibinfo {title} {{Binary Black
  Hole Mergers in the first Advanced LIGO Observing Run}},}\ }\href {\doibase
  10.1103/PhysRevX.6.041015} {\bibfield  {journal} {\bibinfo  {journal} {Phys.
  Rev.}\ }\textbf {\bibinfo {volume} {X6}},\ \bibinfo {pages} {041015}
  (\bibinfo {year} {2016}{\natexlab{d}})},\ \Eprint
  {http://arxiv.org/abs/1606.04856} {arXiv:1606.04856 [gr-qc]} \BibitemShut
  {NoStop}%
\bibitem [{\citenamefont {Gibbons}\ and\ \citenamefont
  {Maeda}(1988)}]{Gibbons:1987ps}%
  \BibitemOpen
  \bibfield  {author} {\bibinfo {author} {\bibfnamefont {G.~W.}\ \bibnamefont
  {Gibbons}}\ and\ \bibinfo {author} {\bibfnamefont {K.-i.}\ \bibnamefont
  {Maeda}},\ }\bibfield  {title} {\enquote {\bibinfo {title} {{Black Holes and
  Membranes in Higher Dimensional Theories with Dilaton Fields}},}\ }\href
  {\doibase 10.1016/0550-3213(88)90006-5} {\bibfield  {journal} {\bibinfo
  {journal} {Nucl. Phys.}\ }\textbf {\bibinfo {volume} {B298}},\ \bibinfo
  {pages} {741--775} (\bibinfo {year} {1988})}\BibitemShut {NoStop}%
\bibitem [{\citenamefont {Garfinkle}\ \emph {et~al.}(1991)\citenamefont
  {Garfinkle}, \citenamefont {Horowitz},\ and\ \citenamefont
  {Strominger}}]{Garfinkle:1990qj}%
  \BibitemOpen
  \bibfield  {author} {\bibinfo {author} {\bibfnamefont {D.}~\bibnamefont
  {Garfinkle}}, \bibinfo {author} {\bibfnamefont {G.~T.}\ \bibnamefont
  {Horowitz}}, \ and\ \bibinfo {author} {\bibfnamefont {A.}~\bibnamefont
  {Strominger}},\ }\bibfield  {title} {\enquote {\bibinfo {title} {{Charged
  black holes in string theory}},}\ }\href {\doibase 10.1103/PhysRevD.43.3140,
  10.1103/PhysRevD.45.3888} {\bibfield  {journal} {\bibinfo  {journal} {Phys.
  Rev.}\ }\textbf {\bibinfo {volume} {D43}},\ \bibinfo {pages} {3140} (\bibinfo
  {year} {1991})},\ \bibinfo {note} {[Erratum: Phys.
  Rev.D45,3888(1992)]}\BibitemShut {NoStop}%
\bibitem [{\citenamefont {Hawking}(1972)}]{Hawking:1972qk}%
  \BibitemOpen
  \bibfield  {author} {\bibinfo {author} {\bibfnamefont {S.~W.}\ \bibnamefont
  {Hawking}},\ }\bibfield  {title} {\enquote {\bibinfo {title} {{Black holes in
  the Brans-Dicke theory of gravitation}},}\ }\href {\doibase
  10.1007/BF01877518} {\bibfield  {journal} {\bibinfo  {journal} {Commun. Math.
  Phys.}\ }\textbf {\bibinfo {volume} {25}},\ \bibinfo {pages} {167--171}
  (\bibinfo {year} {1972})}\BibitemShut {NoStop}%
\bibitem [{\citenamefont {Sotiriou}\ and\ \citenamefont
  {Faraoni}(2012)}]{Sotiriou:2011dz}%
  \BibitemOpen
  \bibfield  {author} {\bibinfo {author} {\bibfnamefont {T.~P.}\ \bibnamefont
  {Sotiriou}}\ and\ \bibinfo {author} {\bibfnamefont {V.}~\bibnamefont
  {Faraoni}},\ }\bibfield  {title} {\enquote {\bibinfo {title} {{Black holes in
  scalar-tensor gravity}},}\ }\href {\doibase 10.1103/PhysRevLett.108.081103}
  {\bibfield  {journal} {\bibinfo  {journal} {Phys. Rev. Lett.}\ }\textbf
  {\bibinfo {volume} {108}},\ \bibinfo {pages} {081103} (\bibinfo {year}
  {2012})},\ \Eprint {http://arxiv.org/abs/1109.6324} {arXiv:1109.6324 [gr-qc]}
  \BibitemShut {NoStop}%
\bibitem [{\citenamefont {Horne}\ and\ \citenamefont
  {Horowitz}(1992)}]{Horne:1992zy}%
  \BibitemOpen
  \bibfield  {author} {\bibinfo {author} {\bibfnamefont {J.~H.}\ \bibnamefont
  {Horne}}\ and\ \bibinfo {author} {\bibfnamefont {G.~T.}\ \bibnamefont
  {Horowitz}},\ }\bibfield  {title} {\enquote {\bibinfo {title} {{Rotating
  dilaton black holes}},}\ }\href {\doibase 10.1103/PhysRevD.46.1340}
  {\bibfield  {journal} {\bibinfo  {journal} {Phys. Rev.}\ }\textbf {\bibinfo
  {volume} {D46}},\ \bibinfo {pages} {1340--1346} (\bibinfo {year} {1992})},\
  \Eprint {http://arxiv.org/abs/hep-th/9203083} {arXiv:hep-th/9203083 [hep-th]}
  \BibitemShut {NoStop}%
\bibitem [{\citenamefont {Frolov}\ \emph {et~al.}(1987)\citenamefont {Frolov},
  \citenamefont {Zelnikov},\ and\ \citenamefont {Bleyer}}]{Frolov:1987rj}%
  \BibitemOpen
  \bibfield  {author} {\bibinfo {author} {\bibfnamefont {V.~P.}\ \bibnamefont
  {Frolov}}, \bibinfo {author} {\bibfnamefont {A.~I.}\ \bibnamefont
  {Zelnikov}}, \ and\ \bibinfo {author} {\bibfnamefont {U.}~\bibnamefont
  {Bleyer}},\ }\bibfield  {title} {\enquote {\bibinfo {title} {{Charged
  Rotating Black Hole From Five-dimensional Point of View}},}\ }\href@noop {}
  {\bibfield  {journal} {\bibinfo  {journal} {Annalen Phys.}\ }\textbf
  {\bibinfo {volume} {44}},\ \bibinfo {pages} {371--377} (\bibinfo {year}
  {1987})}\BibitemShut {NoStop}%
\bibitem [{\citenamefont {Gibbons}(1975)}]{Gibbons:1975kk}%
  \BibitemOpen
  \bibfield  {author} {\bibinfo {author} {\bibfnamefont {G.~W.}\ \bibnamefont
  {Gibbons}},\ }\bibfield  {title} {\enquote {\bibinfo {title} {{Vacuum
  Polarization and the Spontaneous Loss of Charge by Black Holes}},}\ }\href
  {\doibase 10.1007/BF01609829} {\bibfield  {journal} {\bibinfo  {journal}
  {Commun. Math. Phys.}\ }\textbf {\bibinfo {volume} {44}},\ \bibinfo {pages}
  {245--264} (\bibinfo {year} {1975})}\BibitemShut {NoStop}%
\bibitem [{\citenamefont {Eardley}\ and\ \citenamefont
  {Press}(1975)}]{Eardley:1975kp}%
  \BibitemOpen
  \bibfield  {author} {\bibinfo {author} {\bibfnamefont {D.~M.}\ \bibnamefont
  {Eardley}}\ and\ \bibinfo {author} {\bibfnamefont {W.~H.}\ \bibnamefont
  {Press}},\ }\bibfield  {title} {\enquote {\bibinfo {title} {{Astrophysical
  processes near black holes}},}\ }\href {\doibase
  10.1146/annurev.aa.13.090175.002121} {\bibfield  {journal} {\bibinfo
  {journal} {Ann. Rev. Astron. Astrophys.}\ }\textbf {\bibinfo {volume} {13}},\
  \bibinfo {pages} {381--422} (\bibinfo {year} {1975})}\BibitemShut {NoStop}%
\bibitem [{\citenamefont {De~Rujula}\ \emph {et~al.}(1990)\citenamefont
  {De~Rujula}, \citenamefont {Glashow},\ and\ \citenamefont
  {Sarid}}]{DeRujula:1989fe}%
  \BibitemOpen
  \bibfield  {author} {\bibinfo {author} {\bibfnamefont {A.}~\bibnamefont
  {De~Rujula}}, \bibinfo {author} {\bibfnamefont {S.~L.}\ \bibnamefont
  {Glashow}}, \ and\ \bibinfo {author} {\bibfnamefont {U.}~\bibnamefont
  {Sarid}},\ }\bibfield  {title} {\enquote {\bibinfo {title} {Charged dark
  matter},}\ }\href {\doibase 10.1016/0550-3213(90)90227-5} {\bibfield
  {journal} {\bibinfo  {journal} {Nucl. Phys.}\ }\textbf {\bibinfo {volume}
  {B333}},\ \bibinfo {pages} {173--194} (\bibinfo {year} {1990})}\BibitemShut
  {NoStop}%
\bibitem [{\citenamefont {Perl}\ and\ \citenamefont {Lee}(1997)}]{Perl:1997nd}%
  \BibitemOpen
  \bibfield  {author} {\bibinfo {author} {\bibfnamefont {M.~L.}\ \bibnamefont
  {Perl}}\ and\ \bibinfo {author} {\bibfnamefont {E.~R.}\ \bibnamefont {Lee}},\
  }\bibfield  {title} {\enquote {\bibinfo {title} {{The search for elementary
  particles with fractional electric charge and the philosophy of speculative
  experiments}},}\ }\href {\doibase 10.1119/1.18641} {\bibfield  {journal}
  {\bibinfo  {journal} {Am. J. Phys.}\ }\textbf {\bibinfo {volume} {65}},\
  \bibinfo {pages} {698--706} (\bibinfo {year} {1997})}\BibitemShut {NoStop}%
\bibitem [{\citenamefont {McDermott}\ \emph {et~al.}(2011)\citenamefont
  {McDermott}, \citenamefont {Yu},\ and\ \citenamefont
  {Zurek}}]{McDermott:2010pa}%
  \BibitemOpen
  \bibfield  {author} {\bibinfo {author} {\bibfnamefont {S.~D.}\ \bibnamefont
  {McDermott}}, \bibinfo {author} {\bibfnamefont {H.-B.}\ \bibnamefont {Yu}}, \
  and\ \bibinfo {author} {\bibfnamefont {K.~M.}\ \bibnamefont {Zurek}},\
  }\bibfield  {title} {\enquote {\bibinfo {title} {{Turning off the Lights: How
  Dark is Dark Matter?}}}\ }\href {\doibase 10.1103/PhysRevD.83.063509}
  {\bibfield  {journal} {\bibinfo  {journal} {Phys. Rev.}\ }\textbf {\bibinfo
  {volume} {D83}},\ \bibinfo {pages} {063509} (\bibinfo {year} {2011})},\
  \Eprint {http://arxiv.org/abs/1011.2907} {arXiv:1011.2907 [hep-ph]}
  \BibitemShut {NoStop}%
\bibitem [{\citenamefont {Cardoso}\ \emph {et~al.}(2016)\citenamefont
  {Cardoso}, \citenamefont {Macedo}, \citenamefont {Pani},\ and\ \citenamefont
  {Ferrari}}]{Cardoso:2016olt}%
  \BibitemOpen
  \bibfield  {author} {\bibinfo {author} {\bibfnamefont {V.}~\bibnamefont
  {Cardoso}}, \bibinfo {author} {\bibfnamefont {C.~F.~B.}\ \bibnamefont
  {Macedo}}, \bibinfo {author} {\bibfnamefont {P.}~\bibnamefont {Pani}}, \ and\
  \bibinfo {author} {\bibfnamefont {V.}~\bibnamefont {Ferrari}},\ }\bibfield
  {title} {\enquote {\bibinfo {title} {{Black holes and gravitational waves in
  models of minicharged dark matter}},}\ }\href {\doibase
  10.1088/1475-7516/2016/05/054} {\bibfield  {journal} {\bibinfo  {journal}
  {JCAP}\ }\textbf {\bibinfo {volume} {1605}},\ \bibinfo {pages} {054}
  (\bibinfo {year} {2016})},\ \Eprint {http://arxiv.org/abs/1604.07845}
  {arXiv:1604.07845 [hep-ph]} \BibitemShut {NoStop}%
\bibitem [{\citenamefont {Ackerman}\ \emph {et~al.}(2009)\citenamefont
  {Ackerman}, \citenamefont {Buckley}, \citenamefont {Carroll},\ and\
  \citenamefont {Kamionkowski}}]{Ackerman:mha}%
  \BibitemOpen
  \bibfield  {author} {\bibinfo {author} {\bibfnamefont {L.}~\bibnamefont
  {Ackerman}}, \bibinfo {author} {\bibfnamefont {M.~R.}\ \bibnamefont
  {Buckley}}, \bibinfo {author} {\bibfnamefont {S.~M.}\ \bibnamefont
  {Carroll}}, \ and\ \bibinfo {author} {\bibfnamefont {M.}~\bibnamefont
  {Kamionkowski}},\ }\bibfield  {title} {\enquote {\bibinfo {title} {{Dark
  Matter and Dark Radiation}},}\ }\bibfield  {booktitle} {\emph {\bibinfo
  {booktitle} {{Proceedings, 7th International Heidelberg Conference on Dark
  Matter in Astro and Particle Physics (DARK 2009): Christchurch, New Zealand,
  January 18-24, 2009}}},\ }\href {\doibase 10.1103/PhysRevD.79.023519,
  10.1142/9789814293792_0021} {\bibfield  {journal} {\bibinfo  {journal} {Phys.
  Rev.}\ }\textbf {\bibinfo {volume} {D79}},\ \bibinfo {pages} {023519}
  (\bibinfo {year} {2009})},\ \bibinfo {note} {[,277(2008)]},\ \Eprint
  {http://arxiv.org/abs/0810.5126} {arXiv:0810.5126 [hep-ph]} \BibitemShut
  {NoStop}%
\bibitem [{\citenamefont {Feng}\ \emph {et~al.}(2008)\citenamefont {Feng},
  \citenamefont {Tu},\ and\ \citenamefont {Yu}}]{Feng:2008mu}%
  \BibitemOpen
  \bibfield  {author} {\bibinfo {author} {\bibfnamefont {J.~L.}\ \bibnamefont
  {Feng}}, \bibinfo {author} {\bibfnamefont {H.}~\bibnamefont {Tu}}, \ and\
  \bibinfo {author} {\bibfnamefont {H.-B.}\ \bibnamefont {Yu}},\ }\bibfield
  {title} {\enquote {\bibinfo {title} {{Thermal Relics in Hidden Sectors}},}\
  }\href {\doibase 10.1088/1475-7516/2008/10/043} {\bibfield  {journal}
  {\bibinfo  {journal} {JCAP}\ }\textbf {\bibinfo {volume} {0810}},\ \bibinfo
  {pages} {043} (\bibinfo {year} {2008})},\ \Eprint
  {http://arxiv.org/abs/0808.2318} {arXiv:0808.2318 [hep-ph]} \BibitemShut
  {NoStop}%
\bibitem [{\citenamefont {Feng}\ \emph {et~al.}(2009)\citenamefont {Feng},
  \citenamefont {Kaplinghat}, \citenamefont {Tu},\ and\ \citenamefont
  {Yu}}]{Feng:2009mn}%
  \BibitemOpen
  \bibfield  {author} {\bibinfo {author} {\bibfnamefont {J.~L.}\ \bibnamefont
  {Feng}}, \bibinfo {author} {\bibfnamefont {M.}~\bibnamefont {Kaplinghat}},
  \bibinfo {author} {\bibfnamefont {H.}~\bibnamefont {Tu}}, \ and\ \bibinfo
  {author} {\bibfnamefont {H.-B.}\ \bibnamefont {Yu}},\ }\bibfield  {title}
  {\enquote {\bibinfo {title} {{Hidden Charged Dark Matter}},}\ }\href
  {\doibase 10.1088/1475-7516/2009/07/004} {\bibfield  {journal} {\bibinfo
  {journal} {JCAP}\ }\textbf {\bibinfo {volume} {0907}},\ \bibinfo {pages}
  {004} (\bibinfo {year} {2009})},\ \Eprint {http://arxiv.org/abs/0905.3039}
  {arXiv:0905.3039 [hep-ph]} \BibitemShut {NoStop}%
\bibitem [{\citenamefont {Davidson}\ \emph {et~al.}(2000)\citenamefont
  {Davidson}, \citenamefont {Hannestad},\ and\ \citenamefont
  {Raffelt}}]{Davidson:2000hf}%
  \BibitemOpen
  \bibfield  {author} {\bibinfo {author} {\bibfnamefont {S.}~\bibnamefont
  {Davidson}}, \bibinfo {author} {\bibfnamefont {S.}~\bibnamefont {Hannestad}},
  \ and\ \bibinfo {author} {\bibfnamefont {G.}~\bibnamefont {Raffelt}},\
  }\bibfield  {title} {\enquote {\bibinfo {title} {{Updated bounds on
  millicharged particles}},}\ }\href {\doibase 10.1088/1126-6708/2000/05/003}
  {\bibfield  {journal} {\bibinfo  {journal} {JHEP}\ }\textbf {\bibinfo
  {volume} {05}},\ \bibinfo {pages} {003} (\bibinfo {year} {2000})},\ \Eprint
  {http://arxiv.org/abs/hep-ph/0001179} {arXiv:hep-ph/0001179 [hep-ph]}
  \BibitemShut {NoStop}%
\bibitem [{\citenamefont {Burrage}\ \emph {et~al.}(2009)\citenamefont
  {Burrage}, \citenamefont {Jaeckel}, \citenamefont {Redondo},\ and\
  \citenamefont {Ringwald}}]{Burrage:2009yz}%
  \BibitemOpen
  \bibfield  {author} {\bibinfo {author} {\bibfnamefont {C.}~\bibnamefont
  {Burrage}}, \bibinfo {author} {\bibfnamefont {J.}~\bibnamefont {Jaeckel}},
  \bibinfo {author} {\bibfnamefont {J.}~\bibnamefont {Redondo}}, \ and\
  \bibinfo {author} {\bibfnamefont {A.}~\bibnamefont {Ringwald}},\ }\bibfield
  {title} {\enquote {\bibinfo {title} {{Late time CMB anisotropies constrain
  mini-charged particles}},}\ }\href {\doibase 10.1088/1475-7516/2009/11/002}
  {\bibfield  {journal} {\bibinfo  {journal} {JCAP}\ }\textbf {\bibinfo
  {volume} {0911}},\ \bibinfo {pages} {002} (\bibinfo {year} {2009})},\ \Eprint
  {http://arxiv.org/abs/0909.0649} {arXiv:0909.0649 [astro-ph.CO]} \BibitemShut
  {NoStop}%
\bibitem [{\citenamefont {Ahlers}(2009)}]{Ahlers:2009kh}%
  \BibitemOpen
  \bibfield  {author} {\bibinfo {author} {\bibfnamefont {M.}~\bibnamefont
  {Ahlers}},\ }\bibfield  {title} {\enquote {\bibinfo {title} {{The Hubble
  diagram as a probe of mini-charged particles}},}\ }\href {\doibase
  10.1103/PhysRevD.80.023513} {\bibfield  {journal} {\bibinfo  {journal} {Phys.
  Rev.}\ }\textbf {\bibinfo {volume} {D80}},\ \bibinfo {pages} {023513}
  (\bibinfo {year} {2009})},\ \Eprint {http://arxiv.org/abs/0904.0998}
  {arXiv:0904.0998 [hep-ph]} \BibitemShut {NoStop}%
\bibitem [{\citenamefont {Kadota}\ \emph {et~al.}(2016)\citenamefont {Kadota},
  \citenamefont {Sekiguchi},\ and\ \citenamefont {Tashiro}}]{Kadota:2016tqq}%
  \BibitemOpen
  \bibfield  {author} {\bibinfo {author} {\bibfnamefont {K.}~\bibnamefont
  {Kadota}}, \bibinfo {author} {\bibfnamefont {T.}~\bibnamefont {Sekiguchi}}, \
  and\ \bibinfo {author} {\bibfnamefont {H.}~\bibnamefont {Tashiro}},\
  }\bibfield  {title} {\enquote {\bibinfo {title} {{A new constraint on
  millicharged dark matter from galaxy clusters}},}\ }\href@noop {} {\
  (\bibinfo {year} {2016})},\ \Eprint {http://arxiv.org/abs/1602.04009}
  {arXiv:1602.04009 [astro-ph.CO]} \BibitemShut {NoStop}%
\bibitem [{\citenamefont {Hirschmann}\ \emph {et~al.}(2018)\citenamefont
  {Hirschmann}, \citenamefont {Lehner}, \citenamefont {Liebling},\ and\
  \citenamefont {Palenzuela}}]{Hirschmann:2017psw}%
  \BibitemOpen
  \bibfield  {author} {\bibinfo {author} {\bibfnamefont {E.~W.}\ \bibnamefont
  {Hirschmann}}, \bibinfo {author} {\bibfnamefont {L.}~\bibnamefont {Lehner}},
  \bibinfo {author} {\bibfnamefont {S.~L.}\ \bibnamefont {Liebling}}, \ and\
  \bibinfo {author} {\bibfnamefont {C.}~\bibnamefont {Palenzuela}},\ }\bibfield
   {title} {\enquote {\bibinfo {title} {{Black Hole Dynamics in
  Einstein-Maxwell-Dilaton Theory}},}\ }\href {\doibase
  10.1103/PhysRevD.97.064032} {\bibfield  {journal} {\bibinfo  {journal} {Phys.
  Rev.}\ }\textbf {\bibinfo {volume} {D97}},\ \bibinfo {pages} {064032}
  (\bibinfo {year} {2018})},\ \Eprint {http://arxiv.org/abs/1706.09875}
  {arXiv:1706.09875 [gr-qc]} \BibitemShut {NoStop}%
\bibitem [{\citenamefont {Zilhao}\ \emph {et~al.}(2012)\citenamefont {Zilhao},
  \citenamefont {Cardoso}, \citenamefont {Herdeiro}, \citenamefont {Lehner},\
  and\ \citenamefont {Sperhake}}]{Zilhao:2012gp}%
  \BibitemOpen
  \bibfield  {author} {\bibinfo {author} {\bibfnamefont {M.}~\bibnamefont
  {Zilhao}}, \bibinfo {author} {\bibfnamefont {V.}~\bibnamefont {Cardoso}},
  \bibinfo {author} {\bibfnamefont {C.}~\bibnamefont {Herdeiro}}, \bibinfo
  {author} {\bibfnamefont {L.}~\bibnamefont {Lehner}}, \ and\ \bibinfo {author}
  {\bibfnamefont {U.}~\bibnamefont {Sperhake}},\ }\bibfield  {title} {\enquote
  {\bibinfo {title} {{Collisions of charged black holes}},}\ }\href {\doibase
  10.1103/PhysRevD.85.124062} {\bibfield  {journal} {\bibinfo  {journal} {Phys.
  Rev.}\ }\textbf {\bibinfo {volume} {D85}},\ \bibinfo {pages} {124062}
  (\bibinfo {year} {2012})},\ \Eprint {http://arxiv.org/abs/1205.1063}
  {arXiv:1205.1063 [gr-qc]} \BibitemShut {NoStop}%
\bibitem [{\citenamefont {Zilhão}\ \emph {et~al.}(2014)\citenamefont
  {Zilhão}, \citenamefont {Cardoso}, \citenamefont {Herdeiro}, \citenamefont
  {Lehner},\ and\ \citenamefont {Sperhake}}]{Zilhao:2013nda}%
  \BibitemOpen
  \bibfield  {author} {\bibinfo {author} {\bibfnamefont {M.}~\bibnamefont
  {Zilhão}}, \bibinfo {author} {\bibfnamefont {V.}~\bibnamefont {Cardoso}},
  \bibinfo {author} {\bibfnamefont {C.}~\bibnamefont {Herdeiro}}, \bibinfo
  {author} {\bibfnamefont {L.}~\bibnamefont {Lehner}}, \ and\ \bibinfo {author}
  {\bibfnamefont {U.}~\bibnamefont {Sperhake}},\ }\bibfield  {title} {\enquote
  {\bibinfo {title} {{Collisions of oppositely charged black holes}},}\ }\href
  {\doibase 10.1103/PhysRevD.89.044008} {\bibfield  {journal} {\bibinfo
  {journal} {Phys. Rev.}\ }\textbf {\bibinfo {volume} {D89}},\ \bibinfo {pages}
  {044008} (\bibinfo {year} {2014})},\ \Eprint {http://arxiv.org/abs/1311.6483}
  {arXiv:1311.6483 [gr-qc]} \BibitemShut {NoStop}%
\bibitem [{\citenamefont {Buonanno}\ and\ \citenamefont
  {Damour}(1999)}]{Buonanno:1998gg}%
  \BibitemOpen
  \bibfield  {author} {\bibinfo {author} {\bibfnamefont {A.}~\bibnamefont
  {Buonanno}}\ and\ \bibinfo {author} {\bibfnamefont {T.}~\bibnamefont
  {Damour}},\ }\bibfield  {title} {\enquote {\bibinfo {title} {{Effective
  one-body approach to general relativistic two-body dynamics}},}\ }\href
  {\doibase 10.1103/PhysRevD.59.084006} {\bibfield  {journal} {\bibinfo
  {journal} {Phys. Rev.}\ }\textbf {\bibinfo {volume} {D59}},\ \bibinfo {pages}
  {084006} (\bibinfo {year} {1999})},\ \Eprint
  {http://arxiv.org/abs/gr-qc/9811091} {arXiv:gr-qc/9811091 [gr-qc]}
  \BibitemShut {NoStop}%
\bibitem [{\citenamefont {Buonanno}\ and\ \citenamefont
  {Damour}(2000)}]{Buonanno:2000ef}%
  \BibitemOpen
  \bibfield  {author} {\bibinfo {author} {\bibfnamefont {A.}~\bibnamefont
  {Buonanno}}\ and\ \bibinfo {author} {\bibfnamefont {T.}~\bibnamefont
  {Damour}},\ }\bibfield  {title} {\enquote {\bibinfo {title} {{Transition from
  inspiral to plunge in binary black hole coalescences}},}\ }\href {\doibase
  10.1103/PhysRevD.62.064015} {\bibfield  {journal} {\bibinfo  {journal} {Phys.
  Rev.}\ }\textbf {\bibinfo {volume} {D62}},\ \bibinfo {pages} {064015}
  (\bibinfo {year} {2000})},\ \Eprint {http://arxiv.org/abs/gr-qc/0001013}
  {arXiv:gr-qc/0001013 [gr-qc]} \BibitemShut {NoStop}%
\bibitem [{\citenamefont {Julié}(2018{\natexlab{a}})}]{Julie:2017rpw}%
  \BibitemOpen
  \bibfield  {author} {\bibinfo {author} {\bibfnamefont {F.-L.}\ \bibnamefont
  {Julié}},\ }\bibfield  {title} {\enquote {\bibinfo {title} {{On the motion
  of hairy black holes in Einstein-Maxwell-dilaton theories}},}\ }\href
  {\doibase 10.1088/1475-7516/2018/01/026} {\bibfield  {journal} {\bibinfo
  {journal} {JCAP}\ }\textbf {\bibinfo {volume} {1801}},\ \bibinfo {pages}
  {026} (\bibinfo {year} {2018}{\natexlab{a}})},\ \Eprint
  {http://arxiv.org/abs/1711.10769} {arXiv:1711.10769 [gr-qc]} \BibitemShut
  {NoStop}%
\bibitem [{\citenamefont {Flanagan}(2004)}]{Flanagan:2004bz}%
  \BibitemOpen
  \bibfield  {author} {\bibinfo {author} {\bibfnamefont {E.~E.}\ \bibnamefont
  {Flanagan}},\ }\bibfield  {title} {\enquote {\bibinfo {title} {{The Conformal
  frame freedom in theories of gravitation}},}\ }\href {\doibase
  10.1088/0264-9381/21/15/N02} {\bibfield  {journal} {\bibinfo  {journal}
  {Class. Quant. Grav.}\ }\textbf {\bibinfo {volume} {21}},\ \bibinfo {pages}
  {3817} (\bibinfo {year} {2004})},\ \Eprint
  {http://arxiv.org/abs/gr-qc/0403063} {arXiv:gr-qc/0403063 [gr-qc]}
  \BibitemShut {NoStop}%
\bibitem [{\citenamefont {Eardley}(1975)}]{eardley1975observable}%
  \BibitemOpen
  \bibfield  {author} {\bibinfo {author} {\bibfnamefont {D.~M.}\ \bibnamefont
  {Eardley}},\ }\bibfield  {title} {\enquote {\bibinfo {title} {Observable
  effects of a scalar gravitational field in a binary pulsar},}\ }\href@noop {}
  {\bibfield  {journal} {\bibinfo  {journal} {The Astrophysical Journal}\
  }\textbf {\bibinfo {volume} {196}},\ \bibinfo {pages} {L59--L62} (\bibinfo
  {year} {1975})}\BibitemShut {NoStop}%
\bibitem [{\citenamefont {Damour}\ and\ \citenamefont
  {Esposito-Farese}(1992)}]{Damour:1992we}%
  \BibitemOpen
  \bibfield  {author} {\bibinfo {author} {\bibfnamefont {T.}~\bibnamefont
  {Damour}}\ and\ \bibinfo {author} {\bibfnamefont {G.}~\bibnamefont
  {Esposito-Farese}},\ }\bibfield  {title} {\enquote {\bibinfo {title} {{Tensor
  multiscalar theories of gravitation}},}\ }\href {\doibase
  10.1088/0264-9381/9/9/015} {\bibfield  {journal} {\bibinfo  {journal} {Class.
  Quant. Grav.}\ }\textbf {\bibinfo {volume} {9}},\ \bibinfo {pages}
  {2093--2176} (\bibinfo {year} {1992})}\BibitemShut {NoStop}%
\bibitem [{\citenamefont {Taracchini}\ \emph {et~al.}(2014)\citenamefont
  {Taracchini}, \citenamefont {Buonanno}, \citenamefont {Khanna},\ and\
  \citenamefont {Hughes}}]{Taracchini:2014zpa}%
  \BibitemOpen
  \bibfield  {author} {\bibinfo {author} {\bibfnamefont {A.}~\bibnamefont
  {Taracchini}}, \bibinfo {author} {\bibfnamefont {A.}~\bibnamefont
  {Buonanno}}, \bibinfo {author} {\bibfnamefont {G.}~\bibnamefont {Khanna}}, \
  and\ \bibinfo {author} {\bibfnamefont {S.~A.}\ \bibnamefont {Hughes}},\
  }\bibfield  {title} {\enquote {\bibinfo {title} {{Small mass plunging into a
  Kerr black hole: Anatomy of the inspiral-merger-ringdown waveforms}},}\
  }\href {\doibase 10.1103/PhysRevD.90.084025} {\bibfield  {journal} {\bibinfo
  {journal} {Phys. Rev.}\ }\textbf {\bibinfo {volume} {D90}},\ \bibinfo {pages}
  {084025} (\bibinfo {year} {2014})},\ \Eprint {http://arxiv.org/abs/1404.1819}
  {arXiv:1404.1819 [gr-qc]} \BibitemShut {NoStop}%
\bibitem [{\citenamefont {Barausse}\ \emph {et~al.}(2013)\citenamefont
  {Barausse}, \citenamefont {Palenzuela}, \citenamefont {Ponce},\ and\
  \citenamefont {Lehner}}]{Barausse:2012da}%
  \BibitemOpen
  \bibfield  {author} {\bibinfo {author} {\bibfnamefont {E.}~\bibnamefont
  {Barausse}}, \bibinfo {author} {\bibfnamefont {C.}~\bibnamefont
  {Palenzuela}}, \bibinfo {author} {\bibfnamefont {M.}~\bibnamefont {Ponce}}, \
  and\ \bibinfo {author} {\bibfnamefont {L.}~\bibnamefont {Lehner}},\
  }\bibfield  {title} {\enquote {\bibinfo {title} {{Neutron-star mergers in
  scalar-tensor theories of gravity}},}\ }\href {\doibase
  10.1103/PhysRevD.87.081506} {\bibfield  {journal} {\bibinfo  {journal} {Phys.
  Rev.}\ }\textbf {\bibinfo {volume} {D87}},\ \bibinfo {pages} {081506}
  (\bibinfo {year} {2013})},\ \Eprint {http://arxiv.org/abs/1212.5053}
  {arXiv:1212.5053 [gr-qc]} \BibitemShut {NoStop}%
\bibitem [{\citenamefont {Shibata}\ \emph {et~al.}(2014)\citenamefont
  {Shibata}, \citenamefont {Taniguchi}, \citenamefont {Okawa},\ and\
  \citenamefont {Buonanno}}]{Shibata:2013pra}%
  \BibitemOpen
  \bibfield  {author} {\bibinfo {author} {\bibfnamefont {M.}~\bibnamefont
  {Shibata}}, \bibinfo {author} {\bibfnamefont {K.}~\bibnamefont {Taniguchi}},
  \bibinfo {author} {\bibfnamefont {H.}~\bibnamefont {Okawa}}, \ and\ \bibinfo
  {author} {\bibfnamefont {A.}~\bibnamefont {Buonanno}},\ }\bibfield  {title}
  {\enquote {\bibinfo {title} {{Coalescence of binary neutron stars in a
  scalar-tensor theory of gravity}},}\ }\href {\doibase
  10.1103/PhysRevD.89.084005} {\bibfield  {journal} {\bibinfo  {journal} {Phys.
  Rev.}\ }\textbf {\bibinfo {volume} {D89}},\ \bibinfo {pages} {084005}
  (\bibinfo {year} {2014})},\ \Eprint {http://arxiv.org/abs/1310.0627}
  {arXiv:1310.0627 [gr-qc]} \BibitemShut {NoStop}%
\bibitem [{\citenamefont {Palenzuela}\ \emph {et~al.}(2014)\citenamefont
  {Palenzuela}, \citenamefont {Barausse}, \citenamefont {Ponce},\ and\
  \citenamefont {Lehner}}]{Palenzuela:2013hsa}%
  \BibitemOpen
  \bibfield  {author} {\bibinfo {author} {\bibfnamefont {C.}~\bibnamefont
  {Palenzuela}}, \bibinfo {author} {\bibfnamefont {E.}~\bibnamefont
  {Barausse}}, \bibinfo {author} {\bibfnamefont {M.}~\bibnamefont {Ponce}}, \
  and\ \bibinfo {author} {\bibfnamefont {L.}~\bibnamefont {Lehner}},\
  }\bibfield  {title} {\enquote {\bibinfo {title} {{Dynamical scalarization of
  neutron stars in scalar-tensor gravity theories}},}\ }\href {\doibase
  10.1103/PhysRevD.89.044024} {\bibfield  {journal} {\bibinfo  {journal} {Phys.
  Rev.}\ }\textbf {\bibinfo {volume} {D89}},\ \bibinfo {pages} {044024}
  (\bibinfo {year} {2014})},\ \Eprint {http://arxiv.org/abs/1310.4481}
  {arXiv:1310.4481 [gr-qc]} \BibitemShut {NoStop}%
\bibitem [{\citenamefont {Taniguchi}\ \emph {et~al.}(2015)\citenamefont
  {Taniguchi}, \citenamefont {Shibata},\ and\ \citenamefont
  {Buonanno}}]{Taniguchi:2014fqa}%
  \BibitemOpen
  \bibfield  {author} {\bibinfo {author} {\bibfnamefont {K.}~\bibnamefont
  {Taniguchi}}, \bibinfo {author} {\bibfnamefont {M.}~\bibnamefont {Shibata}},
  \ and\ \bibinfo {author} {\bibfnamefont {A.}~\bibnamefont {Buonanno}},\
  }\bibfield  {title} {\enquote {\bibinfo {title} {{Quasiequilibrium sequences
  of binary neutron stars undergoing dynamical scalarization}},}\ }\href
  {\doibase 10.1103/PhysRevD.91.024033} {\bibfield  {journal} {\bibinfo
  {journal} {Phys. Rev.}\ }\textbf {\bibinfo {volume} {D91}},\ \bibinfo {pages}
  {024033} (\bibinfo {year} {2015})},\ \Eprint {http://arxiv.org/abs/1410.0738}
  {arXiv:1410.0738 [gr-qc]} \BibitemShut {NoStop}%
\bibitem [{\citenamefont {Sennett}\ and\ \citenamefont
  {Buonanno}(2016)}]{Sennett:2016rwa}%
  \BibitemOpen
  \bibfield  {author} {\bibinfo {author} {\bibfnamefont {N.}~\bibnamefont
  {Sennett}}\ and\ \bibinfo {author} {\bibfnamefont {A.}~\bibnamefont
  {Buonanno}},\ }\bibfield  {title} {\enquote {\bibinfo {title} {{Modeling
  dynamical scalarization with a resummed post-Newtonian expansion}},}\ }\href
  {\doibase 10.1103/PhysRevD.93.124004} {\bibfield  {journal} {\bibinfo
  {journal} {Phys. Rev.}\ }\textbf {\bibinfo {volume} {D93}},\ \bibinfo {pages}
  {124004} (\bibinfo {year} {2016})},\ \Eprint
  {http://arxiv.org/abs/1603.03300} {arXiv:1603.03300 [gr-qc]} \BibitemShut
  {NoStop}%
\bibitem [{\citenamefont {Jordan}(1955)}]{Jordan:1955}%
  \BibitemOpen
  \bibfield  {author} {\bibinfo {author} {\bibfnamefont {P.}~\bibnamefont
  {Jordan}},\ }\href@noop {} {\emph {\bibinfo {title} {Schwerkraft und
  Weltall}}}\ (\bibinfo  {publisher} {F. Vieweg},\ \bibinfo {address}
  {Braunschweig},\ \bibinfo {year} {1955})\BibitemShut {NoStop}%
\bibitem [{\citenamefont {Fierz}(1956)}]{Fierz:1956zz}%
  \BibitemOpen
  \bibfield  {author} {\bibinfo {author} {\bibfnamefont {M.}~\bibnamefont
  {Fierz}},\ }\bibfield  {title} {\enquote {\bibinfo {title} {{On the physical
  interpretation of P.Jordan's extended theory of gravitation}},}\ }\href@noop
  {} {\bibfield  {journal} {\bibinfo  {journal} {Helv. Phys. Acta}\ }\textbf
  {\bibinfo {volume} {29}},\ \bibinfo {pages} {128--134} (\bibinfo {year}
  {1956})}\BibitemShut {NoStop}%
\bibitem [{\citenamefont {Brans}\ and\ \citenamefont
  {Dicke}(1961)}]{Brans:1961sx}%
  \BibitemOpen
  \bibfield  {author} {\bibinfo {author} {\bibfnamefont {C.}~\bibnamefont
  {Brans}}\ and\ \bibinfo {author} {\bibfnamefont {R.~H.}\ \bibnamefont
  {Dicke}},\ }\bibfield  {title} {\enquote {\bibinfo {title} {{Mach's principle
  and a relativistic theory of gravitation}},}\ }\href {\doibase
  10.1103/PhysRev.124.925} {\bibfield  {journal} {\bibinfo  {journal} {Phys.
  Rev.}\ }\textbf {\bibinfo {volume} {124}},\ \bibinfo {pages} {925--935}
  (\bibinfo {year} {1961})}\BibitemShut {NoStop}%
\bibitem [{\citenamefont {Damour}\ and\ \citenamefont
  {Esposito-Farese}(1993)}]{Damour:1993hw}%
  \BibitemOpen
  \bibfield  {author} {\bibinfo {author} {\bibfnamefont {T.}~\bibnamefont
  {Damour}}\ and\ \bibinfo {author} {\bibfnamefont {G.}~\bibnamefont
  {Esposito-Farese}},\ }\bibfield  {title} {\enquote {\bibinfo {title}
  {{Nonperturbative strong field effects in tensor - scalar theories of
  gravitation}},}\ }\href {\doibase 10.1103/PhysRevLett.70.2220} {\bibfield
  {journal} {\bibinfo  {journal} {Phys. Rev. Lett.}\ }\textbf {\bibinfo
  {volume} {70}},\ \bibinfo {pages} {2220--2223} (\bibinfo {year}
  {1993})}\BibitemShut {NoStop}%
\bibitem [{\citenamefont {Read}\ \emph {et~al.}(2009)\citenamefont {Read},
  \citenamefont {Lackey}, \citenamefont {Owen},\ and\ \citenamefont
  {Friedman}}]{Read:2008iy}%
  \BibitemOpen
  \bibfield  {author} {\bibinfo {author} {\bibfnamefont {J.~S.}\ \bibnamefont
  {Read}}, \bibinfo {author} {\bibfnamefont {B.~D.}\ \bibnamefont {Lackey}},
  \bibinfo {author} {\bibfnamefont {B.~J.}\ \bibnamefont {Owen}}, \ and\
  \bibinfo {author} {\bibfnamefont {J.~L.}\ \bibnamefont {Friedman}},\
  }\bibfield  {title} {\enquote {\bibinfo {title} {{Constraints on a
  phenomenologically parameterized neutron-star equation of state}},}\ }\href
  {\doibase 10.1103/PhysRevD.79.124032} {\bibfield  {journal} {\bibinfo
  {journal} {Phys. Rev.}\ }\textbf {\bibinfo {volume} {D79}},\ \bibinfo {pages}
  {124032} (\bibinfo {year} {2009})},\ \Eprint {http://arxiv.org/abs/0812.2163}
  {arXiv:0812.2163 [astro-ph]} \BibitemShut {NoStop}%
\bibitem [{\citenamefont {Blanchet}(2014)}]{Blanchet:2013haa}%
  \BibitemOpen
  \bibfield  {author} {\bibinfo {author} {\bibfnamefont {L.}~\bibnamefont
  {Blanchet}},\ }\bibfield  {title} {\enquote {\bibinfo {title} {{Gravitational
  Radiation from Post-Newtonian Sources and Inspiralling Compact Binaries}},}\
  }\href {\doibase 10.12942/lrr-2014-2} {\bibfield  {journal} {\bibinfo
  {journal} {Living Rev. Rel.}\ }\textbf {\bibinfo {volume} {17}},\ \bibinfo
  {pages} {2} (\bibinfo {year} {2014})},\ \Eprint
  {http://arxiv.org/abs/1310.1528} {arXiv:1310.1528 [gr-qc]} \BibitemShut
  {NoStop}%
\bibitem [{\citenamefont {Fokker}(1929)}]{Fokker:1929}%
  \BibitemOpen
  \bibfield  {author} {\bibinfo {author} {\bibfnamefont {A.~D.}\ \bibnamefont
  {Fokker}},\ }\bibfield  {title} {\enquote {\bibinfo {title} {{Ein invarianter
  Variationssatz für die Bewegung mehrerer elektrischer Massenteilchen}},}\
  }\href {\doibase 10.1007/BF01340389} {\bibfield  {journal} {\bibinfo
  {journal} {Z. Phys.}\ }\textbf {\bibinfo {volume} {58}},\ \bibinfo {pages}
  {386--393} (\bibinfo {year} {1929})}\BibitemShut {NoStop}%
\bibitem [{\citenamefont {Damour}\ and\ \citenamefont
  {Esposito-Farese}(1996)}]{Damour:1995kt}%
  \BibitemOpen
  \bibfield  {author} {\bibinfo {author} {\bibfnamefont {T.}~\bibnamefont
  {Damour}}\ and\ \bibinfo {author} {\bibfnamefont {G.}~\bibnamefont
  {Esposito-Farese}},\ }\bibfield  {title} {\enquote {\bibinfo {title}
  {{Testing gravity to second postNewtonian order: A Field theory approach}},}\
  }\href {\doibase 10.1103/PhysRevD.53.5541} {\bibfield  {journal} {\bibinfo
  {journal} {Phys. Rev.}\ }\textbf {\bibinfo {volume} {D53}},\ \bibinfo {pages}
  {5541--5578} (\bibinfo {year} {1996})},\ \Eprint
  {http://arxiv.org/abs/gr-qc/9506063} {arXiv:gr-qc/9506063 [gr-qc]}
  \BibitemShut {NoStop}%
\bibitem [{\citenamefont {Bernard}\ \emph {et~al.}(2016)\citenamefont
  {Bernard}, \citenamefont {Blanchet}, \citenamefont {Bohé}, \citenamefont
  {Faye},\ and\ \citenamefont {Marsat}}]{Bernard:2015njp}%
  \BibitemOpen
  \bibfield  {author} {\bibinfo {author} {\bibfnamefont {L.}~\bibnamefont
  {Bernard}}, \bibinfo {author} {\bibfnamefont {L.}~\bibnamefont {Blanchet}},
  \bibinfo {author} {\bibfnamefont {A.}~\bibnamefont {Bohé}}, \bibinfo
  {author} {\bibfnamefont {G.}~\bibnamefont {Faye}}, \ and\ \bibinfo {author}
  {\bibfnamefont {S.}~\bibnamefont {Marsat}},\ }\bibfield  {title} {\enquote
  {\bibinfo {title} {{Fokker action of nonspinning compact binaries at the
  fourth post-Newtonian approximation}},}\ }\href {\doibase
  10.1103/PhysRevD.93.084037} {\bibfield  {journal} {\bibinfo  {journal} {Phys.
  Rev.}\ }\textbf {\bibinfo {volume} {D93}},\ \bibinfo {pages} {084037}
  (\bibinfo {year} {2016})},\ \Eprint {http://arxiv.org/abs/1512.02876}
  {arXiv:1512.02876 [gr-qc]} \BibitemShut {NoStop}%
\bibitem [{\citenamefont {Bernard}(2018)}]{Bernard:2018hta}%
  \BibitemOpen
  \bibfield  {author} {\bibinfo {author} {\bibfnamefont {L.}~\bibnamefont
  {Bernard}},\ }\bibfield  {title} {\enquote {\bibinfo {title} {{Dynamics of
  compact binary systems in scalar-tensor theories: I. Equations of motion to
  the third post-Newtonian order}},}\ }\href@noop {} {\  (\bibinfo {year}
  {2018})},\ \Eprint {http://arxiv.org/abs/1802.10201} {arXiv:1802.10201
  [gr-qc]} \BibitemShut {NoStop}%
\bibitem [{\citenamefont {Blanchet}\ and\ \citenamefont
  {Iyer}(2003)}]{Blanchet:2002mb}%
  \BibitemOpen
  \bibfield  {author} {\bibinfo {author} {\bibfnamefont {L.}~\bibnamefont
  {Blanchet}}\ and\ \bibinfo {author} {\bibfnamefont {B.~R.}\ \bibnamefont
  {Iyer}},\ }\bibfield  {title} {\enquote {\bibinfo {title} {{Third
  postNewtonian dynamics of compact binaries: Equations of motion in the
  center-of-mass frame}},}\ }\href {\doibase 10.1088/0264-9381/20/4/309}
  {\bibfield  {journal} {\bibinfo  {journal} {Class. Quant. Grav.}\ }\textbf
  {\bibinfo {volume} {20}},\ \bibinfo {pages} {755} (\bibinfo {year} {2003})},\
  \Eprint {http://arxiv.org/abs/gr-qc/0209089} {arXiv:gr-qc/0209089 [gr-qc]}
  \BibitemShut {NoStop}%
\bibitem [{\citenamefont {Yunes}\ and\ \citenamefont
  {Hughes}(2010)}]{Yunes:2010qb}%
  \BibitemOpen
  \bibfield  {author} {\bibinfo {author} {\bibfnamefont {N.}~\bibnamefont
  {Yunes}}\ and\ \bibinfo {author} {\bibfnamefont {S.~A.}\ \bibnamefont
  {Hughes}},\ }\bibfield  {title} {\enquote {\bibinfo {title} {{Binary Pulsar
  Constraints on the Parameterized post-Einsteinian Framework}},}\ }\href
  {\doibase 10.1103/PhysRevD.82.082002} {\bibfield  {journal} {\bibinfo
  {journal} {Phys. Rev.}\ }\textbf {\bibinfo {volume} {D82}},\ \bibinfo {pages}
  {082002} (\bibinfo {year} {2010})},\ \Eprint {http://arxiv.org/abs/1007.1995}
  {arXiv:1007.1995 [gr-qc]} \BibitemShut {NoStop}%
\bibitem [{\citenamefont {Barausse}\ \emph {et~al.}(2016)\citenamefont
  {Barausse}, \citenamefont {Yunes},\ and\ \citenamefont
  {Chamberlain}}]{Barausse:2016eii}%
  \BibitemOpen
  \bibfield  {author} {\bibinfo {author} {\bibfnamefont {E.}~\bibnamefont
  {Barausse}}, \bibinfo {author} {\bibfnamefont {N.}~\bibnamefont {Yunes}}, \
  and\ \bibinfo {author} {\bibfnamefont {K.}~\bibnamefont {Chamberlain}},\
  }\bibfield  {title} {\enquote {\bibinfo {title} {{Theory-Agnostic Constraints
  on Black-Hole Dipole Radiation with Multiband Gravitational-Wave
  Astrophysics}},}\ }\href {\doibase 10.1103/PhysRevLett.116.241104} {\bibfield
   {journal} {\bibinfo  {journal} {Phys. Rev. Lett.}\ }\textbf {\bibinfo
  {volume} {116}},\ \bibinfo {pages} {241104} (\bibinfo {year} {2016})},\
  \Eprint {http://arxiv.org/abs/1603.04075} {arXiv:1603.04075 [gr-qc]}
  \BibitemShut {NoStop}%
\bibitem [{\citenamefont {Sennett}\ \emph {et~al.}(2016)\citenamefont
  {Sennett}, \citenamefont {Marsat},\ and\ \citenamefont
  {Buonanno}}]{Sennett:2016klh}%
  \BibitemOpen
  \bibfield  {author} {\bibinfo {author} {\bibfnamefont {N.}~\bibnamefont
  {Sennett}}, \bibinfo {author} {\bibfnamefont {S.}~\bibnamefont {Marsat}}, \
  and\ \bibinfo {author} {\bibfnamefont {A.}~\bibnamefont {Buonanno}},\
  }\bibfield  {title} {\enquote {\bibinfo {title} {{Gravitational waveforms in
  scalar-tensor gravity at 2PN relative order}},}\ }\href {\doibase
  10.1103/PhysRevD.94.084003} {\bibfield  {journal} {\bibinfo  {journal} {Phys.
  Rev.}\ }\textbf {\bibinfo {volume} {D94}},\ \bibinfo {pages} {084003}
  (\bibinfo {year} {2016})},\ \Eprint {http://arxiv.org/abs/1607.01420}
  {arXiv:1607.01420 [gr-qc]} \BibitemShut {NoStop}%
\bibitem [{\citenamefont {Del~Pozzo}\ \emph {et~al.}(2011)\citenamefont
  {Del~Pozzo}, \citenamefont {Veitch},\ and\ \citenamefont
  {Vecchio}}]{DelPozzo:2011pg}%
  \BibitemOpen
  \bibfield  {author} {\bibinfo {author} {\bibfnamefont {W.}~\bibnamefont
  {Del~Pozzo}}, \bibinfo {author} {\bibfnamefont {J.}~\bibnamefont {Veitch}}, \
  and\ \bibinfo {author} {\bibfnamefont {A.}~\bibnamefont {Vecchio}},\
  }\bibfield  {title} {\enquote {\bibinfo {title} {{Testing General Relativity
  using Bayesian model selection: Applications to observations of gravitational
  waves from compact binary systems}},}\ }\href {\doibase
  10.1103/PhysRevD.83.082002} {\bibfield  {journal} {\bibinfo  {journal} {Phys.
  Rev.}\ }\textbf {\bibinfo {volume} {D83}},\ \bibinfo {pages} {082002}
  (\bibinfo {year} {2011})},\ \Eprint {http://arxiv.org/abs/1101.1391}
  {arXiv:1101.1391 [gr-qc]} \BibitemShut {NoStop}%
\bibitem [{\citenamefont {Cornish}\ \emph {et~al.}(2011)\citenamefont
  {Cornish}, \citenamefont {Sampson}, \citenamefont {Yunes},\ and\
  \citenamefont {Pretorius}}]{Cornish:2011ys}%
  \BibitemOpen
  \bibfield  {author} {\bibinfo {author} {\bibfnamefont {N.}~\bibnamefont
  {Cornish}}, \bibinfo {author} {\bibfnamefont {L.}~\bibnamefont {Sampson}},
  \bibinfo {author} {\bibfnamefont {N.}~\bibnamefont {Yunes}}, \ and\ \bibinfo
  {author} {\bibfnamefont {F.}~\bibnamefont {Pretorius}},\ }\bibfield  {title}
  {\enquote {\bibinfo {title} {{Gravitational Wave Tests of General Relativity
  with the Parameterized Post-Einsteinian Framework}},}\ }\href {\doibase
  10.1103/PhysRevD.84.062003} {\bibfield  {journal} {\bibinfo  {journal} {Phys.
  Rev.}\ }\textbf {\bibinfo {volume} {D84}},\ \bibinfo {pages} {062003}
  (\bibinfo {year} {2011})},\ \Eprint {http://arxiv.org/abs/1105.2088}
  {arXiv:1105.2088 [gr-qc]} \BibitemShut {NoStop}%
\bibitem [{\citenamefont {Gossan}\ \emph {et~al.}(2012)\citenamefont {Gossan},
  \citenamefont {Veitch},\ and\ \citenamefont {Sathyaprakash}}]{Gossan:2011ha}%
  \BibitemOpen
  \bibfield  {author} {\bibinfo {author} {\bibfnamefont {S.}~\bibnamefont
  {Gossan}}, \bibinfo {author} {\bibfnamefont {J.}~\bibnamefont {Veitch}}, \
  and\ \bibinfo {author} {\bibfnamefont {B.~S.}\ \bibnamefont
  {Sathyaprakash}},\ }\bibfield  {title} {\enquote {\bibinfo {title} {{Bayesian
  model selection for testing the no-hair theorem with black hole
  ringdowns}},}\ }\href {\doibase 10.1103/PhysRevD.85.124056} {\bibfield
  {journal} {\bibinfo  {journal} {Phys. Rev.}\ }\textbf {\bibinfo {volume}
  {D85}},\ \bibinfo {pages} {124056} (\bibinfo {year} {2012})},\ \Eprint
  {http://arxiv.org/abs/1111.5819} {arXiv:1111.5819 [gr-qc]} \BibitemShut
  {NoStop}%
\bibitem [{\citenamefont {Sampson}\ \emph {et~al.}(2014)\citenamefont
  {Sampson}, \citenamefont {Yunes}, \citenamefont {Cornish}, \citenamefont
  {Ponce}, \citenamefont {Barausse}, \citenamefont {Klein}, \citenamefont
  {Palenzuela},\ and\ \citenamefont {Lehner}}]{Sampson:2014qqa}%
  \BibitemOpen
  \bibfield  {author} {\bibinfo {author} {\bibfnamefont {L.}~\bibnamefont
  {Sampson}}, \bibinfo {author} {\bibfnamefont {N.}~\bibnamefont {Yunes}},
  \bibinfo {author} {\bibfnamefont {N.}~\bibnamefont {Cornish}}, \bibinfo
  {author} {\bibfnamefont {M.}~\bibnamefont {Ponce}}, \bibinfo {author}
  {\bibfnamefont {E.}~\bibnamefont {Barausse}}, \bibinfo {author}
  {\bibfnamefont {A.}~\bibnamefont {Klein}}, \bibinfo {author} {\bibfnamefont
  {C.}~\bibnamefont {Palenzuela}}, \ and\ \bibinfo {author} {\bibfnamefont
  {L.}~\bibnamefont {Lehner}},\ }\bibfield  {title} {\enquote {\bibinfo {title}
  {{Projected Constraints on Scalarization with Gravitational Waves from
  Neutron Star Binaries}},}\ }\href {\doibase 10.1103/PhysRevD.90.124091}
  {\bibfield  {journal} {\bibinfo  {journal} {Phys. Rev.}\ }\textbf {\bibinfo
  {volume} {D90}},\ \bibinfo {pages} {124091} (\bibinfo {year} {2014})},\
  \Eprint {http://arxiv.org/abs/1407.7038} {arXiv:1407.7038 [gr-qc]}
  \BibitemShut {NoStop}%
\bibitem [{\citenamefont {Damour}\ \emph {et~al.}(2000)\citenamefont {Damour},
  \citenamefont {Iyer},\ and\ \citenamefont {Sathyaprakash}}]{Damour:2000gg}%
  \BibitemOpen
  \bibfield  {author} {\bibinfo {author} {\bibfnamefont {T.}~\bibnamefont
  {Damour}}, \bibinfo {author} {\bibfnamefont {B.~R.}\ \bibnamefont {Iyer}}, \
  and\ \bibinfo {author} {\bibfnamefont {B.~S.}\ \bibnamefont
  {Sathyaprakash}},\ }\bibfield  {title} {\enquote {\bibinfo {title}
  {{Frequency domain P approximant filters for time truncated inspiral
  gravitational wave signals from compact binaries}},}\ }\href {\doibase
  10.1103/PhysRevD.62.084036} {\bibfield  {journal} {\bibinfo  {journal} {Phys.
  Rev.}\ }\textbf {\bibinfo {volume} {D62}},\ \bibinfo {pages} {084036}
  (\bibinfo {year} {2000})},\ \Eprint {http://arxiv.org/abs/gr-qc/0001023}
  {arXiv:gr-qc/0001023 [gr-qc]} \BibitemShut {NoStop}%
\bibitem [{LIG()}]{LIGOsensitivity}%
  \BibitemOpen
  \href@noop {} {\enquote {\bibinfo {title} {{LIGO-T1800042-v5: Updated
  Advanced LIGO sensitivity design curve}},}\ }\bibinfo {howpublished}
  {\url{https://dcc.ligo.org/LIGO-T1800044/public}}\BibitemShut {NoStop}%
\bibitem [{\citenamefont {Coquereaux}\ and\ \citenamefont
  {Esposito-Farese}(1990)}]{Coquereaux:1990qs}%
  \BibitemOpen
  \bibfield  {author} {\bibinfo {author} {\bibfnamefont {R.}~\bibnamefont
  {Coquereaux}}\ and\ \bibinfo {author} {\bibfnamefont {G.}~\bibnamefont
  {Esposito-Farese}},\ }\bibfield  {title} {\enquote {\bibinfo {title} {{The
  Theory of Kaluza-Klein-Jordan-Thiry revisited}},}\ }\href@noop {} {\bibfield
  {journal} {\bibinfo  {journal} {Ann. Inst. H. Poincare Phys. Theor.}\
  }\textbf {\bibinfo {volume} {52}},\ \bibinfo {pages} {113--150} (\bibinfo
  {year} {1990})}\BibitemShut {NoStop}%
\bibitem [{\citenamefont {Julié}(2018{\natexlab{b}})}]{Julie:2017ucp}%
  \BibitemOpen
  \bibfield  {author} {\bibinfo {author} {\bibfnamefont {F.-L.}\ \bibnamefont
  {Julié}},\ }\bibfield  {title} {\enquote {\bibinfo {title} {{Reducing the
  two-body problem in scalar-tensor theories to the motion of a test particle :
  a scalar-tensor effective-one-body approach}},}\ }\href {\doibase
  10.1103/PhysRevD.97.024047} {\bibfield  {journal} {\bibinfo  {journal} {Phys.
  Rev.}\ }\textbf {\bibinfo {volume} {D97}},\ \bibinfo {pages} {024047}
  (\bibinfo {year} {2018}{\natexlab{b}})},\ \Eprint
  {http://arxiv.org/abs/1709.09742} {arXiv:1709.09742 [gr-qc]} \BibitemShut
  {NoStop}%
\bibitem [{\citenamefont {Damour}\ and\ \citenamefont
  {Nagar}(2010)}]{Damour:2009wj}%
  \BibitemOpen
  \bibfield  {author} {\bibinfo {author} {\bibfnamefont {T.}~\bibnamefont
  {Damour}}\ and\ \bibinfo {author} {\bibfnamefont {A.}~\bibnamefont {Nagar}},\
  }\bibfield  {title} {\enquote {\bibinfo {title} {{Effective One Body
  description of tidal effects in inspiralling compact binaries}},}\ }\href
  {\doibase 10.1103/PhysRevD.81.084016} {\bibfield  {journal} {\bibinfo
  {journal} {Phys. Rev.}\ }\textbf {\bibinfo {volume} {D81}},\ \bibinfo {pages}
  {084016} (\bibinfo {year} {2010})},\ \Eprint {http://arxiv.org/abs/0911.5041}
  {arXiv:0911.5041 [gr-qc]} \BibitemShut {NoStop}%
\bibitem [{\citenamefont {Julié}\ and\ \citenamefont
  {Deruelle}(2017)}]{Julie:2017pkb}%
  \BibitemOpen
  \bibfield  {author} {\bibinfo {author} {\bibfnamefont {F.-L.}\ \bibnamefont
  {Julié}}\ and\ \bibinfo {author} {\bibfnamefont {N.}~\bibnamefont
  {Deruelle}},\ }\bibfield  {title} {\enquote {\bibinfo {title} {{Two-body
  problem in Scalar-Tensor theories as a deformation of General Relativity : an
  Effective-One-Body approach}},}\ }\href {\doibase 10.1103/PhysRevD.95.124054}
  {\bibfield  {journal} {\bibinfo  {journal} {Phys. Rev.}\ }\textbf {\bibinfo
  {volume} {D95}},\ \bibinfo {pages} {124054} (\bibinfo {year} {2017})},\
  \Eprint {http://arxiv.org/abs/1703.05360} {arXiv:1703.05360 [gr-qc]}
  \BibitemShut {NoStop}%
\bibitem [{\citenamefont {Damour}\ \emph {et~al.}(2015)\citenamefont {Damour},
  \citenamefont {Jaranowski},\ and\ \citenamefont {Schäfer}}]{Damour:2015isa}%
  \BibitemOpen
  \bibfield  {author} {\bibinfo {author} {\bibfnamefont {T.}~\bibnamefont
  {Damour}}, \bibinfo {author} {\bibfnamefont {P.}~\bibnamefont {Jaranowski}},
  \ and\ \bibinfo {author} {\bibfnamefont {G.}~\bibnamefont {Schäfer}},\
  }\bibfield  {title} {\enquote {\bibinfo {title} {{Fourth post-Newtonian
  effective one-body dynamics}},}\ }\href {\doibase 10.1103/PhysRevD.91.084024}
  {\bibfield  {journal} {\bibinfo  {journal} {Phys. Rev.}\ }\textbf {\bibinfo
  {volume} {D91}},\ \bibinfo {pages} {084024} (\bibinfo {year} {2015})},\
  \Eprint {http://arxiv.org/abs/1502.07245} {arXiv:1502.07245 [gr-qc]}
  \BibitemShut {NoStop}%
\bibitem [{\citenamefont {Buonanno}(2000)}]{Buonanno:2000qq}%
  \BibitemOpen
  \bibfield  {author} {\bibinfo {author} {\bibfnamefont {A.}~\bibnamefont
  {Buonanno}},\ }\bibfield  {title} {\enquote {\bibinfo {title} {{Reduction of
  the two-body dynamics to a one-body description in classical
  electrodynamics}},}\ }\href {\doibase 10.1103/PhysRevD.62.104022} {\bibfield
  {journal} {\bibinfo  {journal} {Phys. Rev.}\ }\textbf {\bibinfo {volume}
  {D62}},\ \bibinfo {pages} {104022} (\bibinfo {year} {2000})},\ \Eprint
  {http://arxiv.org/abs/hep-th/0004042} {arXiv:hep-th/0004042 [hep-th]}
  \BibitemShut {NoStop}%
\bibitem [{\citenamefont {Damour}(2016)}]{Damour:2016gwp}%
  \BibitemOpen
  \bibfield  {author} {\bibinfo {author} {\bibfnamefont {T.}~\bibnamefont
  {Damour}},\ }\bibfield  {title} {\enquote {\bibinfo {title} {{Gravitational
  scattering, post-Minkowskian approximation and Effective One-Body theory}},}\
  }\href {\doibase 10.1103/PhysRevD.94.104015} {\bibfield  {journal} {\bibinfo
  {journal} {Phys. Rev.}\ }\textbf {\bibinfo {volume} {D94}},\ \bibinfo {pages}
  {104015} (\bibinfo {year} {2016})},\ \Eprint
  {http://arxiv.org/abs/1609.00354} {arXiv:1609.00354 [gr-qc]} \BibitemShut
  {NoStop}%
\bibitem [{\citenamefont {Vines}(2018)}]{Vines:2017hyw}%
  \BibitemOpen
  \bibfield  {author} {\bibinfo {author} {\bibfnamefont {J.}~\bibnamefont
  {Vines}},\ }\bibfield  {title} {\enquote {\bibinfo {title} {{Scattering of
  two spinning black holes in post-Minkowskian gravity, to all orders in spin,
  and effective-one-body mappings}},}\ }\href {\doibase
  10.1088/1361-6382/aaa3a8} {\bibfield  {journal} {\bibinfo  {journal} {Class.
  Quant. Grav.}\ }\textbf {\bibinfo {volume} {35}},\ \bibinfo {pages} {084002}
  (\bibinfo {year} {2018})},\ \Eprint {http://arxiv.org/abs/1709.06016}
  {arXiv:1709.06016 [gr-qc]} \BibitemShut {NoStop}%
\bibitem [{\citenamefont {Pan}\ \emph {et~al.}(2010)\citenamefont {Pan},
  \citenamefont {Buonanno}, \citenamefont {Buchman}, \citenamefont {Chu},
  \citenamefont {Kidder}, \citenamefont {Pfeiffer},\ and\ \citenamefont
  {Scheel}}]{Pan:2009wj}%
  \BibitemOpen
  \bibfield  {author} {\bibinfo {author} {\bibfnamefont {Y.}~\bibnamefont
  {Pan}}, \bibinfo {author} {\bibfnamefont {A.}~\bibnamefont {Buonanno}},
  \bibinfo {author} {\bibfnamefont {L.~T.}\ \bibnamefont {Buchman}}, \bibinfo
  {author} {\bibfnamefont {T.}~\bibnamefont {Chu}}, \bibinfo {author}
  {\bibfnamefont {L.~E.}\ \bibnamefont {Kidder}}, \bibinfo {author}
  {\bibfnamefont {H.~P.}\ \bibnamefont {Pfeiffer}}, \ and\ \bibinfo {author}
  {\bibfnamefont {M.~A.}\ \bibnamefont {Scheel}},\ }\bibfield  {title}
  {\enquote {\bibinfo {title} {{Effective-one-body waveforms calibrated to
  numerical relativity simulations: coalescence of non-precessing, spinning,
  equal-mass black holes}},}\ }\href {\doibase 10.1103/PhysRevD.81.084041}
  {\bibfield  {journal} {\bibinfo  {journal} {Phys. Rev.}\ }\textbf {\bibinfo
  {volume} {D81}},\ \bibinfo {pages} {084041} (\bibinfo {year} {2010})},\
  \Eprint {http://arxiv.org/abs/0912.3466} {arXiv:0912.3466 [gr-qc]}
  \BibitemShut {NoStop}%
\bibitem [{\citenamefont {Jai-akson}\ \emph {et~al.}(2017)\citenamefont
  {Jai-akson}, \citenamefont {Chatrabhuti}, \citenamefont {Evnin},\ and\
  \citenamefont {Lehner}}]{Jai-akson:2017ldo}%
  \BibitemOpen
  \bibfield  {author} {\bibinfo {author} {\bibfnamefont {P.}~\bibnamefont
  {Jai-akson}}, \bibinfo {author} {\bibfnamefont {A.}~\bibnamefont
  {Chatrabhuti}}, \bibinfo {author} {\bibfnamefont {O.}~\bibnamefont {Evnin}},
  \ and\ \bibinfo {author} {\bibfnamefont {L.}~\bibnamefont {Lehner}},\
  }\bibfield  {title} {\enquote {\bibinfo {title} {{Black hole merger estimates
  in Einstein-Maxwell and Einstein-Maxwell-dilaton gravity}},}\ }\href
  {\doibase 10.1103/PhysRevD.96.044031} {\bibfield  {journal} {\bibinfo
  {journal} {Phys. Rev.}\ }\textbf {\bibinfo {volume} {D96}},\ \bibinfo {pages}
  {044031} (\bibinfo {year} {2017})},\ \Eprint
  {http://arxiv.org/abs/1706.06519} {arXiv:1706.06519 [gr-qc]} \BibitemShut
  {NoStop}%
\bibitem [{\citenamefont {Ferrari}\ \emph {et~al.}(2001)\citenamefont
  {Ferrari}, \citenamefont {Pauri},\ and\ \citenamefont
  {Piazza}}]{Ferrari:2000ep}%
  \BibitemOpen
  \bibfield  {author} {\bibinfo {author} {\bibfnamefont {V.}~\bibnamefont
  {Ferrari}}, \bibinfo {author} {\bibfnamefont {M.}~\bibnamefont {Pauri}}, \
  and\ \bibinfo {author} {\bibfnamefont {F.}~\bibnamefont {Piazza}},\
  }\bibfield  {title} {\enquote {\bibinfo {title} {{Quasinormal modes of
  charged, dilaton black holes}},}\ }\href {\doibase
  10.1103/PhysRevD.63.064009} {\bibfield  {journal} {\bibinfo  {journal} {Phys.
  Rev.}\ }\textbf {\bibinfo {volume} {D63}},\ \bibinfo {pages} {064009}
  (\bibinfo {year} {2001})},\ \Eprint {http://arxiv.org/abs/gr-qc/0005125}
  {arXiv:gr-qc/0005125 [gr-qc]} \BibitemShut {NoStop}%
\bibitem [{\citenamefont {Damour}\ \emph {et~al.}(1991)\citenamefont {Damour},
  \citenamefont {Soffel},\ and\ \citenamefont {Xu}}]{Damour:1990pi}%
  \BibitemOpen
  \bibfield  {author} {\bibinfo {author} {\bibfnamefont {T.}~\bibnamefont
  {Damour}}, \bibinfo {author} {\bibfnamefont {M.}~\bibnamefont {Soffel}}, \
  and\ \bibinfo {author} {\bibfnamefont {C.}~\bibnamefont {Xu}},\ }\bibfield
  {title} {\enquote {\bibinfo {title} {{General relativistic celestial
  mechanics. 1. Method and definition of reference systems}},}\ }\href
  {\doibase 10.1103/PhysRevD.43.3273} {\bibfield  {journal} {\bibinfo
  {journal} {Phys. Rev.}\ }\textbf {\bibinfo {volume} {D43}},\ \bibinfo {pages}
  {3273--3307} (\bibinfo {year} {1991})}\BibitemShut {NoStop}%
\bibitem [{\citenamefont {Blanchet}\ and\ \citenamefont
  {Damour}(1989)}]{Blanchet:1989ki}%
  \BibitemOpen
  \bibfield  {author} {\bibinfo {author} {\bibfnamefont {L.}~\bibnamefont
  {Blanchet}}\ and\ \bibinfo {author} {\bibfnamefont {T.}~\bibnamefont
  {Damour}},\ }\bibfield  {title} {\enquote {\bibinfo {title} {{Postnewtonian
  Generation of Gravitational Waves}},}\ }\href@noop {} {\bibfield  {journal}
  {\bibinfo  {journal} {Ann. Inst. H. Poincare Phys. Theor.}\ }\textbf
  {\bibinfo {volume} {50}},\ \bibinfo {pages} {377--408} (\bibinfo {year}
  {1989})}\BibitemShut {NoStop}%
\bibitem [{\citenamefont {Blanchet}\ and\ \citenamefont
  {Damour}(1986)}]{blanchet1986radiative}%
  \BibitemOpen
  \bibfield  {author} {\bibinfo {author} {\bibfnamefont {L.}~\bibnamefont
  {Blanchet}}\ and\ \bibinfo {author} {\bibfnamefont {T.}~\bibnamefont
  {Damour}},\ }\bibfield  {title} {\enquote {\bibinfo {title} {Radiative
  gravitational fields in general relativity i. general structure of the field
  outside the source},}\ }\href@noop {} {\bibfield  {journal} {\bibinfo
  {journal} {Phil. Trans. R. Soc. Lond. A}\ }\textbf {\bibinfo {volume}
  {320}},\ \bibinfo {pages} {379--430} (\bibinfo {year} {1986})}\BibitemShut
  {NoStop}%
\bibitem [{\citenamefont {Damour}\ and\ \citenamefont
  {Iyer}(1991)}]{Damour:1990gj}%
  \BibitemOpen
  \bibfield  {author} {\bibinfo {author} {\bibfnamefont {T.}~\bibnamefont
  {Damour}}\ and\ \bibinfo {author} {\bibfnamefont {B.~R.}\ \bibnamefont
  {Iyer}},\ }\bibfield  {title} {\enquote {\bibinfo {title} {{Multipole
  analysis for electromagnetism and linearized gravity with irreducible
  cartesian tensors}},}\ }\href {\doibase 10.1103/PhysRevD.43.3259} {\bibfield
  {journal} {\bibinfo  {journal} {Phys. Rev.}\ }\textbf {\bibinfo {volume}
  {D43}},\ \bibinfo {pages} {3259--3272} (\bibinfo {year} {1991})}\BibitemShut
  {NoStop}%
\bibitem [{\citenamefont {Lang}(2015)}]{Lang:2014osa}%
  \BibitemOpen
  \bibfield  {author} {\bibinfo {author} {\bibfnamefont {R.~N.}\ \bibnamefont
  {Lang}},\ }\bibfield  {title} {\enquote {\bibinfo {title} {{Compact binary
  systems in scalar-tensor gravity. III. Scalar waves and energy flux}},}\
  }\href {\doibase 10.1103/PhysRevD.91.084027} {\bibfield  {journal} {\bibinfo
  {journal} {Phys. Rev.}\ }\textbf {\bibinfo {volume} {D91}},\ \bibinfo {pages}
  {084027} (\bibinfo {year} {2015})},\ \Eprint {http://arxiv.org/abs/1411.3073}
  {arXiv:1411.3073 [gr-qc]} \BibitemShut {NoStop}%
\end{thebibliography}

%

\end{document}